\documentclass[useAMS,usenatbib]{mn2e}

\usepackage{amsmath}
\usepackage{graphicx}
\usepackage{txfonts}
\usepackage{epsfig}
\usepackage{lscape}
\usepackage{yfonts}
\usepackage{natbib}
\usepackage{longtable}
\usepackage{subfigure}
\def\static{\Delta F^{\mbox{\scriptsize stat}}}
\def\rotating{\Delta F^{\mbox{\scriptsize rot}}}

\title[Removal of systematics in photometric data]{Removal of
  systematics in photometric measurements: static and rotating
  illumination corrections in FORS2@VLT data} \author[Coccato et
  al.]{L. Coccato$^{1}$\thanks{E-mail: lcoccato@gmail.com},
  D.M. Bramich$^{1,2}$, W. Freudling$^{1}$, and S. Moehler$^{1}$
  \\ $^{1}$European Southern Observatory,
  Karl-Schwarzschild-Stra$\beta$e 2, D-85748 Garching bei M\"unchen,
  Germany.\\
$^{2}$Qatar Environment and Energy Research Institute, Qatar Foundation, Tornado Tower, Floor 19, P.O. Box 5825, Doha, Qatar.\\}
\begin{document}

\date{Accepted xxx . Received yyy}

\pagerange{\pageref{firstpage}--\pageref{lastpage}} \pubyear{2013}

\maketitle

\label{firstpage}

\begin{abstract}
Images taken with modern detectors require calibration via flat
fielding to obtain the same flux scale across the whole image. One
method for obtaining the best possible flat fielding accuracy is to
derive a photometric model from dithered stellar observations. A large
variety of effects have been taken into account in such modelling.
Recently, \citet{moe2010} discovered systematic variations in
available flat frames for the European Southern Observatory's FORS
instrument that change with the orientation of the projected image on
the sky. The effect on photometry is large compared to other
systematic effects that have already been taken into account.  In this
paper, we present a correction method for this effect: a
generalization of the fitting procedure of \citet{bra2012} to include
a polynomial representation of rotating flat fields. We then applied
the method to the specific case of FORS2 photometric observations of a
series of standard star fields, and provide parametrised solutions
that can be applied by the users.  We found
polynomial coefficients to describe the static and rotating
large-scale systematic flat-field variations across the FORS2 field of
view. Applying these coefficients to FORS2 data, the systematic
changes in the flux scale across FORS2 images can be improved by
$\sim$1\% to $\sim$2\% of the total flux. This represents a
significant improvement in the era of large-scale surveys, which
require homogeneous photometry at the 1\% level or better.
\end{abstract}


\section{Introduction}
\label{sec:intro}

The problem of the calibration of an ensemble of photometric data can
be divided into two parts (\citealt{pad2008}; \citealt{bet2013}):
namely, a {\it relative} calibration, where all of the data are
calibrated to the same photometric scale with arbitrary flux units,
and an {\it absolute} calibration, where the photometric scale is
calibrated into physical flux units (e.g. J~m$^{-2}$~s$^{-1}$) via
comparison to a particular object (or set of objects) with known
magnitudes.  The relative calibration can depend on a slew of
parameters pertinent to how the photometric data were obtained
(e.g. colour dependence, atmospheric transparency, detector
coordinates, image quality, etc.)  and, a priori, it is not always
clear which set of parameters best describe the relative
calibration. The additional step of absolute calibration boils down to
the determination of an extra parameter (zero-point), to convert the
measurements into physical flux units.
 

Early papers on photometric calibration methodology defined the
modelling procedures required for a relative calibration
(\citealt{har1981}; \citealt{pop1982}; \citealt{ree1982};
\citealt{man1983}; \citealt{hon1992}). These works considered
photometric calibration models with relative zero-points,
extinction, and colour terms combined with time dependence of the
coefficients. They also considered the analysis of inhomogeneous
observations of overlapping fields from potentially different
detectors for a mix of standard and non-standard stars (i.e. non-variable
stars with and without standard magnitudes, respectively). 

The process of flat fielding is required to calibrate images taken
with modern detectors. This is achieved in practice by observing a
uniformly illuminated source such as the twilight sky or a flat-field
screen. Images of the uniform source, called flat frames, then record
the sensitivity variations of the combined telescope/camera/detector
system including the high-spatial-frequency pixel-to-pixel variations
and the lower-spatial-frequency variations caused by out-of-focus dust
shadows etc. However, flat fielding performed in this way does not
correct perfectly for the sensitivity variations, especially for the
larger scale variations. This is usually due to undesired effects
present in the flat frames such as central light (or sky)
concentration or non-uniform illumination, but it may also be due to
the different spectral energy distributions of the flat field source
and the astronomical objects which are to be measured.

\citet{man1995} introduced the concept of performing a correction to
the flat field calibration (referred to as an ``illumination
correction'') by using the measured star magnitudes. By using the
stars to perform (at least part of) the flat fielding, many of the
problems with using flat frames are avoided. The illumination
correction introduces terms into the photometric calibration model
that are a function of detector coordinates. For the purposes of this
paper, we refer to this as a {\it static illumination correction}.

With the advent of large-scale surveys which require homogeneous
photometry at the 1\% level or better, the size of photometric
modelling problems has exploded. \citet{pad2008} presented a
photometric calibration model for Sloan Digital Sky Survey data on a
scale many times larger than had been seen before with
$\sim$3$\times$10$^{7}$ calibration data points, $\sim$10$^{7}$
star mean magnitudes to be determined, and $\sim$2000
calibration parameters. The Supernova Legacy Survey followed suit
(\citealt{reg2009}) and they also developed a tractable way of solving
for all of the parameters in the photometric models, including the
star mean magnitudes for the non-standard stars, by using a
single-step in the least squares solution (see also
\citealt{sch2012}). Most large surveys now regularly include all of
the terms mentioned previously in the photometric model.
In terms of photometric calibration modelling, a wide range of
effects have already been considered in the literature, including
relatively obscure effects that are only significant at the milli-magnitude
level (e.g. intrapixel photometric sensitivity maps - \citealt{pit2002}).

However, what has not been discussed in the literature is how to
include in the photometric calibration model a flat field correction
that rotates with an instrument component. From now on we refer to
this as a {\it rotating illumination correction}. The motivation for
modelling a rotating illumination correction comes from the work of
\citet[Paper I]{moe2010} where they found that both of ESO's FOcal
Reducer and low-dispersion Spectrographs (FORS,
\citealt{Appenzeller+98}) instruments at the VLT produce twilight
flats with a fixed large-scale structure component plus another
large-scale structure that rotates with the field rotator; this
rotating component was further tracked down to being caused by the
Linear Atmospheric Dispersion Corrector (LADC). The rotating flat
field structure has the potential to degrade the photometric accuracy
by adding a large systematic error of up to $\sim$4\% to broadband
observations in the case of FORS.  Although in Paper I we suggested
several ad-hoc solutions to this problem, we did not consider
including the correction in the photometric model.

Our paper details how to include a static and rotating flat-field
illumination correction in the photometric model for the first time
while noting an important degeneracy for polynomials (Section
\ref{sec:model}). In Section \ref{sec:application} we use this
photometric model to analyse FORS2 observations of standard star
fields and derive flat-field correction maps that include both static
and rotating illumination corrections. In Section \ref{sec:improving},
we describe a recipe for the user to follow, which accounts for the
static and rotating illumination corrections determined in
Section~\ref{sec:application}, in order to obtain improved FORS2
photometric measurements. We summarise our results in
Section~\ref{sec:summary}.

\section{Modelling static and rotating illumination corrections}
\label{sec:model}

In order to derive an illumination correction, the photometric data of
a large number of stars observed many times need to be analysed. A fit
to such data is relatively easy when a linear model is sufficient to
describe the relevant effects.  Our goal is to expand the model of
\citet{bra2012} to accommodate static and rotating illumination
corrections.  A feature of the FORS2 instrument is that it includes
two detectors that are simultaneously exposed and they rotate in the
focal plane about a common axis. For our discussion below, we use the
straight forward generalisation to an arbitrary number of
detectors. However, we note that our method is limited to the case
that the axis of rotation is known a priori. A fit of the rotation
centre would lead to a non-linear model and is beyond the scope of our
discussion.

We therefore consider a set of $N_{\mbox{\scriptsize data}}$ magnitude
measurements $m_{i}$ (indexed by $i$) taken from a set of
$N_{\mbox{\scriptsize im}}$ images (indexed by $r$) that are obtained
with detectors that are rigidly mounted on a common structure within
an instrument. The $N_{\mbox{\scriptsize det}}$ detectors (indexed by
$k$) of the instrument look through the same optical elements and are
similar in terms of spectral sensitivity, and each image may come from
any one detector. We assume that each observation obtained by the
instrument generates $N_{\mbox{\scriptsize det}}$ images at the same
epoch, and that the full set of images may include different pointings
in the sky. The observations will include measurements of
$N_{\mbox{\scriptsize obj}}$ objects (indexed by $j$). Our adopted
indexing implies that the $i$th magnitude measurement in our
photometric data sample belongs to the $j(i)$th object, the $k(i)$th
detector, and the $r(i)$th image, where the adopted notation for $j$,
$k$ and $r$ reflects the fact that these indices are functions of the
index $i$. However, in the rest of this paper we use the simplified
notation $j$, $k$ and $r$ for $j(i)$, $k(i)$ and $r(i)$, respectively,
in order to avoid confusion in our subscript notation.

Let $M_{j}$ denote the true instrumental magnitude of the $j$th object.
We also adopt a single reference coordinate system $(\eta,\xi)$ in the
instrument focal plane that applies to all detectors so that each
magnitude measurement has associated coordinates
$(\eta_{i},\xi_{i})$. 

Then we may write our photometric model as:

\begin{equation}
\overline{m}_{i} = M_{j} + Z + \Delta Z^{\mbox{\scriptsize det}}_{k} + \Delta Z^{\mbox{\scriptsize im}}_{r} 
                   + \Delta F^{\mbox{\scriptsize stat}}(\eta_{i},\xi_{i})
                   + \Delta F^{\mbox{\scriptsize rot}}(\eta^{\prime}_{i},\xi^{\prime}_{i})
\label{eqn:phot_model_simple}
\end{equation}

where $\overline{m}_{i}$ is the model magnitude for the $i$th
magnitude measurement and $Z$ is the overall system zero-point.  By
setting $\Delta Z^{\mbox{\scriptsize det}}_{1} = 0$, the quantity
$\Delta Z^{\mbox{\scriptsize det}}_{k}$ accounts for the sensitivity
variation of the $k$th detector relative to the first detector. By
setting $\Delta Z^{\mbox{\scriptsize im}}_{1} = 0$, the quantity
$\Delta Z^{\mbox{\scriptsize det}}_{r}$ accounts for any variations in
the throughput for the light's journey from the top of the atmosphere,
through the telescope and instrument, and to the detector surface, for
the $r$th image relative to the first image. The quantity $\static
(\eta_{i},\xi_{i})$ represents the sensitivity variation across the
focal plane relative to some arbitrary fiducial coordinates
$(\eta_{c},\xi_{c})$ where $\static (\eta_{c},\xi_{c}) = 0$ and this
term is what we refer to as the static illumination correction. The
quantity $\rotating (\eta^{\prime}_{i},\xi^{\prime}_{i})$ represents the
sensitivity variation across the de-rotated focal plane relative to
fiducial coordinates coincident with the centre of rotation
$(\eta^{\prime}_{c},\xi^{\prime}_{c})$ and this term is what we refer
to as the rotating illumination correction. For simplicity and without
loss of generality, we set $(\eta^{\prime}_{c},\xi^{\prime}_{c}) =
(\eta_{c},\xi_{c}) = (0,0)$. Given a rotation angle $\theta$
anti-clockwise around the origin from the $(\eta,\xi)$ coordinate
system to the $(\eta^{\prime},\xi^{\prime})$ system, we have:
\begin{eqnarray}
\eta^{\prime} &=& \eta \cos \theta + \xi \sin \theta  \label{eqn:eta_rot} \\ 
\xi^{\prime}  &=& -\eta \sin \theta + \xi \cos \theta \label{eqn:xi_rot} 
\end{eqnarray}

Our photometric model in Equation~\ref{eqn:phot_model_simple} has
numerous potential degeneracies. For instance, if none of the $M_{j}$
are fixed (or known), then the overall zero-point $Z$ becomes
degenerate with the $M_{j}$ because adding a constant $C$ to $Z$ may
be offset by subtracting $C$ from all of the $M_{j}$. Hence, one
$M_{j}$ for one star should be fixed to an arbitrary value to set the
(arbitrary) zero-point of the magnitude scale for the relative
calibration. A convenient way to do this is to use the standard
magnitude of a known standard star for the value of $M_{j}$, which has
the advantage that if other $M_{j}$ values need to be defined rather
than fit in order to avoid model degeneracies, then this is possible
in a consistent way by using more standard stars. Following on from
this, to avoid degeneracies with the detector zero-point offsets
$\Delta Z^{\mbox{\scriptsize det}}_{k}$, for each detector pair there
should be at least one standard star with known $M_{j}$ observed at
least once on both detectors in the set of observations\footnote{In
  fact, this constraint is somewhat stricter than necessary, but it is
  a simple constraint that serves our purpose later on.}. Degeneracies
in the image zero-point offsets $\Delta Z^{\mbox{\scriptsize im}}_{r}$
may be avoided by 
always ensuring overlapping images and/or the presence
of standard star measurements for each image.

In order to constrain the static illumination correction function
$\static$, multiple observations of the same objects are required with
different spatial offsets that cover a grid-like pattern over the two
spatial dimensions. Similarly, in order to constrain the rotating
illumination correction function $\rotating$,
multiple observations of the same objects are required at different
``rotation angles'' (for whichever instrument/telescope component
rotates relative to the detectors) {\it and at different distances
  from the centre of rotation}. This second requirement is important
but not so obvious. For example, \citet{hru2010} collected an imaging
data set where the photometry reveals a sky-position-angle-dependent
systematic error. However, the observations lacked spatial offsets
relative to the centre of rotation and consequently we found that we
could not apply the photometric modelling methodology outlined in this
paper to their data. With rotations but no offsets, the annuli of the
rotating illumination correction function are constrained, but the
relative corrections between annuli remain unconstrained, and
therefore the photometric modelling is a degenerate problem in this
case.

Polynomials are a sensible choice of model for the spatial variation
of both the static and rotating illumination corrections since these
sensitivity variations are expected to vary smoothly on a relatively
large scale (e.g. Paper~I). Hence we adopt\footnote{Although using
  orthogonal polynomials would help improve the orthogonality of the
  fitting problem, this choice would not render the fitting problem as
  fully orthogonal because the dot products that define orthogonality
  and feature in the normal equations use inverse-variance weights
  (see later).}:

\begin{equation}
\Delta F^{\mbox{\scriptsize stat}}(\eta,\xi) = \sum_{m = 0}^{D_{\mbox{\tiny stat}}} \sum_{n = 0}^{D_{\mbox{\tiny stat}} - m} a_{mn} \, \eta^{m} \, \xi^{n}
\label{eqn:stat_poly}
\end{equation}
\begin{equation}
\Delta F^{\mbox{\scriptsize rot}}(\eta^{\prime},\xi^{\prime}) = \sum_{m = 0}^{D_{\mbox{\tiny rot}}} \sum_{n = 0}^{D_{\mbox{\tiny rot}} - m} b_{mn} \, \left( \eta^{\prime} \right)^{m} \left( \xi^{\prime} \right)^{n}
\label{eqn:rot_poly}
\end{equation}

where $D_{\mbox{\scriptsize stat}}$ and $D_{\mbox{\scriptsize rot}}$
are the degrees of the polynomials for the static and rotating
illumination corrections, respectively, and $a_{mn}$ and $b_{mn}$ are
the polynomial coefficients.

Substituting Equations~\ref{eqn:eta_rot}~\&~\ref{eqn:xi_rot} into
Equation~\ref{eqn:rot_poly}, and then substituting
Equations~\ref{eqn:stat_poly}~\&~\ref{eqn:rot_poly} into
Equation~\ref{eqn:phot_model_simple}, defines the remaining parameters
of our photometric model and renders the model as a linear model.
However, as written above, some of the model parameters are degenerate
because the circularly symmetric component of the illumination
correction can be included in either the static or rotating
correction. Formally, this is a consequence of the Pythagorean
relation: \begin{equation} \eta^{2} + \xi^{2} = \left( \eta^{\prime}
  \right)^{2} + \left( \xi^{\prime} \right)^{2}
        \label{eqn:rot_degeneracy} \end{equation}

To remove this degeneracy without affecting the flexibility of the
photometric model, it is sufficient to drop the polynomial terms
$\left( \eta^{\prime} \right)^{2n}$ for $1 \le n \le \lfloor
D_{\mbox{\scriptsize rot}} / 2 \rfloor$. 

The fitting of the photometric model to the data follows
\citet{bra2012}. Firstly, we write our photometric model using
Kronecker delta functions:

\begin{eqnarray}
\overline{m}_{i} &=& \left( \sum_{p = 1}^{N_{\mbox{\tiny obj}}} \delta_{jp} \, M_{p} \right)
                    + \left( \sum_{p = 1}^{N_{\mbox{\tiny det}}} \delta_{kp} \, \Delta Z^{\mbox{\scriptsize det}}_{p} \right)
                    + \left( \sum_{p = 1}^{N_{\mbox{\tiny im}}} \delta_{rp} \, \Delta Z^{\mbox{\scriptsize im}}_{p} \right) + \nonumber \\
                & & +Z + \Delta F^{\mbox{\scriptsize stat}}(\eta_{i},\xi_{i}) + \Delta F^{\mbox{\scriptsize rot}}(\eta^{\prime}_{i},\xi^{\prime}_{i}) 
\label{eqn:phot_model_complicated}
\end{eqnarray}
where:
\begin{equation}
\delta_{ij} =
\begin{cases}
1 & \mbox{if $i = j$} \\
0 & \mbox{if $i \ne j$} \\
\end{cases}
\label{eqn:kronecker_delta}
\end{equation}

Then, with all of the free parameters present in the photometric
model, we proceed to minimize the chi-squared:

\begin{equation}
\chi^{2} = \sum_{i = 1}^{N_{\mbox{\tiny data}}} \left( \frac{m_{i} - \overline{m}_{i}}{\sigma_{i}} \right)^{2}
\label{eqn:chi2}
\end{equation}

where $\sigma_{i}$ is the uncertainty on the $i$th photometric
measurement. The chi-squared minimization is achieved by constructing
the normal equations for the general linear least squares problem
(\citealt{Press+07}) and solving them using Cholesky factorization
followed by forward and back substitution. Since the size of the
matrix for the normal equations may become very large for large
problems with data from many objects, we use the property that a
sub-matrix in the normal equations is diagonal to render the problem
tractable (see Appendix~A of \citealt{reg2009} or Section~2.3 of
\citealt{bra2012} for details). We also iterate the fitting procedure
by rejecting measurements that lie more than $\sqrt{\ln N_{\rm data}}
\cdot \sigma_i$ away\footnote{This is a suitable $\sigma$-clip
  threshold that can be derived by considering the Bayesian
  Information Criterion \citep{sch78}. $\sigma_i$ is the
  error on $m_i$.}  from the fitted model (and
further drop the object observations for which an object now has only
a single observation). The iterative fitting is necessary because of
the likely presence of variable sources and outlier photometric
measurements, and it generally converges within a couple of
iterations.

The whole fitting procedure is implemented in the {\tt IDL} program
{\tt fit\_photometric\_calibration.pro} available as part of the {\tt
  DanIDL}\footnote{http://www.danidl.co.uk} library of routines.

\section{Application of the photometric model to FORS2 observations of standard star fields}
\label{sec:application}

In the following, we apply the photometric model described in Section
\ref{sec:model} to a set of standard star fields observed with FORS2 as part
of the nightly calibration plan.

\subsection{Observations and reductions}
\label{sec:obs_and_red}

We downloaded from the ESO
archive\footnote{http://archive.eso.org/eso/eso\_archive\_main.html}
FORS2 imaging observations of standard star fields carried out with
the $B$, $V$, $R$, and $I$ ESO filters\footnote{They are recorded as
  b\_HIGH+113, v\_HIGH+114, R\_SPECIAL+76, and I\_BESS+77 in the ESO
  database, see
  http://www.eso.org/sci/facilities/paranal/instruments/fors/inst/Filters/}.
We limited our data selection to the time range between
1st~November~2011 and 7th~July 2013. The starting date coincides with
when the new nightly calibration plan for FORS imaging was put into
operations (see \citealt{Bramich+12}); this plan was indeed defined in
order to ensure that the spatial and angular offsets needed to
constrain the model for the $\static$ and $\rotating$ illumination
correction patterns are performed (see Section \ref{sec:model}). The
time interval was additionally split into two ranges using the epoch
of the UT1-M1 mirror illumination and LADC re-alignment (31st May 2012
- 11th June 2012) as a break-point. We will refer to these two time
ranges as {\it range A} (1st November 2011 -- 30th May 2012) and {\it
  range B} (12th June 2012 -- 7th July 2013). Since such instrument
interventions can modify the overall instrumental response and the
shape of the illumination patterns that we want to study, we processed
the data from the two time ranges independently.

We also imposed the following constraints on the instrument set-up for the data
to be analysed: Standard Resolution Collimator, $2\times 2$ pixel binning
(i.e. spatial scale of $0\farcs25$ per binned pixel), detectors {\tt CCID20-14-5-3}
and {\tt CCID20-14-5-6} for chip 1 and chip 2, respectively.
Collimator, binning mode, and detector name are indicated by the
header keywords {\tt HIERARCH ESO INS COLL NAME}, {\tt HIERARCH ESO
  DET WIN1 BINX/Y}, and {\tt HIERARCH ESO DET CHIP1 ID}, respectively.

According to the calibration plan, at the beginning of each night, two
standard star fields from the Stetson catalogue\footnote{See: 
    http://www3.cadc-ccda.hia-iha.nrc-cnrc.gc.ca/community/\\STETSON/standards/.}
are observed; the first is observed at low airmass, the second at
high airmass. If these observations indicate that the night is
photometric, more standard star fields are observed during the night to
check its photometric stability.
Standard star fields are observed at different position angles on the
sky, and with different telescope offsets from the nominal field coordinates
(from $10''$ to $60''$).
 
The data were reduced using the automatic FORS2 ESORex pipeline
version 4.9.23\footnote{The FORS2 data reduction manual and pipeline,
  and the ESOrex are available at
  http://www.eso.org/sci/software/pipelines/ \label{note:pipeline}}. The
pipeline includes automatic identification of the stars, measurement
of instrumental magnitudes, and cross-checking with the standard star
catalogue.  The identification of sources, and measurement of
instrumental magnitudes within the FORS2 pipeline is performed with
SExtractor \citep{Bertin+96}, which evaluates the local sky background
and computes $10''$ diameter aperture photometry.

One fundamental difference with the standard data reduction cascade is
that in our analysis we divided each master flat by its smooth
large-scale component. This is created by applying a boxcar smoothing
window (200$\times$200 pixels) to the master flat frame while masking 2
sigma outliers (to avoid large variations within few pixels) and
regions close to the edges of the illuminated area (to avoid spurious
edge effects in the filtering).  This new master flat, which contains
only the pixel-to-pixel sensitivity variations, is used to calibrate
the images. In this way, the large-scale sensitivity variations across
the detectors are left unmodified in the photometric data. The aim of
this paper is indeed to study these systematic sensitivity variations
across the field-of-view using the photometric data, and to separate
them into the contributions of the static ($\static$) and rotating
($\rotating$) illumination corrections defined in
Section~\ref{sec:model}.

\subsection{Star identification}

In our analysis we use the information from all detected stars in each
observed field regardless of whether or not they have catalogued
standard magnitudes.

Standard stars are identified by comparing the coordinates of the
standard stars in Stetson's reference catalogue with those of all of
the detected sources obtained using the frame WCS information. Then,
more accurate coordinates and the frame astrometric solution are
re-computed by minimizing the offsets in right ascension and declination
between the standard stars' coordinates in the image and in the
reference catalogue. We refer to the FORS2 pipeline reference manual for
more details (see Footnote \ref{note:pipeline}).

Other objects detected in our data are considered to be useful stars if they
fulfill the following requirements:

\begin{itemize}

\item the object is detected in more than 3 images (to avoid spurious
detections);

\item the differences between the celestial coordinates of the object as
  measured in each image must be smaller than 1\farcs2 on the sky;

\item the object must have no other detected source within $3''$;

\item the object has a measured magnitude that is brighter than the
  1st quartile of the magnitude histogram in each image (to avoid
  objects that are too faint).

\end{itemize} 

The above requirements do not apply to standard stars.

\begin{table*}
\caption{Number of standard and other stars
  detected in all of the observations on photometric nights.}
\begin{center}
\begin{tabular}{l c c c c c c c c}
\hline
\noalign{\smallskip}
Filter & $N_{\rm obj}^{\rm std}$  &$N_{\rm data}^{\rm std}$ &$N_{\rm obj}^{\rm other}$ &$N_{\rm data}^{\rm other}$ &$N_{\rm obj}^{\rm std}$  &$N_{\rm data}^{\rm std}$ &$N_{\rm obj}^{\rm other}$ &$N_{\rm data}^{\rm other}$ \\
       & (range A) &(range A)        &(range A)    &(range A)            &(range B) &(range B)         &(range B)   &(range B)            \\
\noalign{\smallskip}
 (1)   &   (2)     &       (3)       &  (4)        &   (5)               &  (6)     & (7)             &   (8)       &   (9)               \\
\hline
\noalign{\smallskip}
 $B$   & 1894      & 4970            &1236         &   5374              &  2720   & 23543            &   9092      &    55885            \\
 $V$   & 1748      & 4613            &2231         &  9194               &  2650   & 21564            &  14219      &    94401            \\
 $R$   & 1584      & 4304            &2946         &  12047              &  2393   & 18910            &  18185      &   112897            \\
 $I$   & 1680      & 4632            &2454         &   10149             &  2440   & 20579            &  18823      &   116635            \\
\noalign{\smallskip}
\hline
\noalign{\smallskip}
\label{tab:stars}
\end{tabular}
\begin{minipage}{18cm}
Notes-- Column 1: filter identification code: $B$ indicates the ESO
filter named {\tt b\_HIGH+113}, $V$ indicates {\tt v\_HIGH+114}, $R$ indicates {\tt R\_SPECIAL+76},
and $I$ indicates {\tt I\_BESS+77}.
Column 2: Number of (unique) standard stars observed in time range A (1st November 2011 -- 30th May 2012).
Column 3: Number of instrumental magnitude measurements for the stars in column 2.
Column 4: Number of (unique) other stars observed in time range A.
Column 5: Number of instrumental magnitude measurements for the stars in column 4.
Columns 6-9: Same as cols 2-5, but for time range B (12th June 2012 -- 7th July 2013).
\end{minipage}
\end{center}
\end{table*}

\subsection{Reference coordinate system}
\label{sec:reference_system}

The $(\eta,\xi)$ reference coordinate system that we used is defined
as follows. Let $x$ and $y$ be the pixel coordinates on the calibrated
images as reported by the FORS2 data reduction pipeline. The
conversion between these coordinates and those of the archival raw
images (i.e. not yet processed by the pipeline) are $x = x_{\rm raw}$,
$y=y_{\rm raw}-5$, for both detectors, due to the removal of
pre-/overscan regions during processing.

Consistent with the previous analysis of Paper~I, we consider the
centre of the illuminated area in the focal plane to be the centre of
rotation and the origin of our $(\eta,\xi)$ reference coordinate
system adopted in Section~\ref{sec:model}.  The $(x,y)$ coordinates of
this point are:

\[
x_c^{\rm chip1} = 1022; 
y_c^{\rm chip1} = 115
\] 

for chip 1, and 

\[
x_c^{\rm chip2} = 1022; 
y_c^{\rm chip2} = 1157
\] 

for chip 2.

The conversion from the image coordinates $(x,y)$ into the
$(\eta,\xi)$ coordinates is:

\begin{eqnarray}
  \eta &=& \frac{x -x_c^{\rm chip1}}{1700}  \nonumber \\
  \xi &=& \frac{y  -y_c^{\rm chip1}}{1700} 
\label{eqn:coords1a}
\end{eqnarray}
for chip 1, and:
\begin{eqnarray}
  \eta &=&  \frac{x \cdot \cos(\phi) - y \cdot \sin(\phi) - x_c^{\rm chip2}}{1700}  \nonumber \\
  \xi  &=&  \frac{x \cdot \sin(\phi) + y \cdot \cos(\phi) - y_c^{\rm chip2}}{1700}
\label{eqn:coords1b}
 \end{eqnarray}
for chip 2. 

The number 1700 is an arbitrary but fixed normalisation coefficient,
which is comparable to the size of the useful field of view
  ($\sim 1700 \times 1700$ binned pixels); the quantity $\phi$ is the
offset angle between chip 1 and chip 2 and it can be read from the
header keyword {\tt HIERARCH ESO DET CHIP1 RGAP}. For our data it is
$\phi = 0.08278^{\circ}$.

The rotator angle $\theta$ needed to convert between the coordinate
systems $(\eta,\xi)$ and $(\eta^{\prime},\xi^{\prime})$ (see Equations
\ref{eqn:eta_rot} and \ref{eqn:xi_rot}) is computed as the average
of the following entries in the image header {\tt HIERARCH ESO
  ADA ABSROT START} and {\tt HIERARCH ESO ADA ABSROT END}.
The maximum difference between the two entries is 0.77 degrees,
  which corresponds to a shift of $2\farcs5$ at the border of
  the FORS2 field of view.
The impact on the photometry is negligible, as the rotating illumination correction patterns vary at most by a factor of 1.0001 within
  $2\farcs5$.

\subsection{Application of the photometric model}
\label{sec:phot_models}

In our fit we used the measured magnitudes $m_i$ of all of the stars
that were observed and detected in our data set on nights with stable
photometric conditions, as classified by ESO Quality
Control\footnote{http://www.eso.org/observing/dfo/quality/FORS2/qc/zeropoints/\\zeropoints.html\#stable}. The
list of images obtained during photometric conditions is given in
Table~\ref{tab:log}; the number of standard and other stars identified
in these images for the considered time ranges is summarised in
Table~\ref{tab:stars}.

For each time range, we fitted the photometric model defined in
Equation \ref{eqn:phot_model_complicated} to measure: i) the true
instrumental magnitudes $M_j$ of the observed stars, ii) the values of
the coefficients $a_{mn}$ and $b_{mn}$ of the static and rotating
illumination corrections defined in Equations \ref{eqn:stat_poly} and
\ref{eqn:rot_poly}, respectively.  Note that we used the catalogued
standard magnitudes to fix the true instrumental magnitude $M_{j}$ for
one standard star in each standard star field since the standard star
fields that were observed do not overlap. The true instrumental
magnitudes for the remainder of the standard stars, and for the other
detected stars, were left as free parameters in the photometric model.

\begin{figure*}
\subfigure[]{
\vbox{
  \hbox{
     \psfig{file=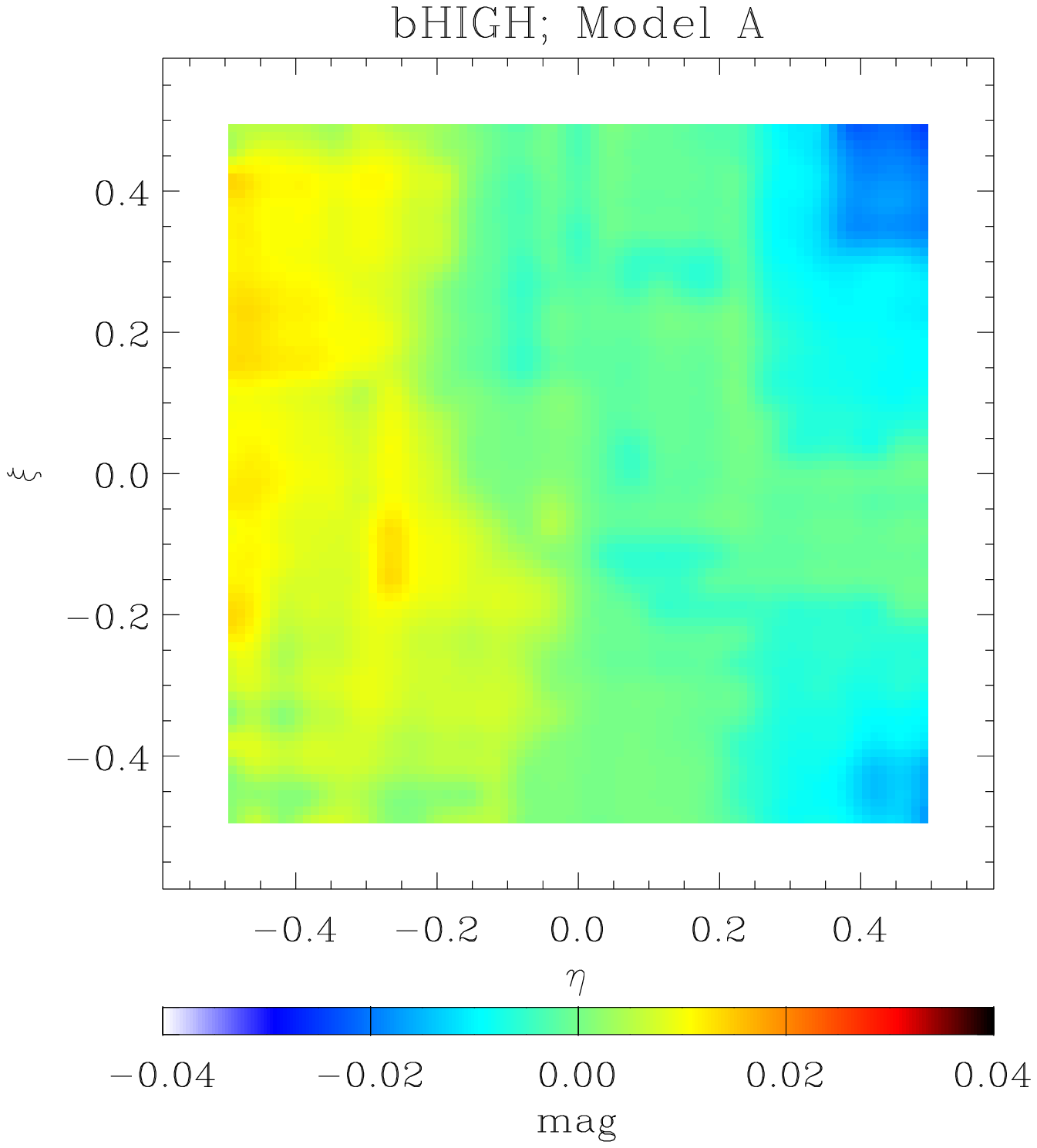,width=4.2cm,clip=,bb=35 413 409 785}
     \psfig{file=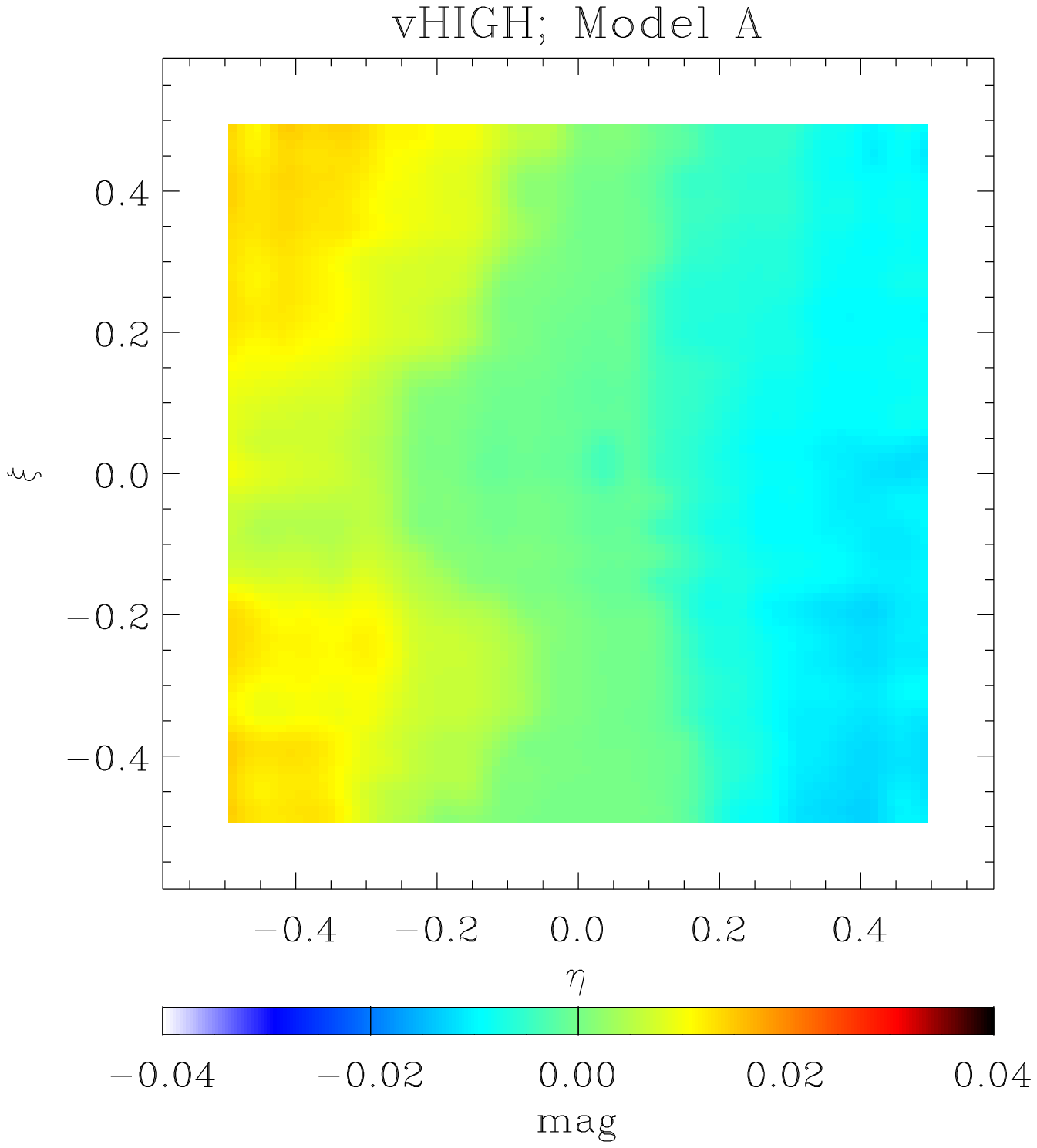,width=4.2cm,clip=,bb=35 413 409 785}
     \psfig{file=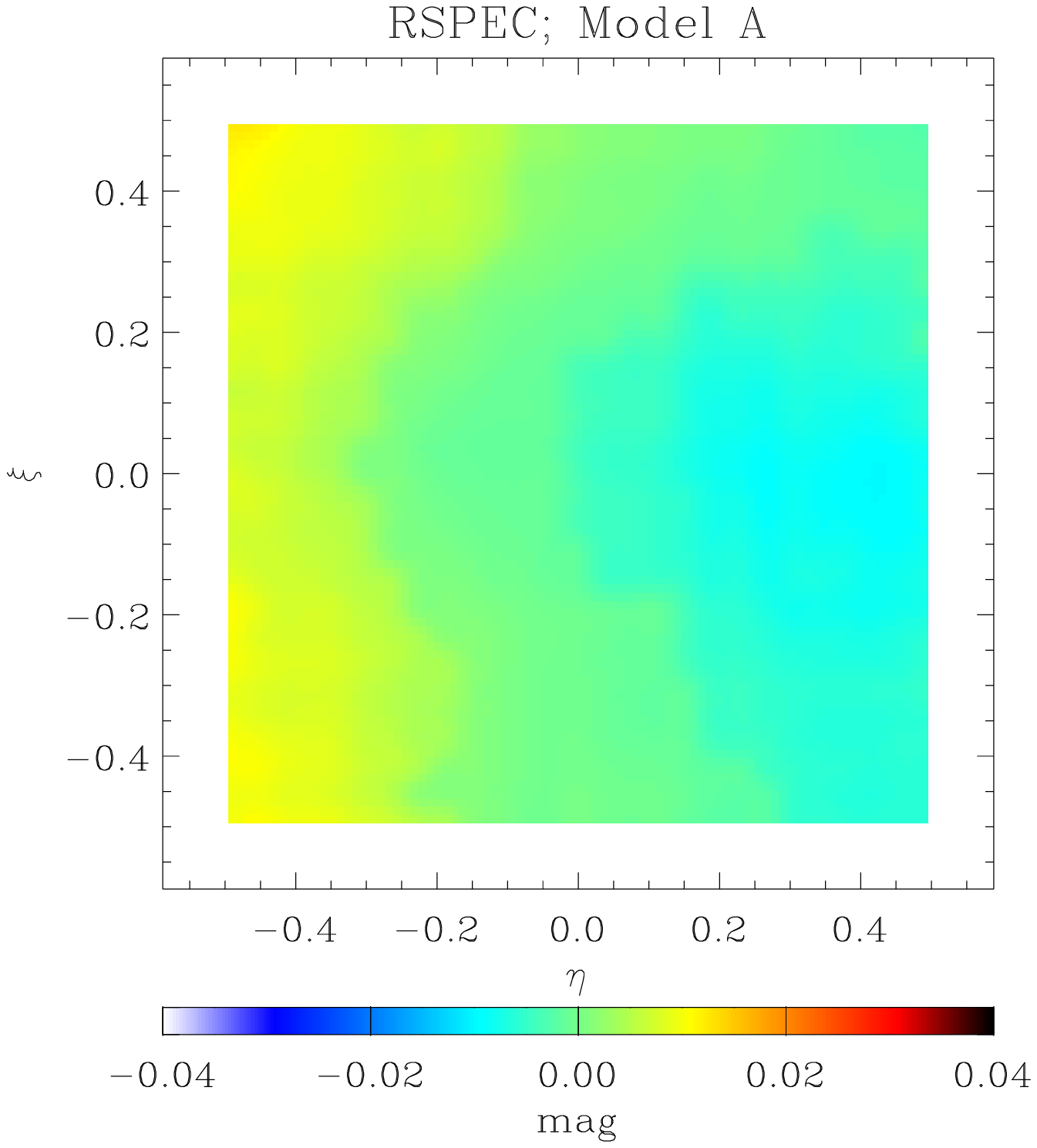,width=4.2cm,clip=,bb=35 413 409 785}
     \psfig{file=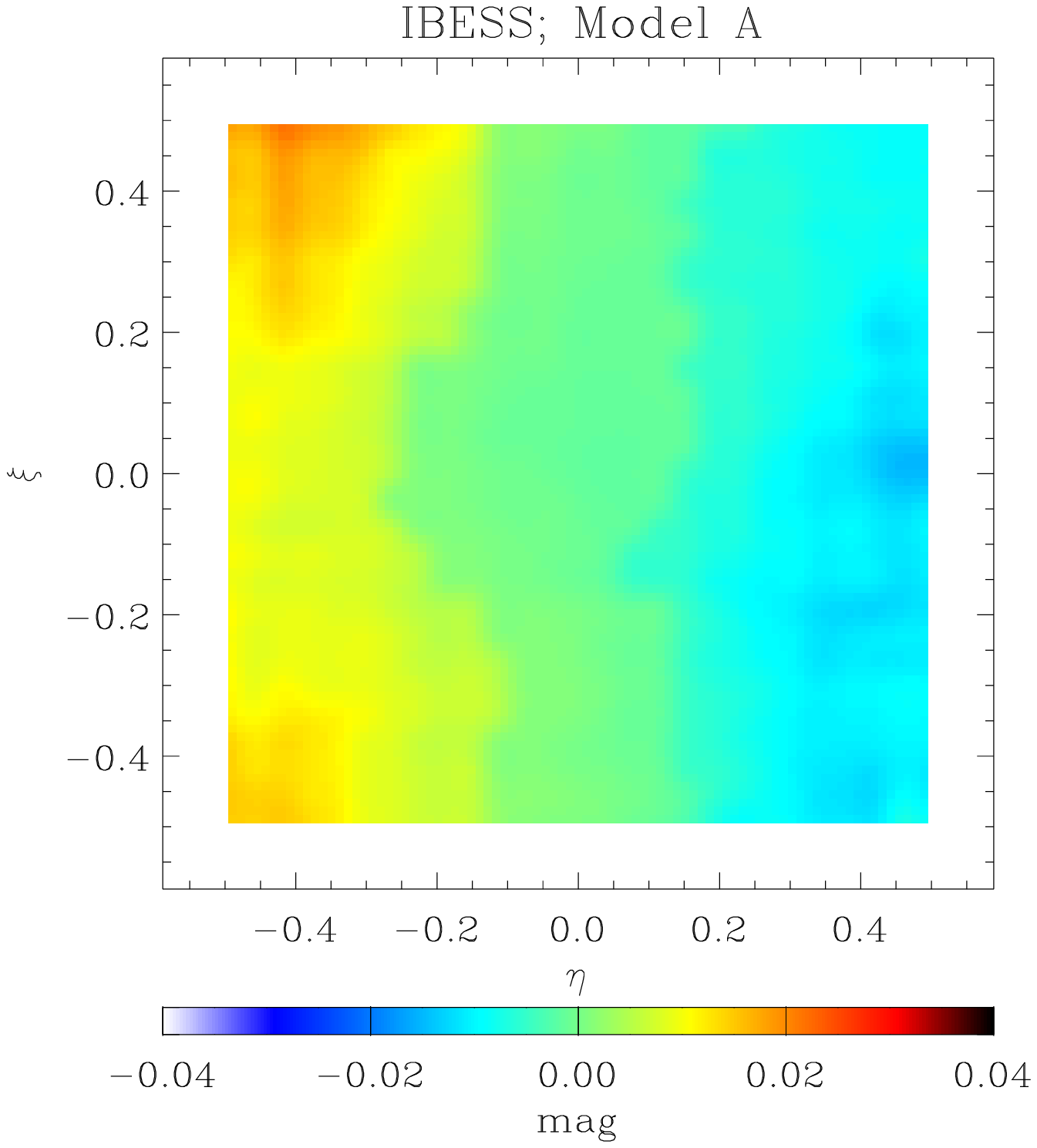,width=4.2cm,clip=,bb=35 413 409 785}
     \psfig{file=residuals_2D_paperbHIGH_2A.ps,width=4.2cm,clip=,bb=75 358 423 414,angle=90}
   }
\vspace{.3cm}
  \hbox{
     \psfig{file=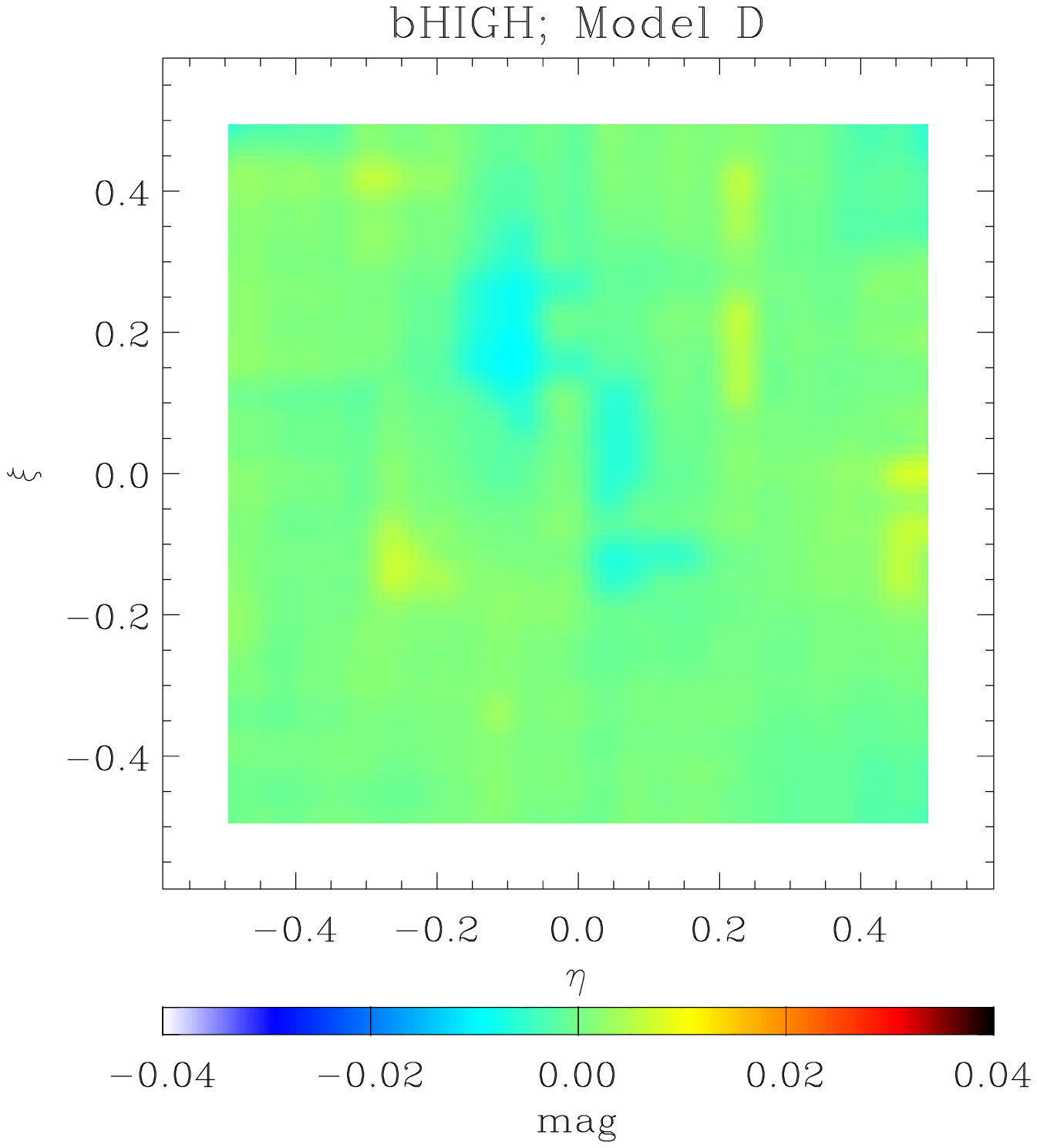,width=4.2cm,clip=,bb=35 413 409 785}
     \psfig{file=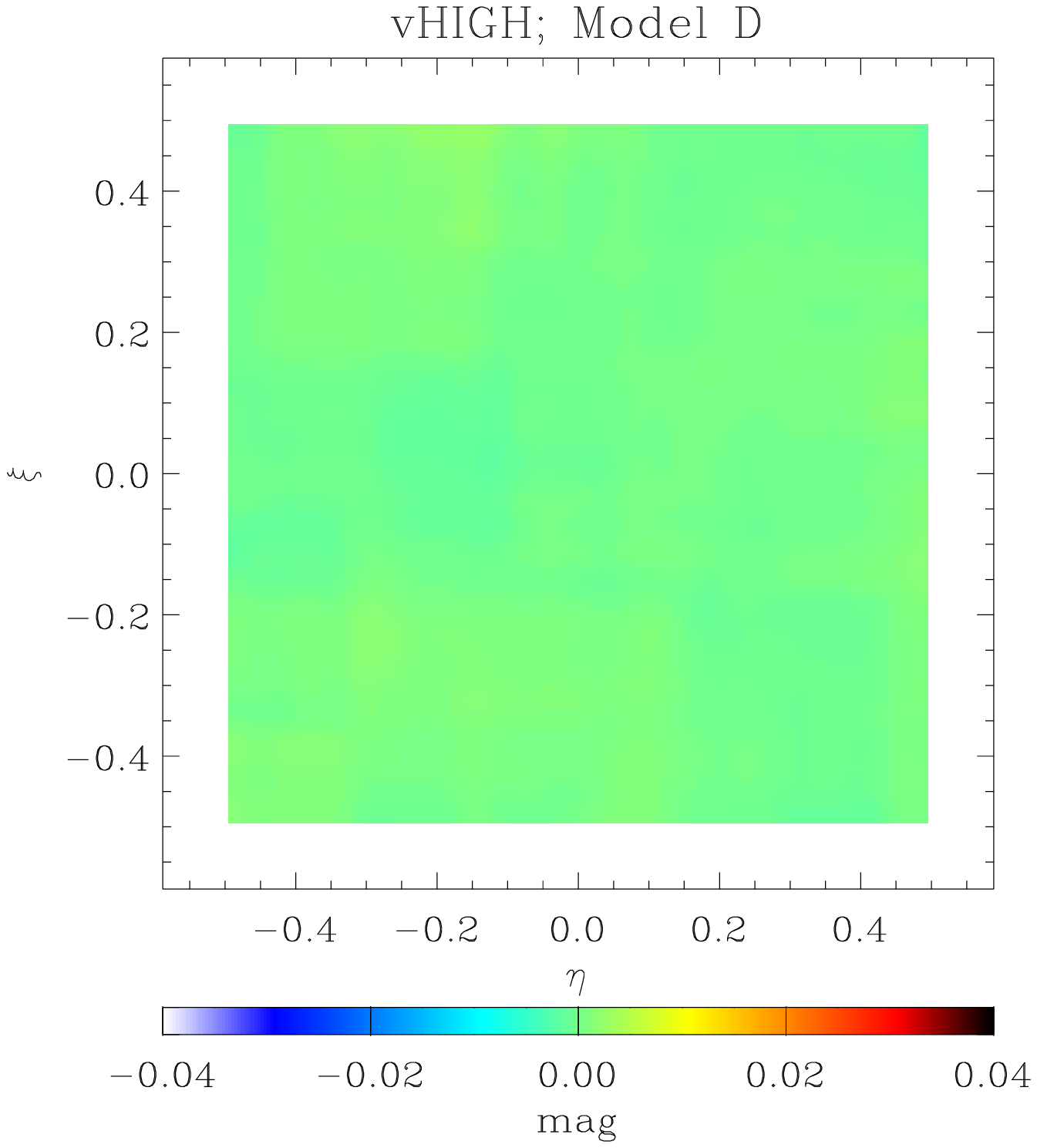,width=4.2cm,clip=,bb=35 413 409 785}
     \psfig{file=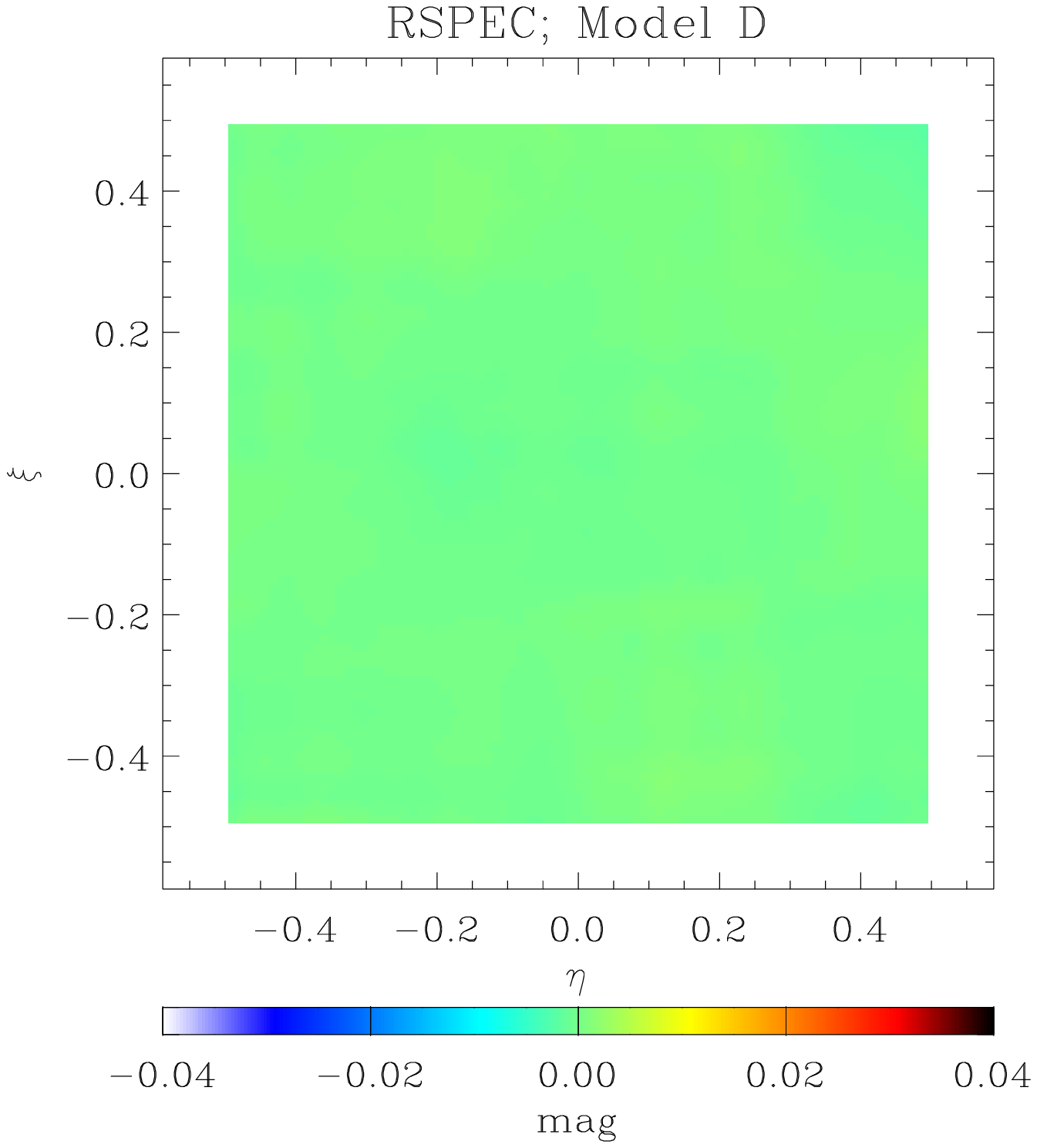,width=4.2cm,clip=,bb=35 413 409 785}
     \psfig{file=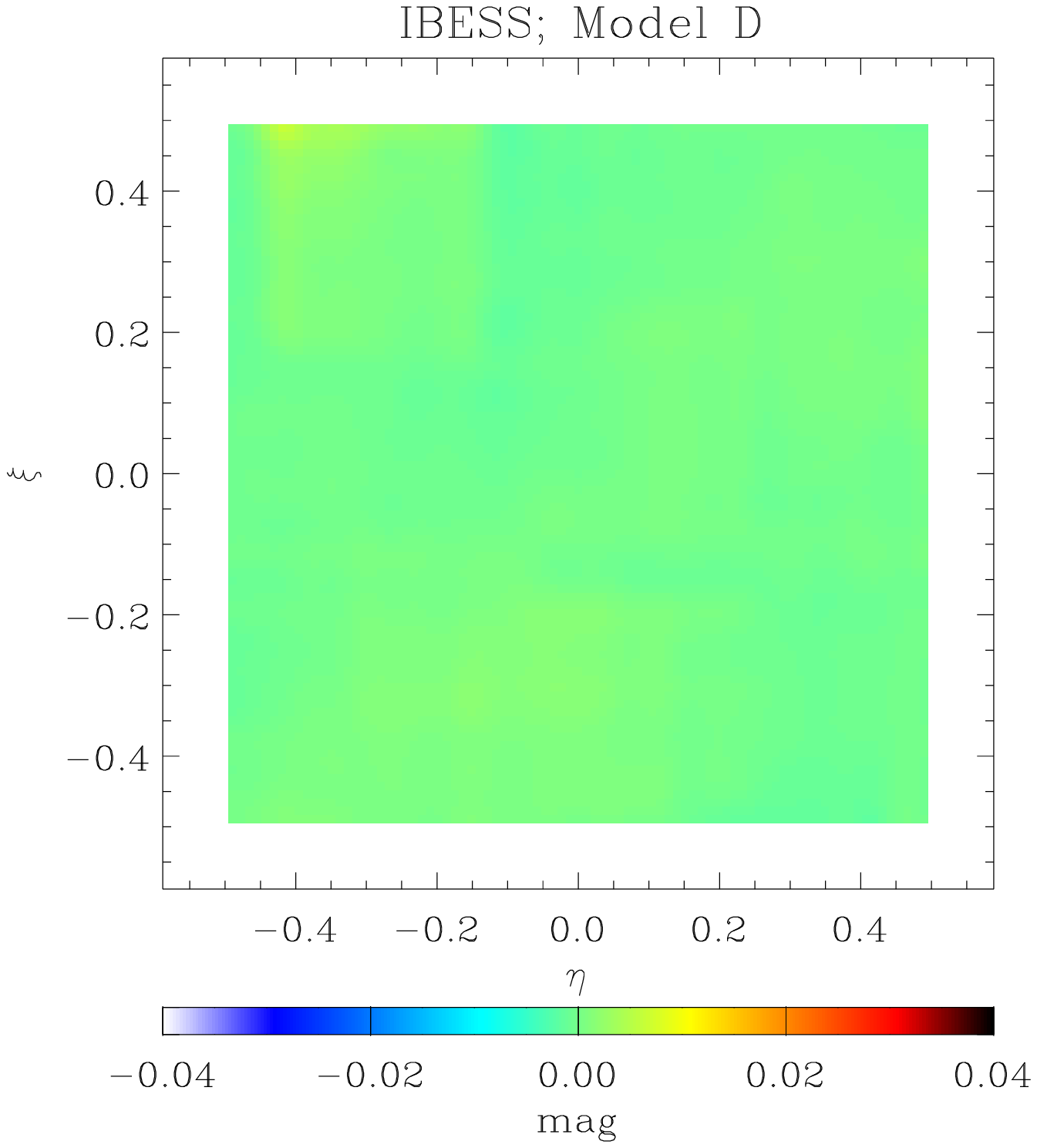,width=4.2cm,clip=,bb=35 413 409 785}
     \psfig{file=residuals_2D_paperbHIGH_2D.ps,width=4.2cm,clip=,bb=75 358 423 414,angle=90}
   }
 }
}
\subfigure[]{
 \vbox{
  \hbox{
     \psfig{file=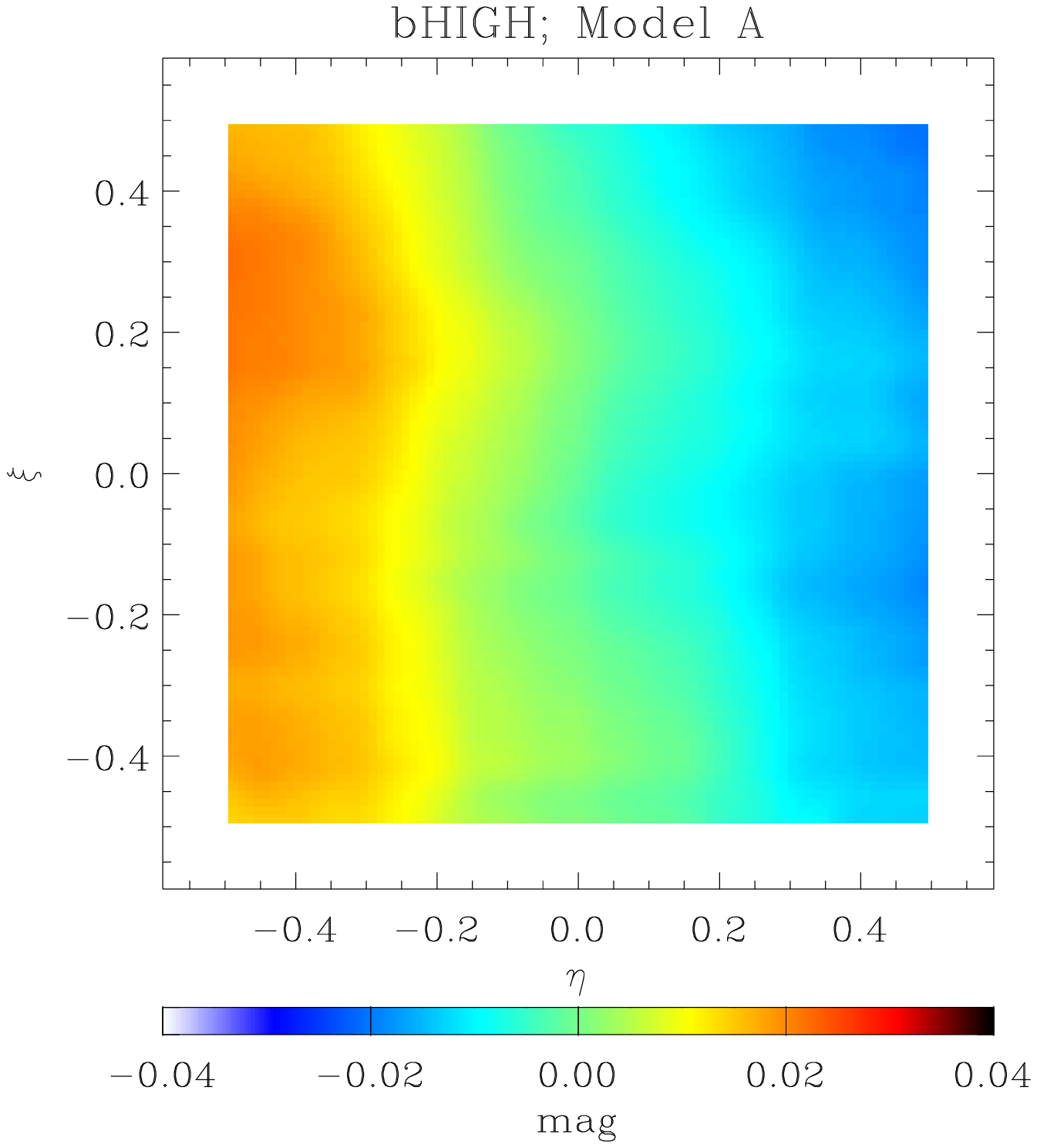,width=4.2cm,clip=,bb=35 413 409 785}
     \psfig{file=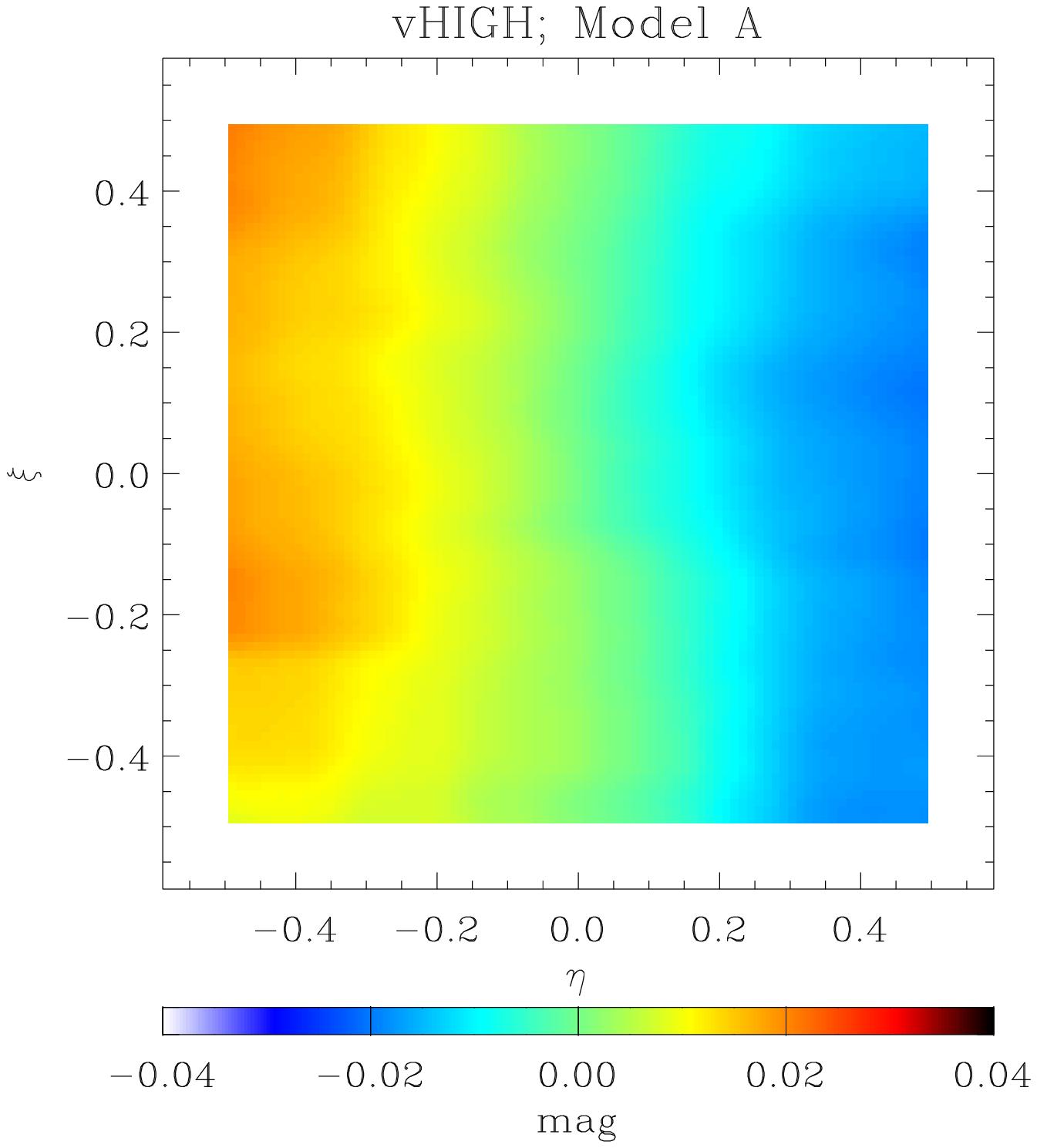,width=4.2cm,clip=,bb=35 413 409 785}
     \psfig{file=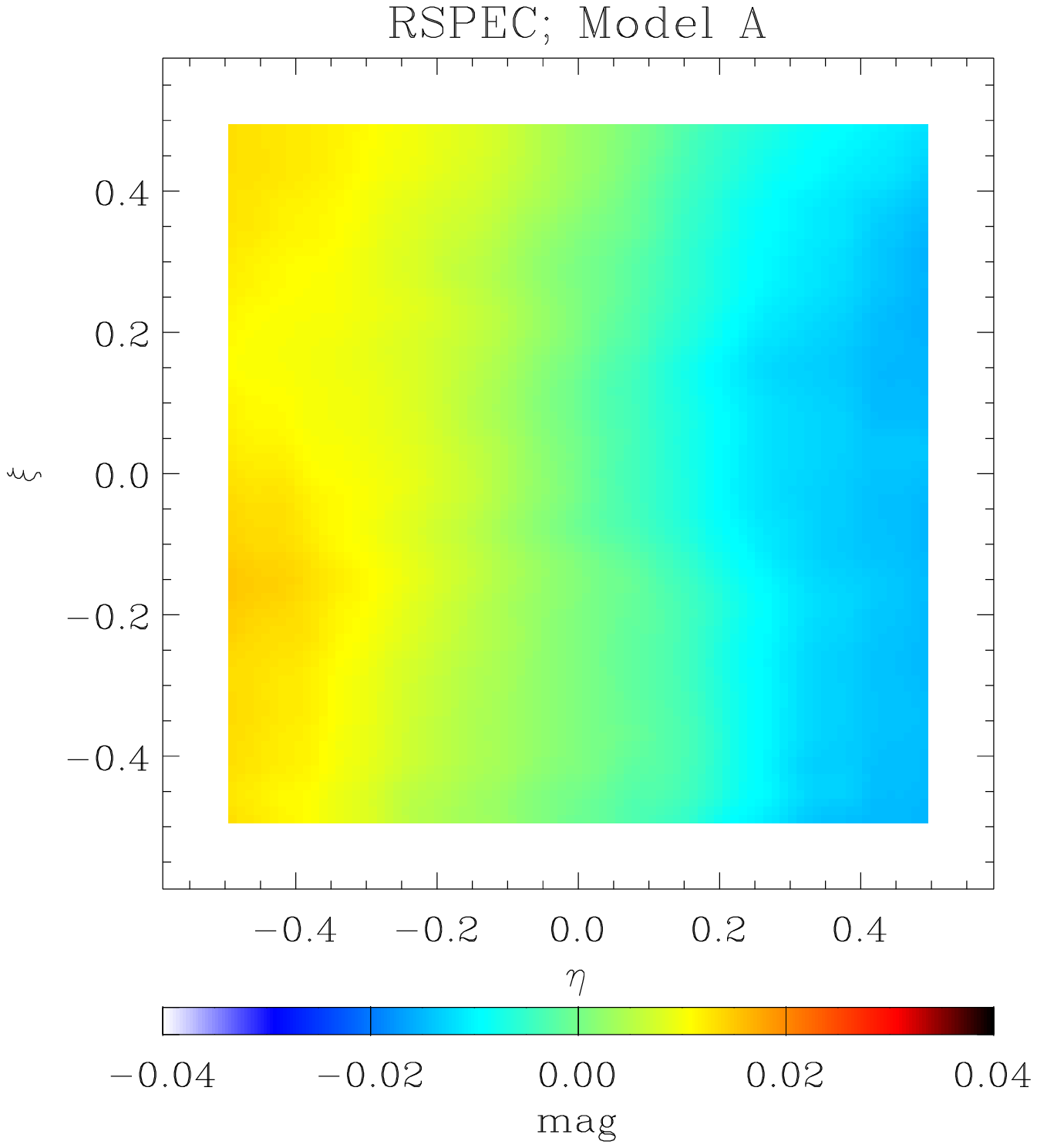,width=4.2cm,clip=,bb=35 413 409 785}
     \psfig{file=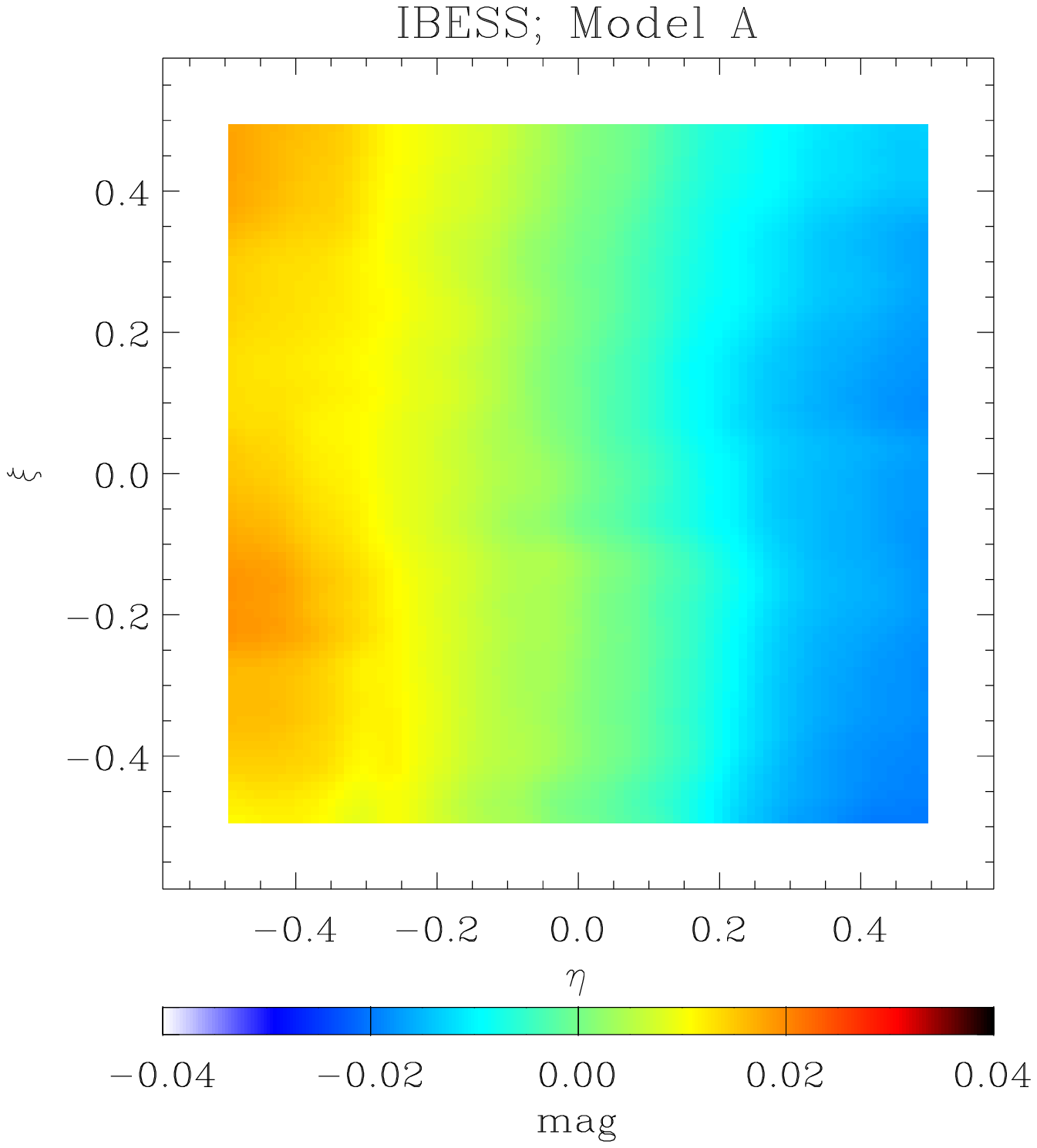,width=4.2cm,clip=,bb=35 413 409 785}
     \psfig{file=residuals_2D_paperbHIGH_2A_time_rangeB.ps,width=4.2cm,clip=,bb=75 358 423 414,angle=90}
   }
\vspace{.3cm}
  \hbox{
     \psfig{file=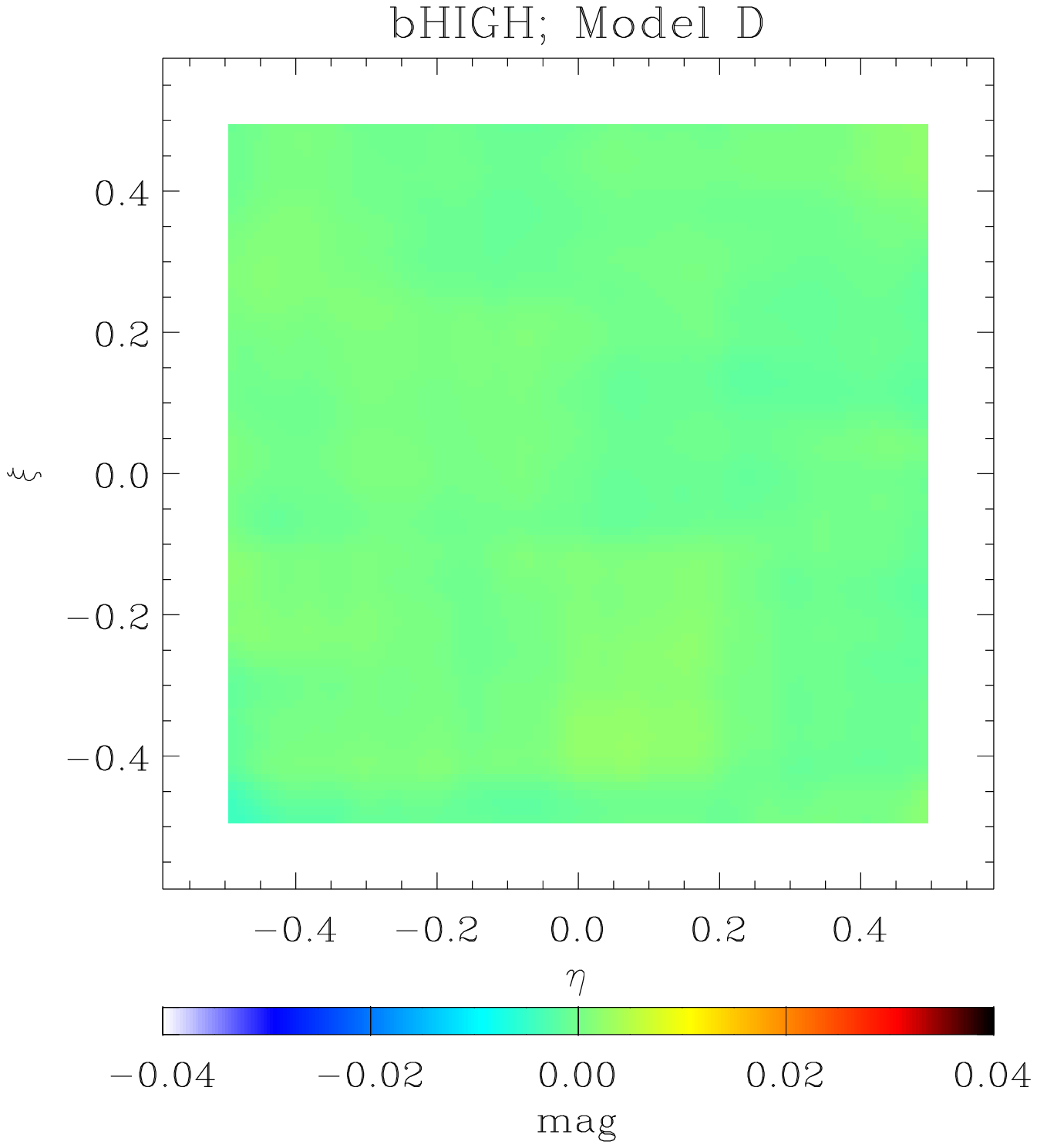,width=4.2cm,clip=,bb=35 413 409 785}
     \psfig{file=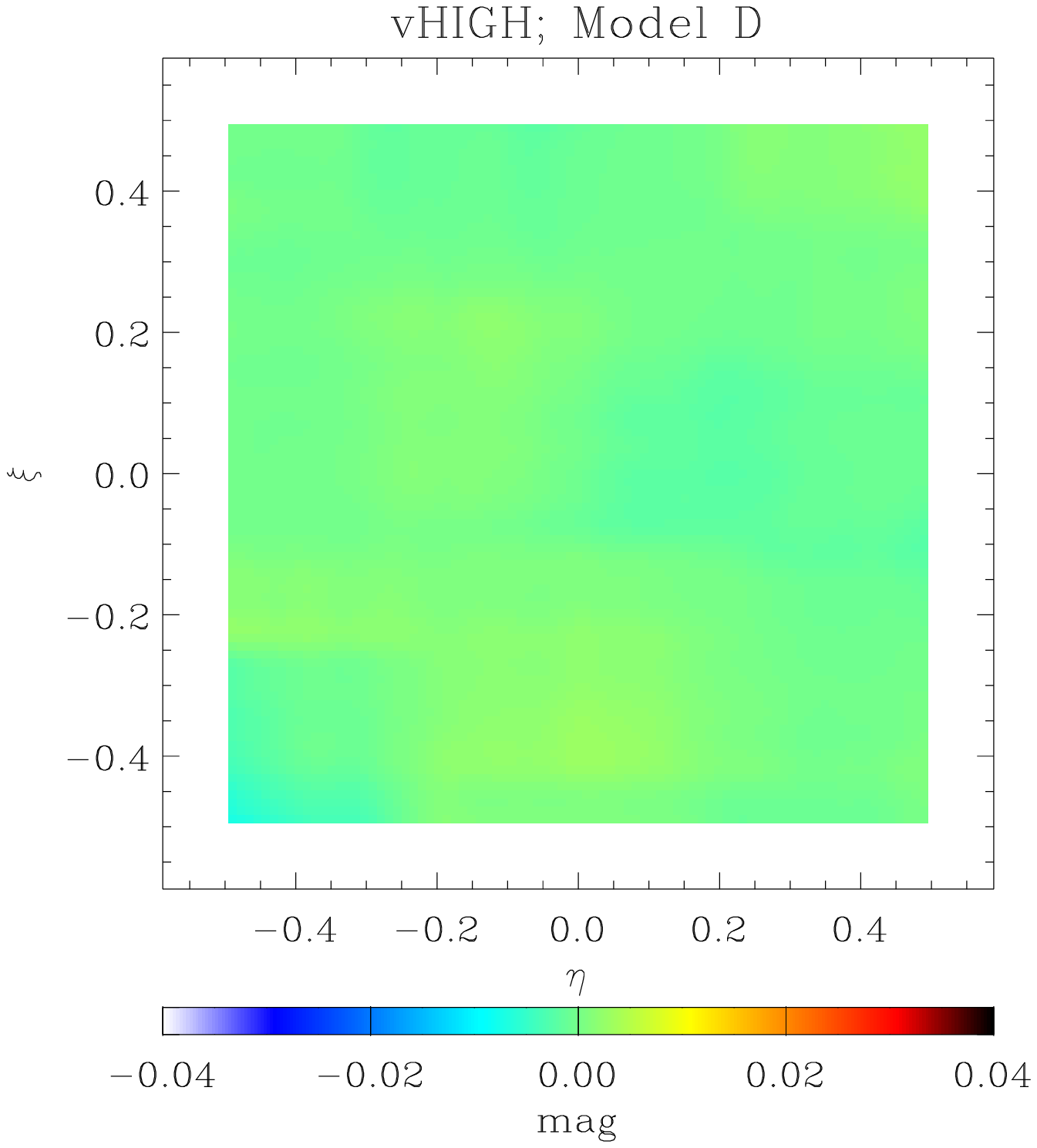,width=4.2cm,clip=,bb=35 413 409 785}
     \psfig{file=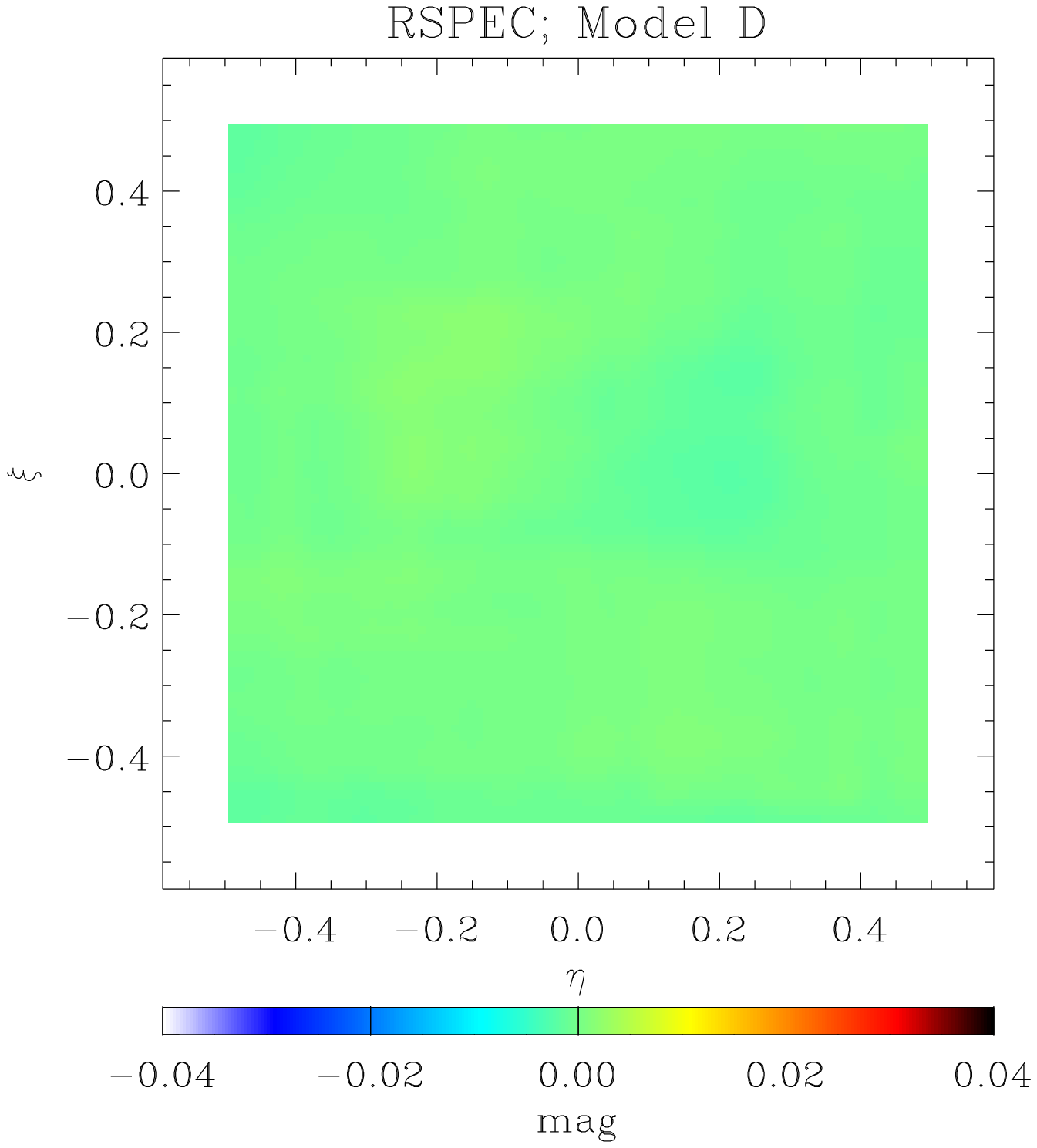,width=4.2cm,clip=,bb=35 413 409 785}
     \psfig{file=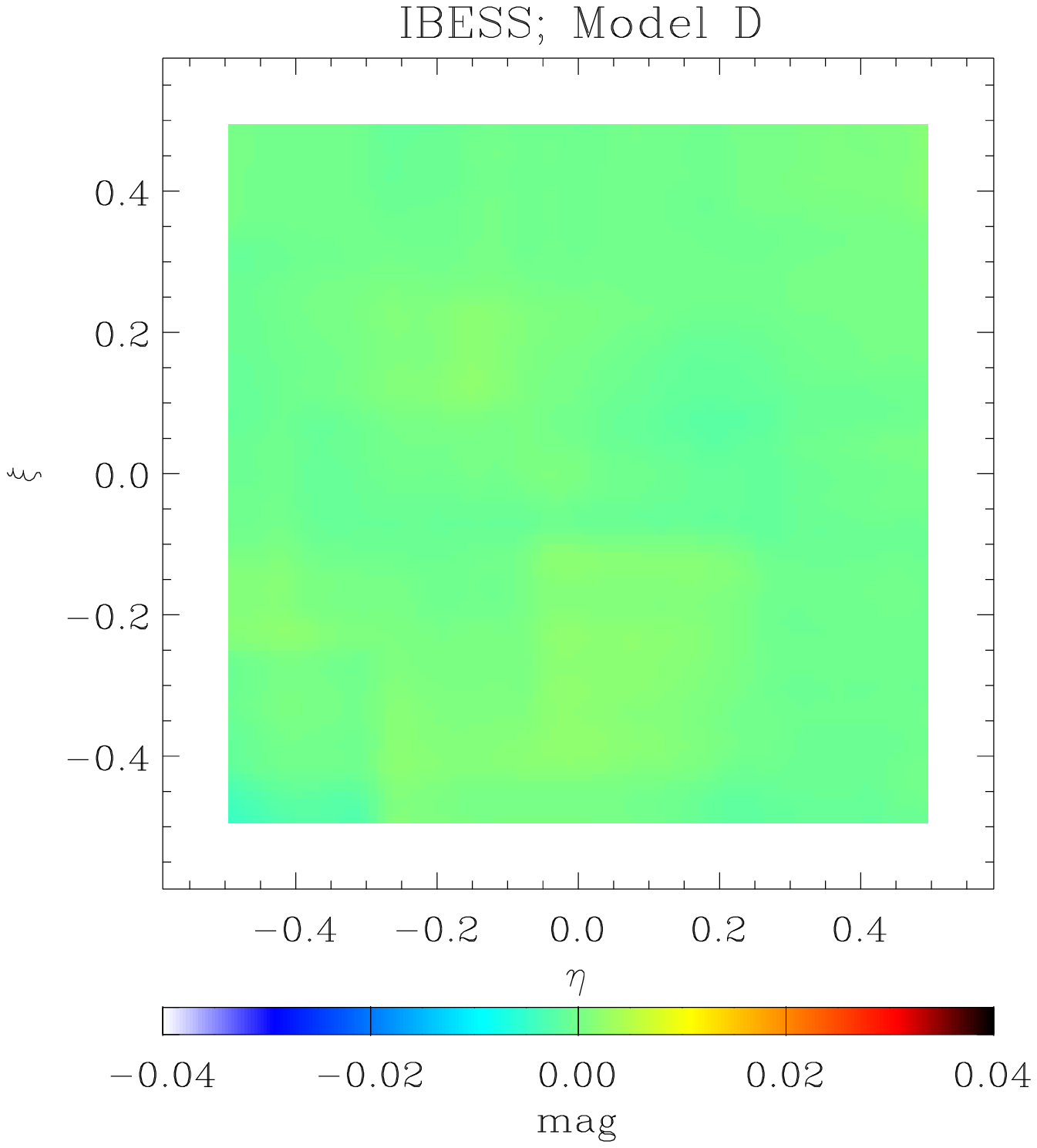,width=4.2cm,clip=,bb=35 413 409 785}
     \psfig{file=residuals_2D_paperbHIGH_2D_time_rangeB.ps,width=4.2cm,clip=,bb=75 358 423 414,angle=90}
   }
 }
}
\caption{Two-dimensional maps of mean residuals; panels (a) refer to
  the data in the time range A (1st November 2011 -- 30th May 2012),
  whereas panels (b) refer to the time range B (12th June 2012 -- 7th
  July 2013).  {\it Both panels}: residuals are computed as the
  difference between each star magnitude and the model
  prediction. Upper panels refer to models with no illumination
  corrections (model A), lower panels refer to the best fit models
  (model D), where both static and rotating illumination corrections
  have been fitted. Plots in different columns represent the results
  from different filters. Observations are distributed approximately
  uniformly across the plotted area.}
\label{fig:res2d_1}
\end{figure*}

\subsection{Results}
\label{sec:results}

For each time range, we fitted 4 models that differ by the degrees of
the polynomials representing the static and rotating illumination
corrections. Table~\ref{tab:models} describes them in detail. Model A
represents the photometric model without fitting any illumination
corrections, model B represents the model that includes the static
illumination correction only (degree 2), model C includes the rotating
illumination correction only (degree 2), and model D includes both static and
rotating illuminating corrections (both of degree 2).

\begin{table*}
\caption{Description and results for the adopted photometric models.}
\begin{center}
\begin{tabular}{c c c c c c c c c c c c}
\hline
\noalign{\smallskip}
Model  & $D_{\rm stat}$ & $ D_{\rm rot}$ & $N$   &D.O.F.& $\chi^2$ & $\tilde\chi^2$& $N$ & D.O.F. & $\chi^2$ & $\tilde\chi^2$\\
(1)    &(2)  & (3)   & (4)   & (5)  &   (6)    &    (7)        &(8)  & (9)    & (11)     & (12)         \\
\hline
\noalign{\smallskip}                                                                                                                                                        
\multicolumn{3}{c}{} & \multicolumn{8}{c}{\bf Time range A (1st November 2011 -- 30th May 2012)}\\
\noalign{\smallskip}                                                                                                                                                        
       &     &       &\multicolumn{4}{c}{\bf B filter}                   & \multicolumn{4}{c}{\bf V filter}               \\
  A    &  0  &   0   & 10344 & 7177 &    41776 &  5.82          & 13807&   9792&   28016 & 2.86  \\
  B    &  0  &   2   & 10344 & 7173 &    41021 &  5.72          & 13807&   9788&   27851 & 2.85  \\
  C    &  2  &   0   & 10344 & 7172 &    40336 &  5.62          & 13807&   9787&   25689 & 2.62  \\
  D    &  2  &   2   & 10344 & 7168 &    39681 &  5.54          & 13807&   9783&   25650 & 2.62  \\                                                                                                                                                       
\noalign{\smallskip}                                                                                                                                                        
       &     &       &   \multicolumn{4}{c}{\bf R filter}                & \multicolumn{4}{c}{\bf I filter} \\
\noalign{\smallskip}                                                                     
  A    &  0  &   0   &16351&   11786  &     23407&  1.99  & 14781 &  10612  &     23488 & 2.21  \\
  B    &  0  &   2   &16351&   11782  &     23127&  1.96  & 14781 &  10608  &     23365 & 2.20  \\
  C    &  2  &   0   &16351&   11781  &     21841&  1.85  & 14781 &  10607  &     21491 & 2.03  \\
  D    &  2  &   2   &16351&   11777  &     21741&  1.85  & 14781 &  10603  &     21460 & 2.02  \\
\noalign{\smallskip}                                                                                                                                                        
\noalign{\smallskip}
\multicolumn{3}{c}{} & \multicolumn{8}{c}{\bf Time range B (12th June 2012 -- 7th July 2013)}\\
\noalign{\smallskip}
       &     &       &\multicolumn{4}{c}{\bf B filter}                   & \multicolumn{4}{c}{\bf V filter}               \\
\noalign{\smallskip}                                                                                                                                                        
  A    &  0  &   0   & 79428 & 67418&   197422 &  2.93          &115965&  98895& 326958  & 3.31  \\
  B    &  0  &   2   & 79428 & 67414&   194185 &  2.88          &115965&  98891& 325010  & 3.29  \\
  C    &  2  &   0   & 79428 & 67413&   179661 &  2.67          &115965&  98890& 302577  & 3.06  \\
  D    &  2  &   2   & 79428 & 67409&   177067 &  2.63          &115965&  98886& 300813  & 3.04  \\
\noalign{\smallskip}                                                                                                                                                        
       &     &       &\multicolumn{4}{c}{\bf R filter}                   & \multicolumn{4}{c}{\bf I filter}               \\
\noalign{\smallskip}                                                                                                                                                        
  A    &  0  &   0   &131807&   111030  &     300992&  2.71  & 137214 &  115752  &  295460  &2.55   \\
  B    &  0  &   2   &131807&   111026  &     299980&  2.70  & 137214 &  115748  &  295050  &2.55   \\
  C    &  2  &   0   &131807&   111025  &     285045&  2.57  & 137214 &  115747  &  277629  &2.40   \\
  D    &  2  &   2   &131807&   111021  &     284093&  2.56  & 137214 &  115743  &  277158  &2.39   \\
\noalign{\smallskip}                                                                                                                                                        
\hline
\label{tab:models}
\end{tabular}
\begin{minipage}{18cm}
Note-- Col. 1: Model label. 
       Cols.2, 3: Degrees of the polynomial surfaces used for the static and rotating illumination corrections.
       Col. 4: Number of data points used in the fit, with reference to the $B$ and $R$ filters.
       Col. 5: Model degrees of freedom (D.O.F. = number of data points - number of model parameters), with reference to the $B$ and $R$ filters.
       Col. 6, 7: $\chi^2$ and reduced $\chi^2$ of the model fit, with reference to the $B$ and $R$ filters.
       Cols.8-12: Same as Cols 4-7, but with reference to the $V$ and $I$ filters.
\end{minipage}
\end{center}
\end{table*}

We repeated the fit many times discarding randomly 40\% of the
measurements to check the robustness of the fit results, finding
consistent static and rotating illumination corrections in all of the random
realizations.

Figure~\ref{fig:res2d_1} shows the maps of the mean residuals of our
models across the focal plane (i.e. in the $(\eta,\xi)$ coordinate
system).  If large-scale spatial sensitivity variations are
negligible, then, for model A, where no static or rotating
illumination correction terms are fitted, there should be no structure
in the residual maps above the noise level. However, we do observe
such structure (see the top row of panels in Figure~\ref{fig:res2d_1})
across the focal plane. The inclusion of polynomial surfaces to
account for the static and rotating illumination corrections clearly
helps to remove these large-scale structures in the residuals and to
decrease the $\chi^2$ (Table~\ref{tab:models}).

\begin{figure*}
\subfigure[]{
\vbox{
  \hbox{
     \psfig{file=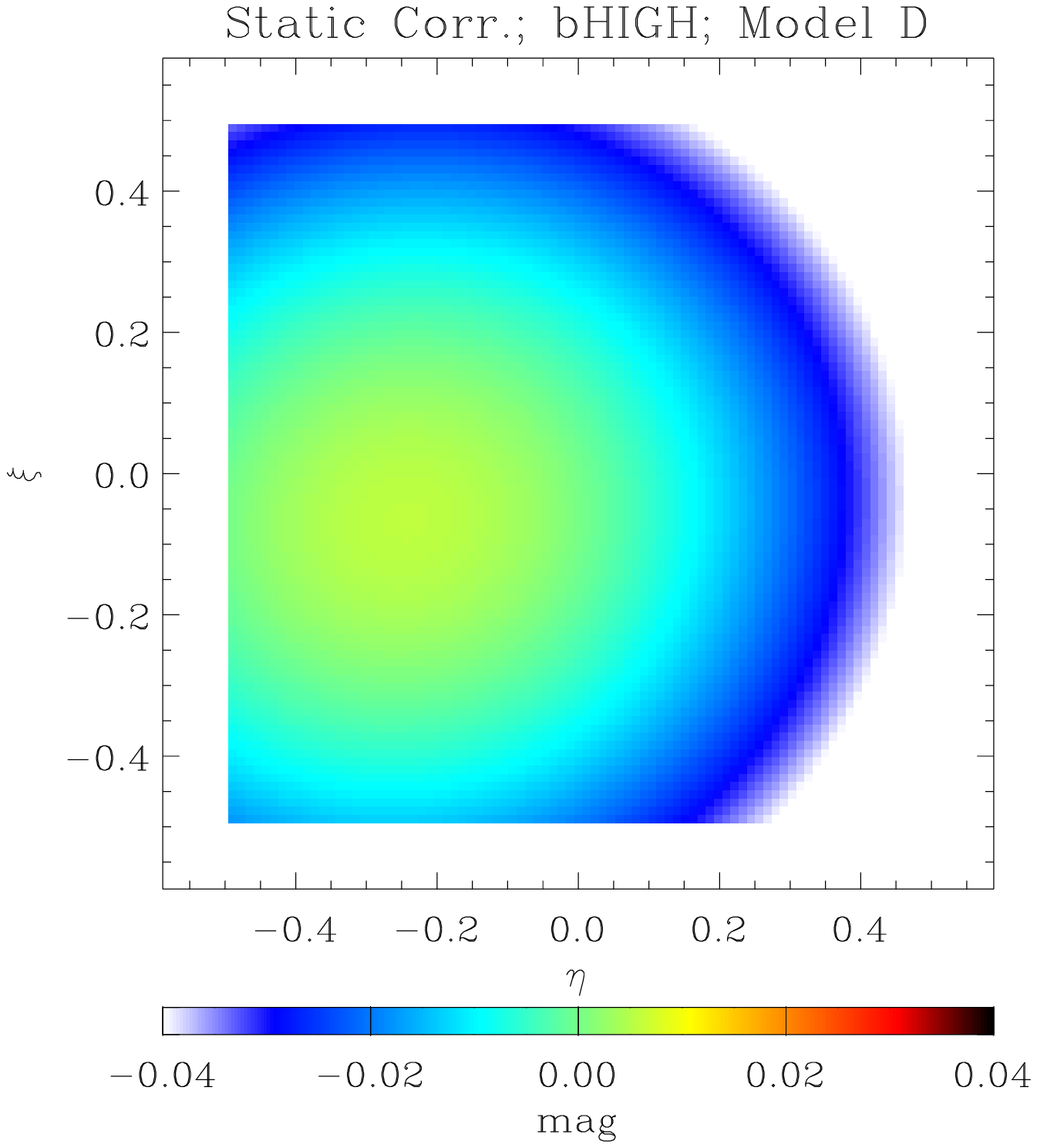,width=4.2cm,clip=,bb=35 413 409 785}
     \psfig{file=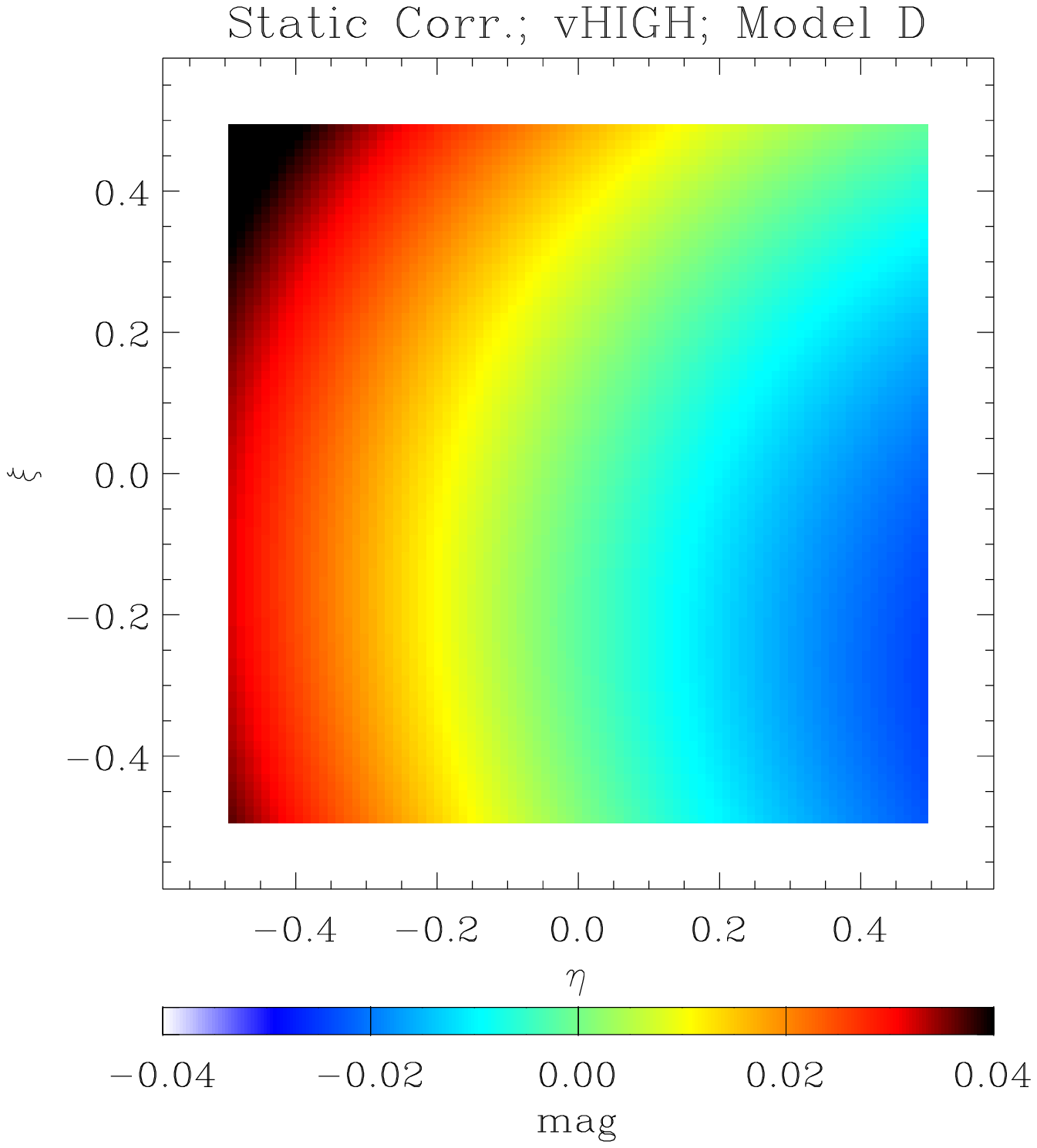,width=4.2cm,clip=,bb=35 413 409 785}
     \psfig{file=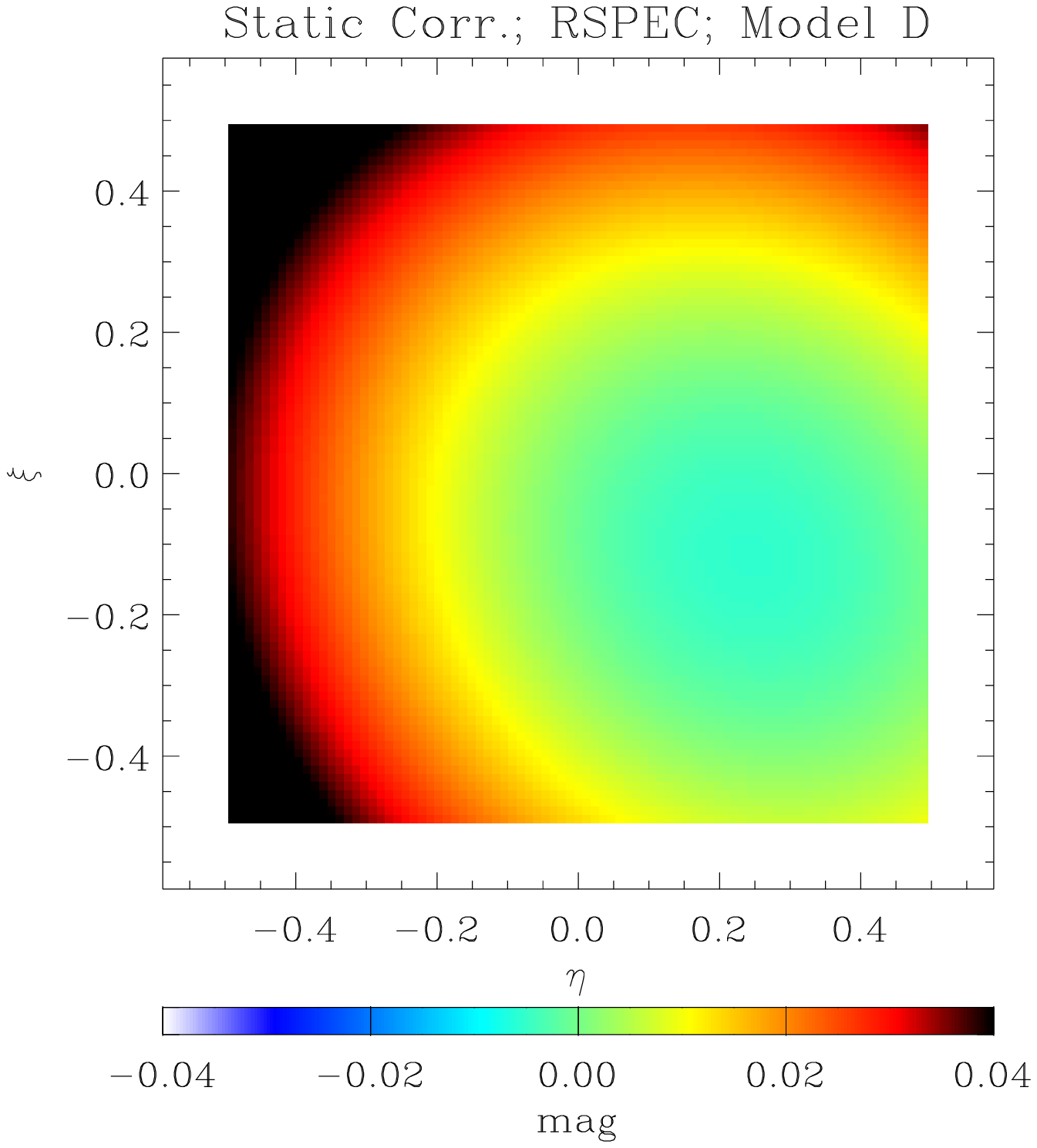,width=4.2cm,clip=,bb=35 413 409 785}
     \psfig{file=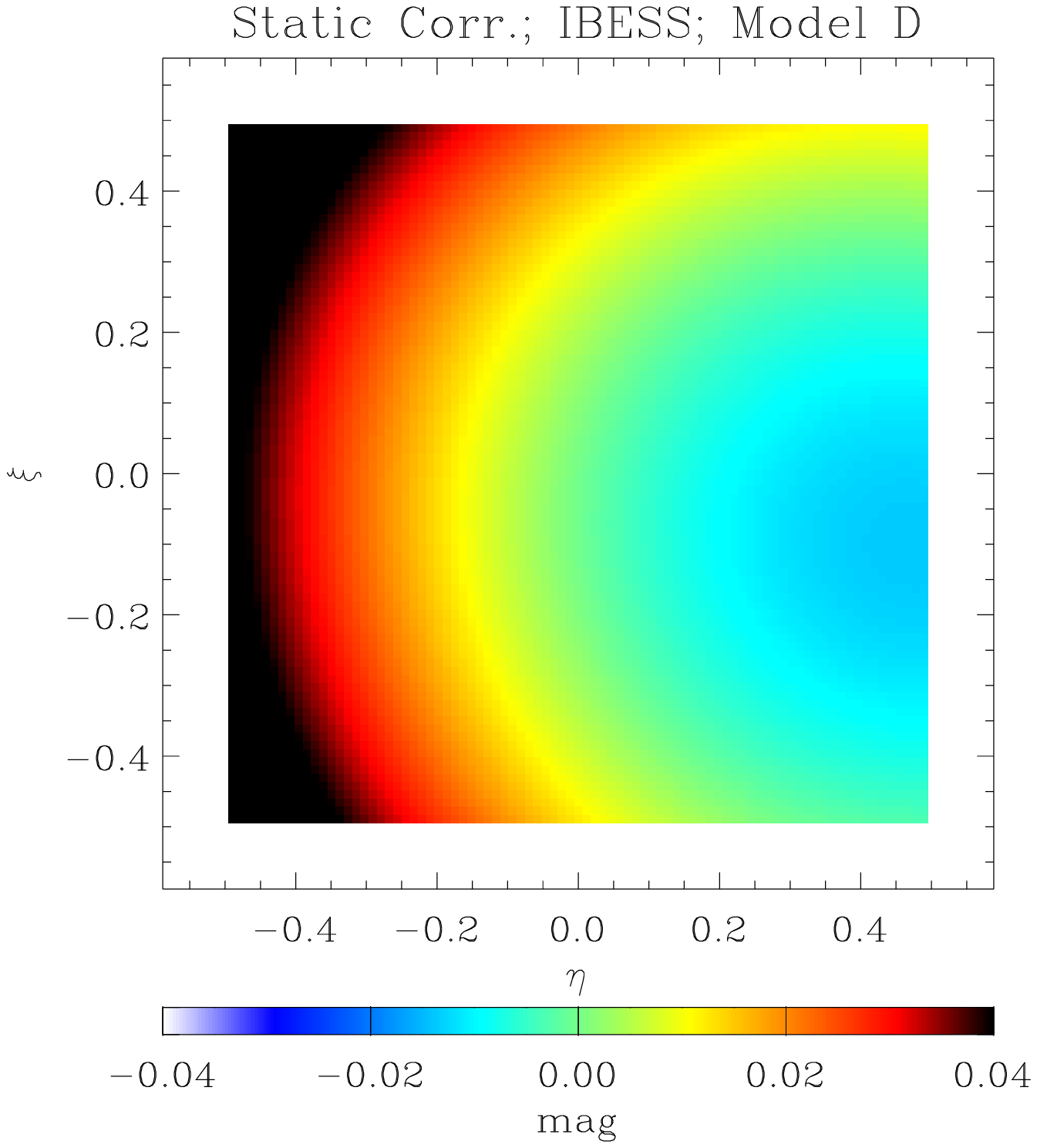,width=4.2cm,clip=,bb=35 413 409 785}
     \psfig{file=ill_patt_2D_paperIBESS_2D.ps,width=4.2cm,clip=,bb=75 358 423 414,angle=90}
   }
\vspace{.3cm}
  \hbox{
     \psfig{file=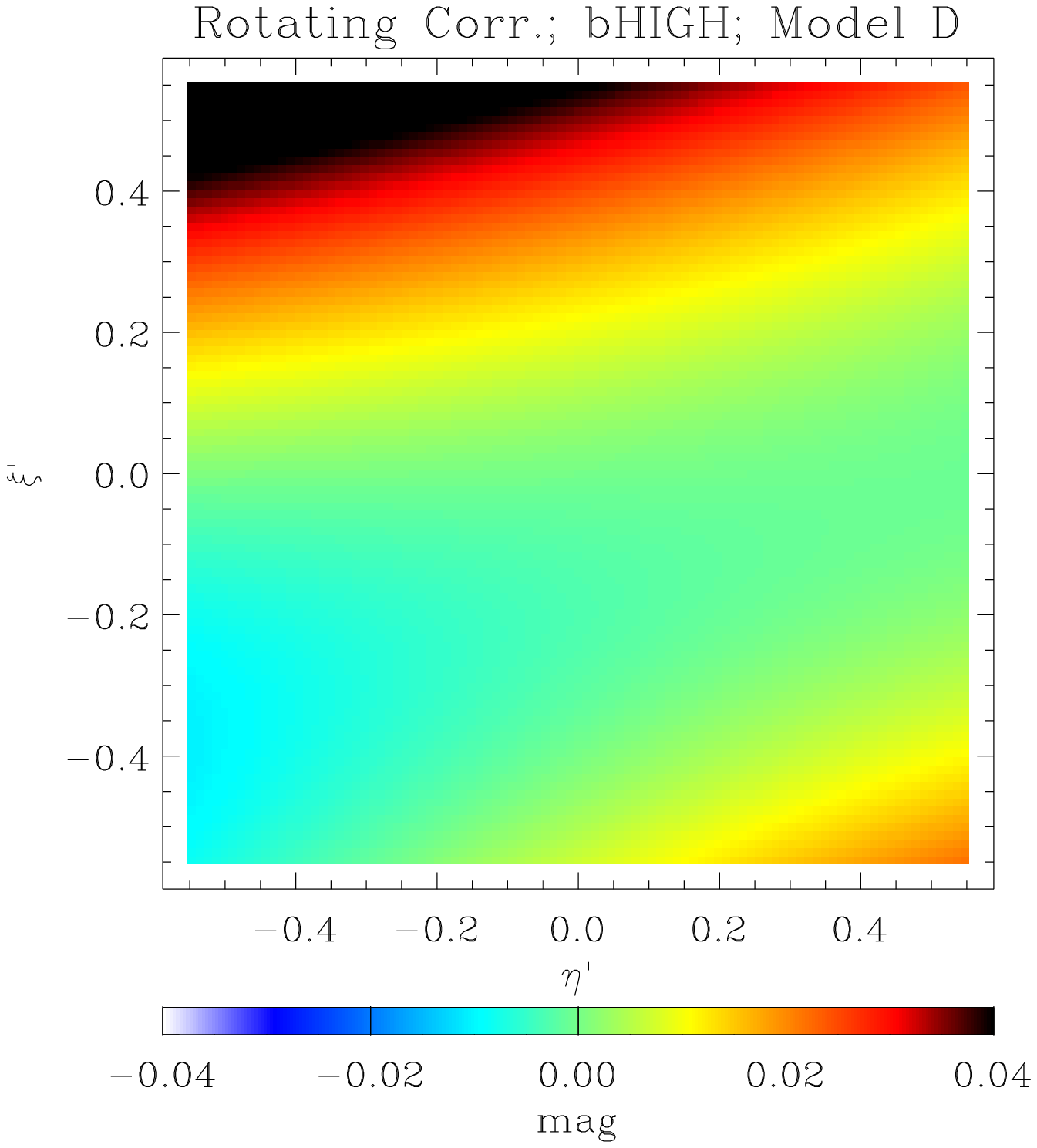,width=4.2cm,clip= ,bb=35 413 409 785}
     \psfig{file=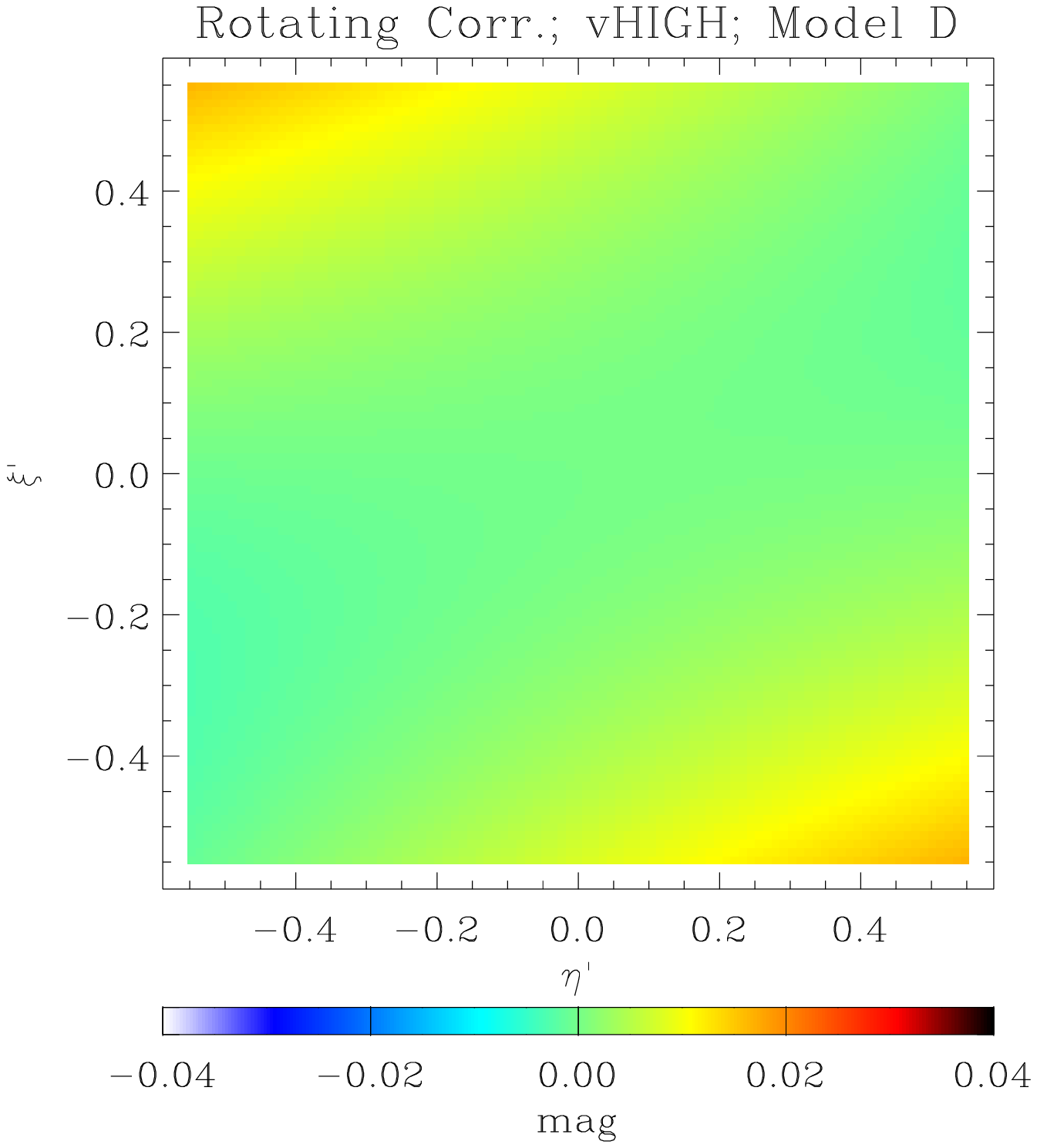,width=4.2cm,clip= ,bb=35 413 409 785}
     \psfig{file=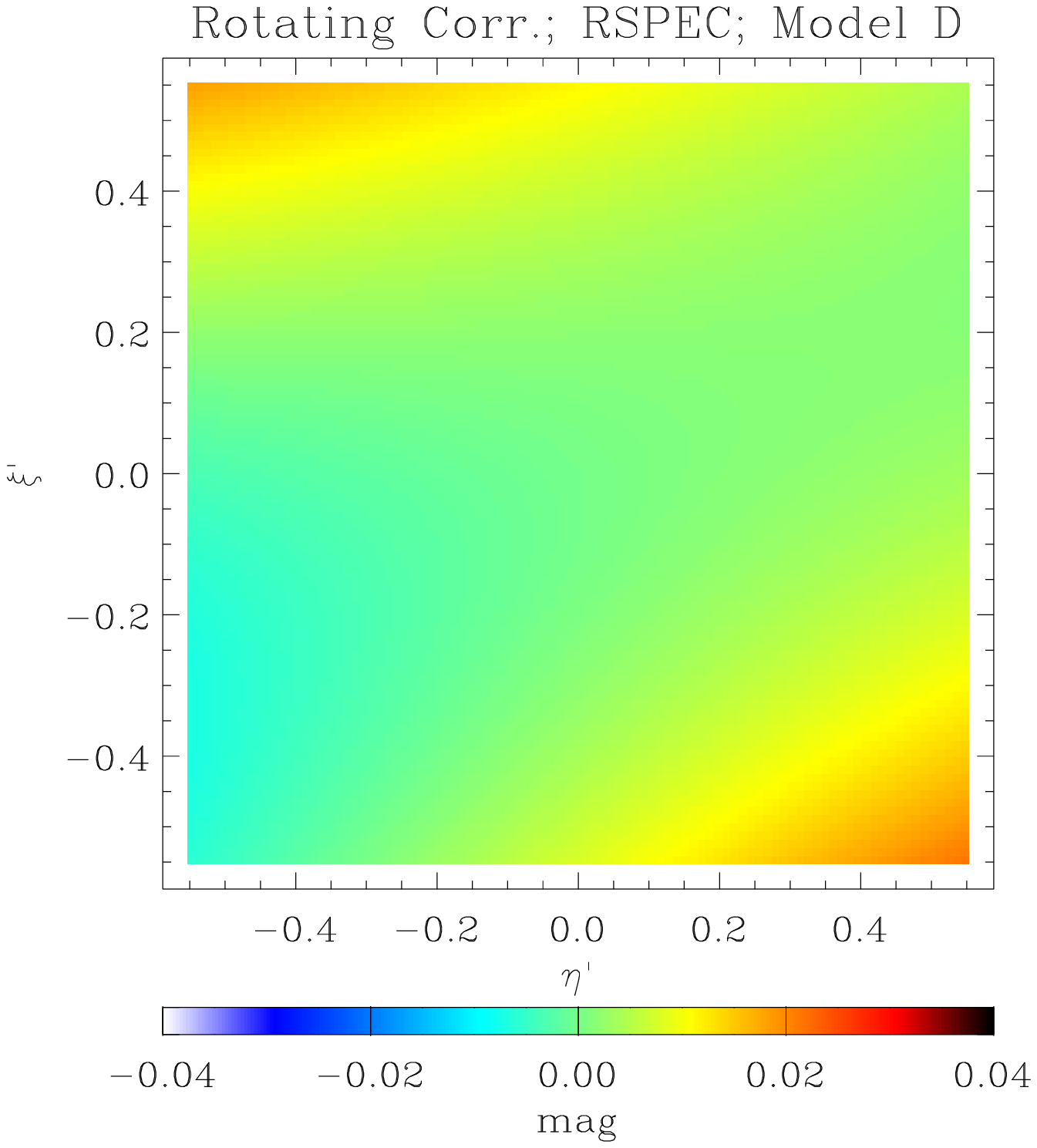,width=4.2cm,clip= ,bb=35 413 409 785}
     \psfig{file=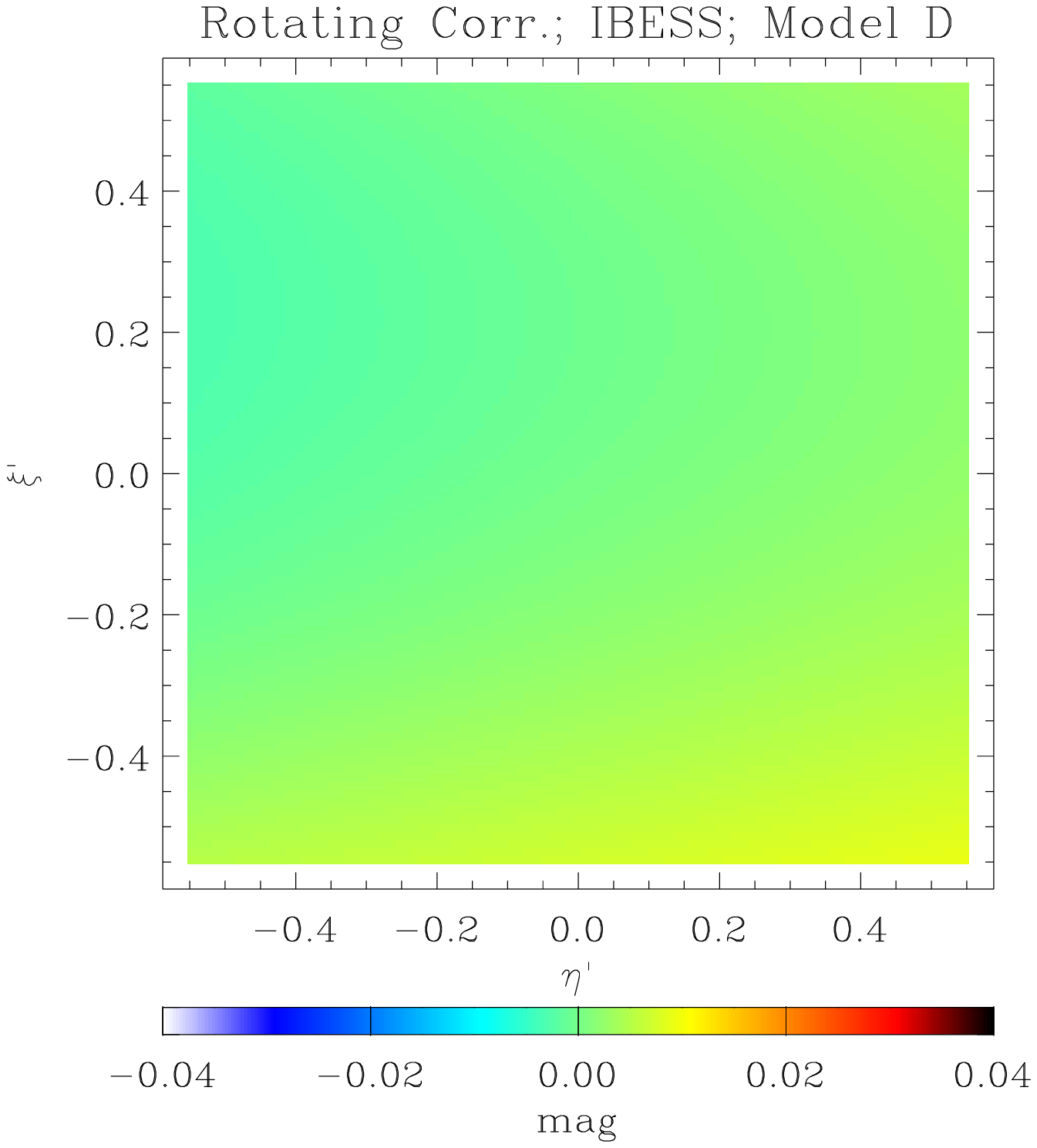,width=4.2cm,clip= ,bb=35 413 409 785}
     \psfig{file=rot_patt_2D_paperIBESS_2D.ps,width=4.2cm,clip= ,bb=75 358 423 414,angle=90}
   }
  }
}
\subfigure[]{
\vbox{
  \hbox{
     \psfig{file=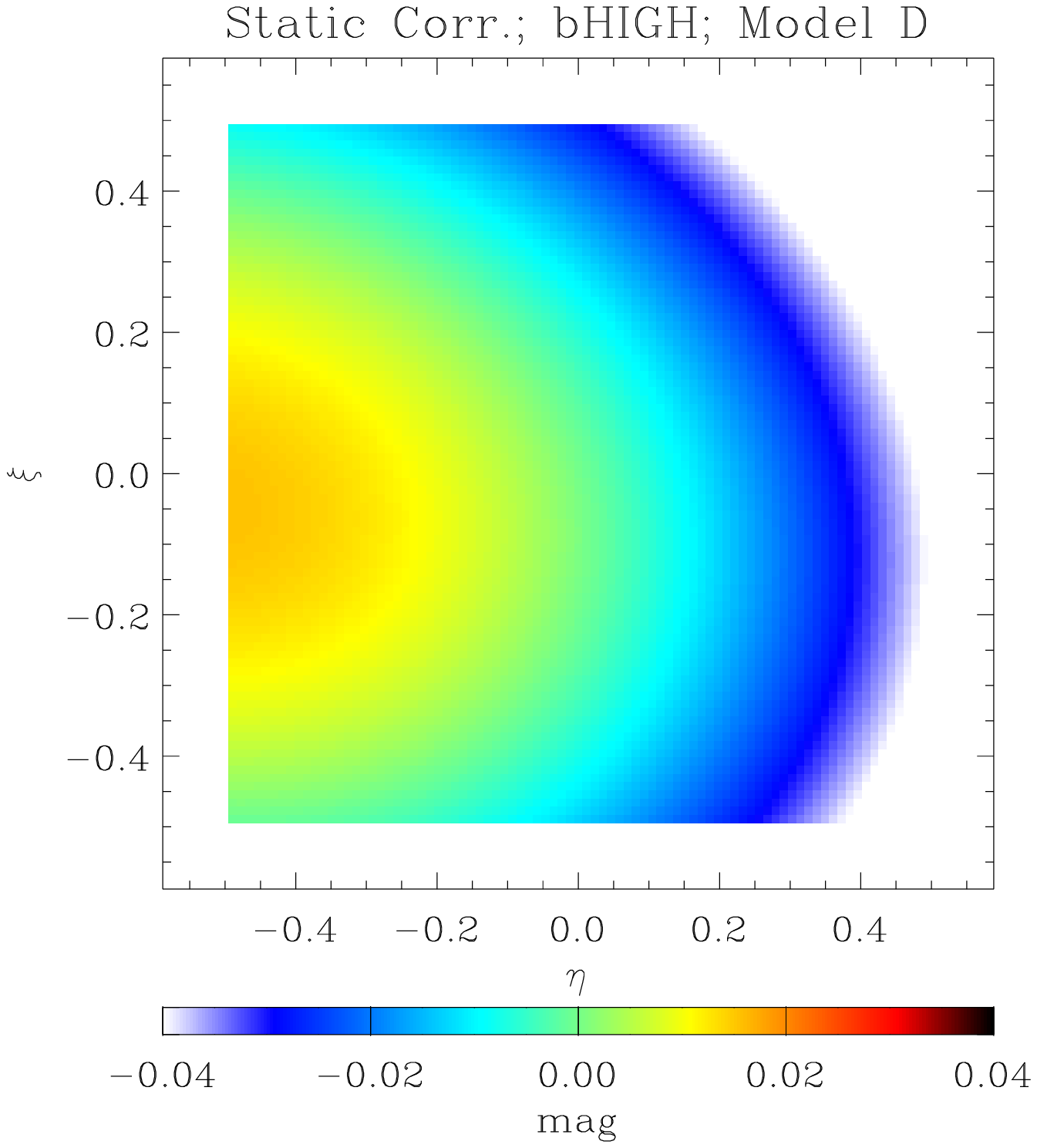,width=4.2cm,clip=,bb=35 413 409 785}
     \psfig{file=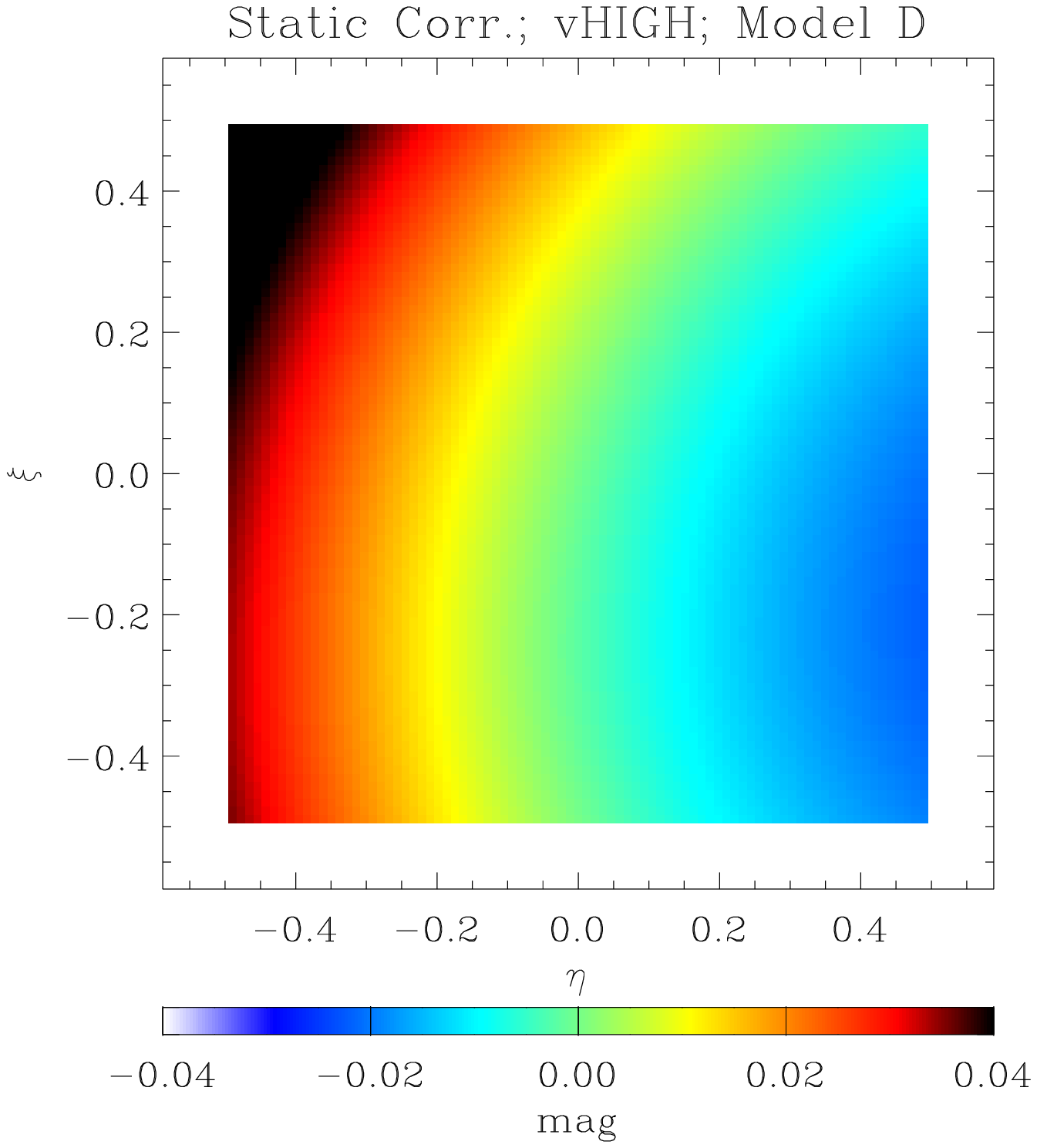,width=4.2cm,clip=,bb=35 413 409 785}
     \psfig{file=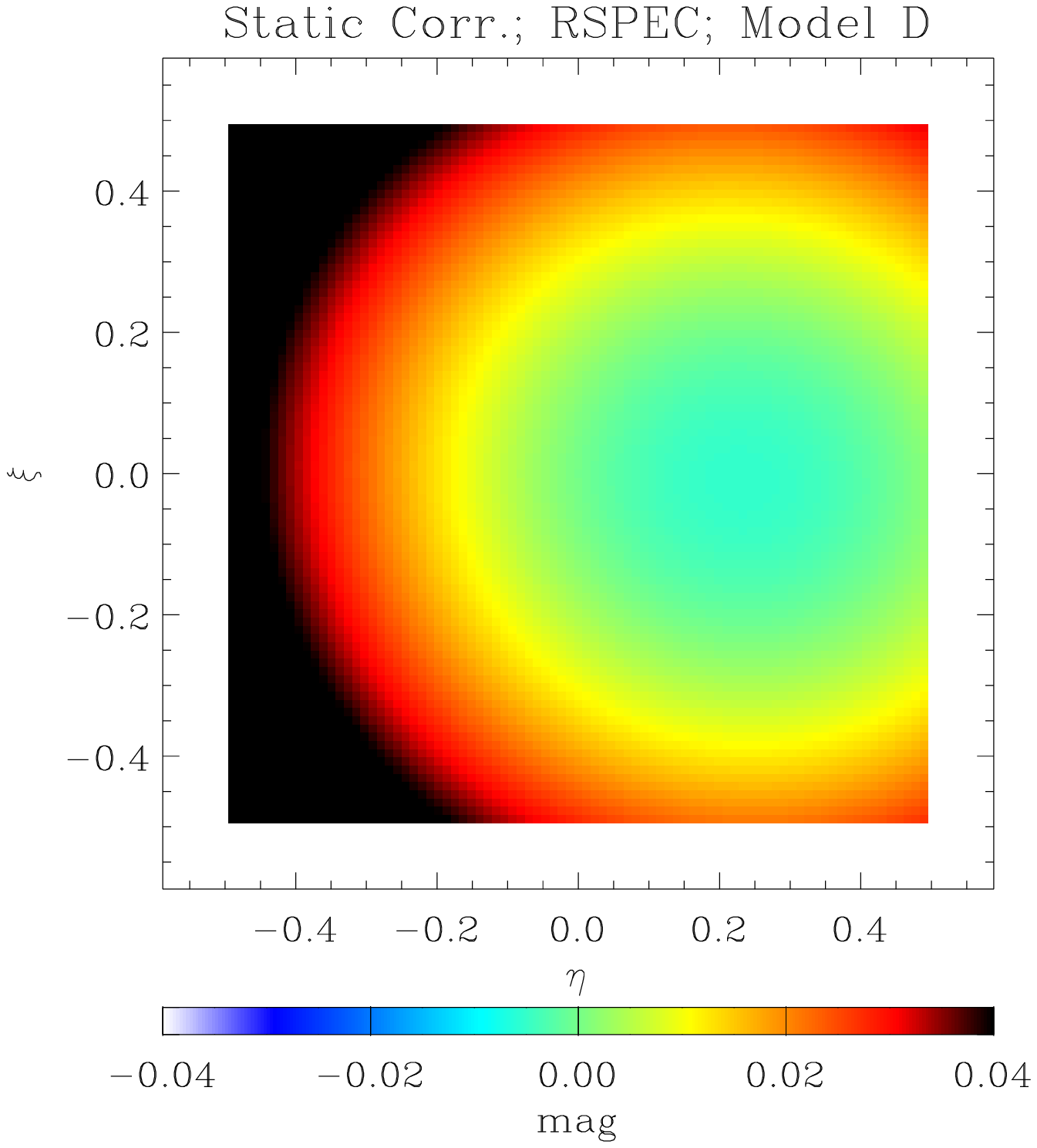,width=4.2cm,clip=,bb=35 413 409 785}
     \psfig{file=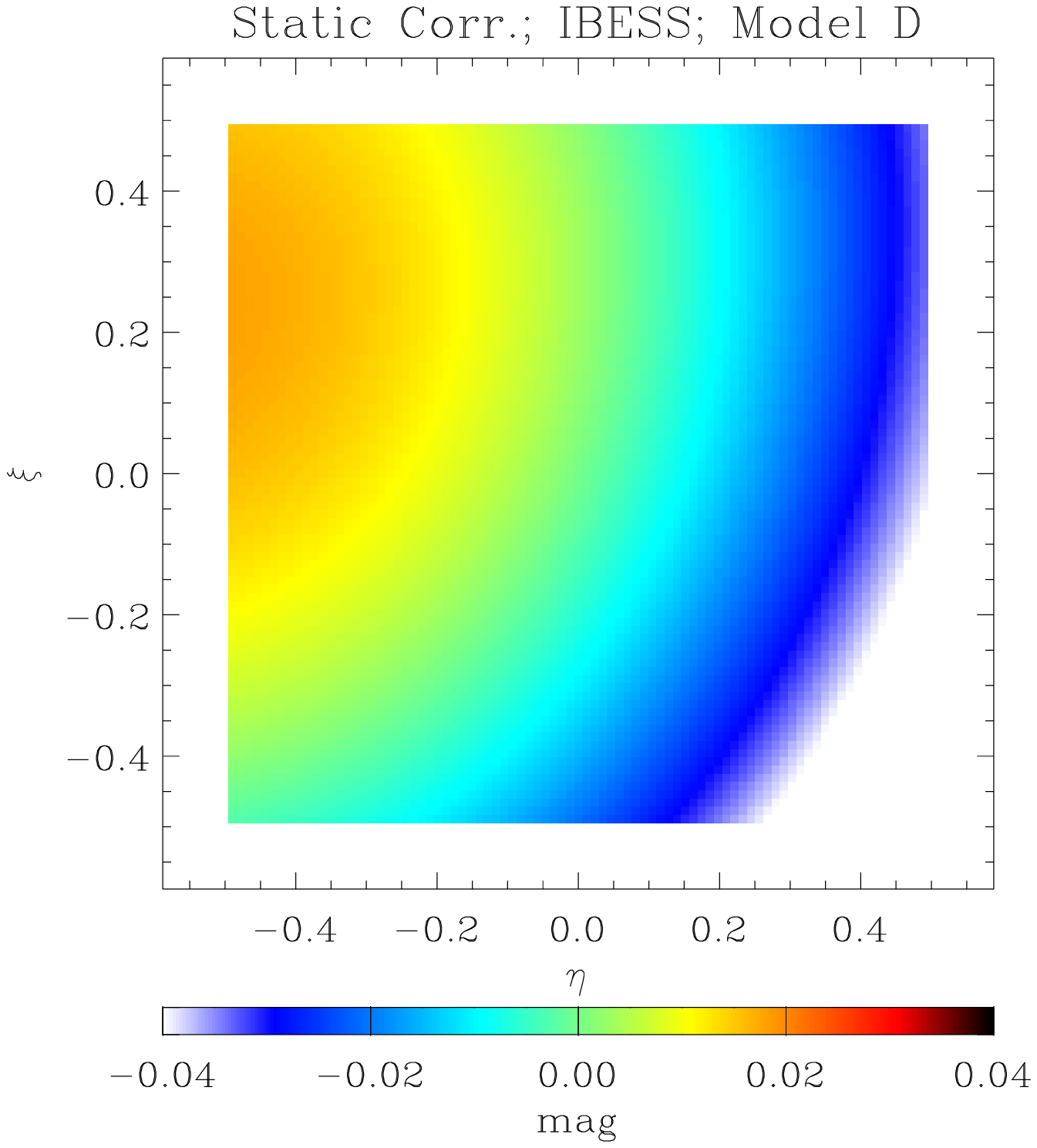,width=4.2cm,clip=,bb=35 413 409 785}
     \psfig{file=ill_patt_2D_paperIBESS_2D_time_rangeB.ps,width=4.2cm,clip=,bb=75 358 423 414,angle=90}
   }
\vspace{.3cm}
  \hbox{
     \psfig{file=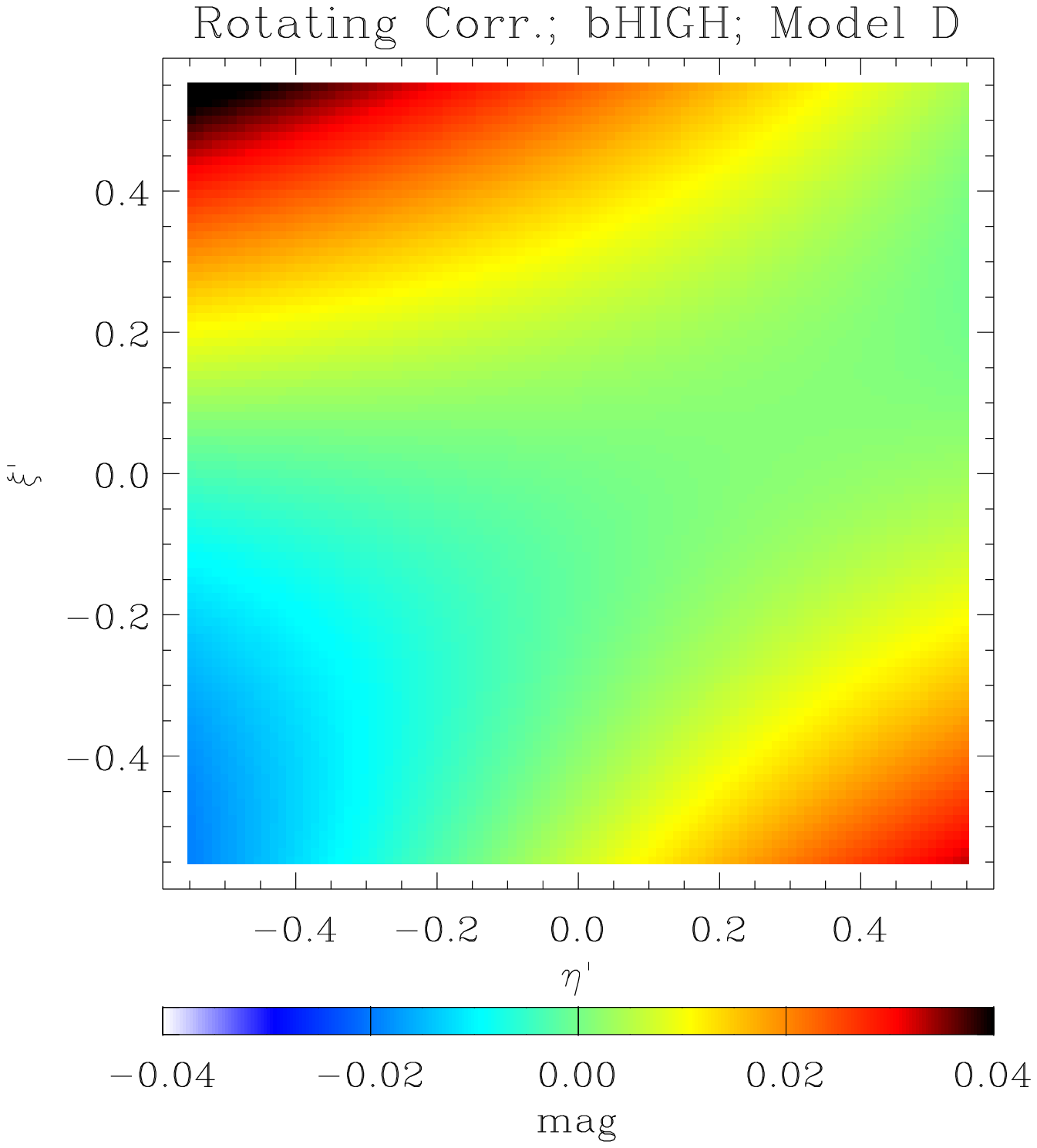,width=4.2cm,clip= ,bb=35 413 409 785}
     \psfig{file=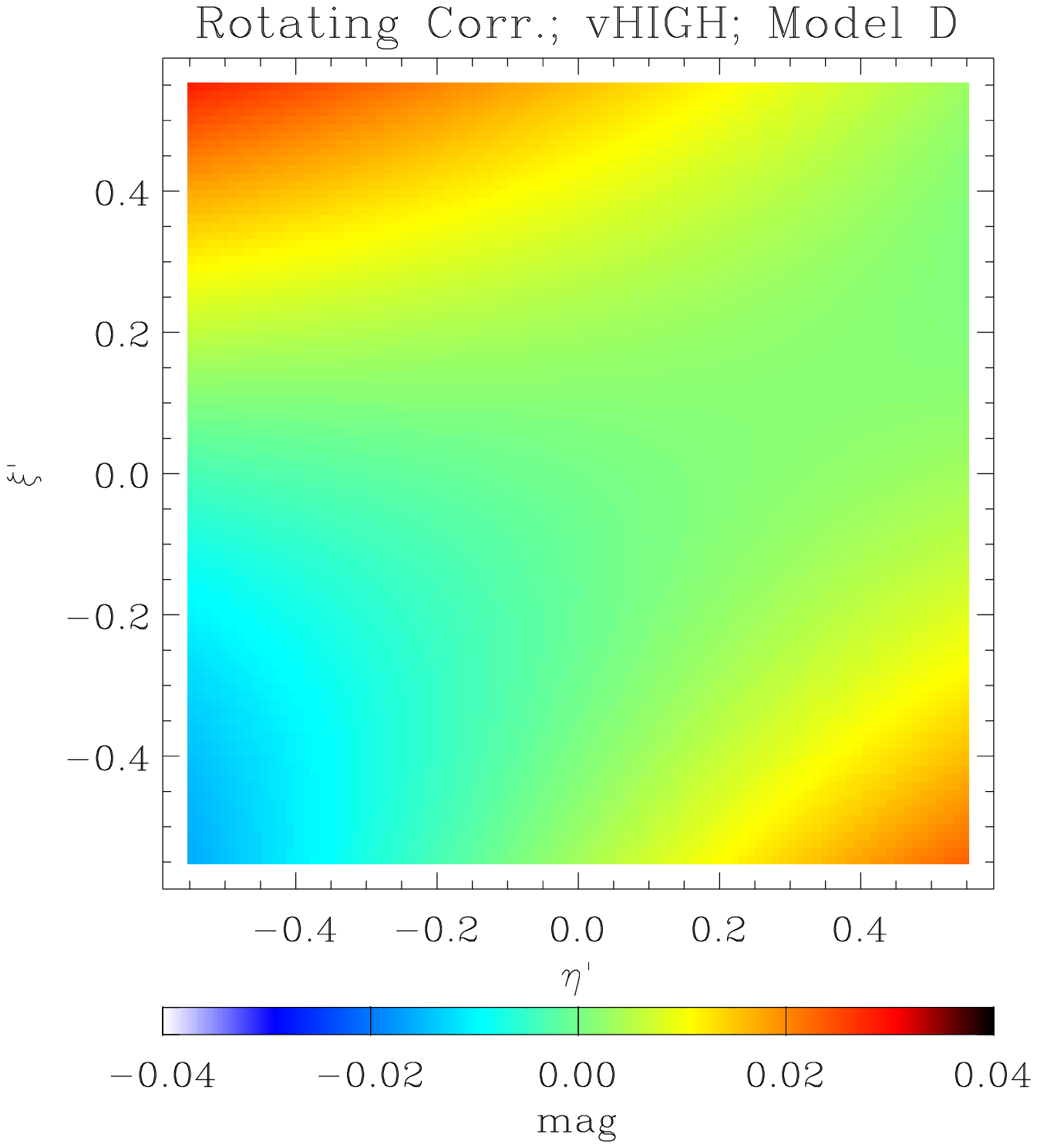,width=4.2cm,clip= ,bb=35 413 409 785}
     \psfig{file=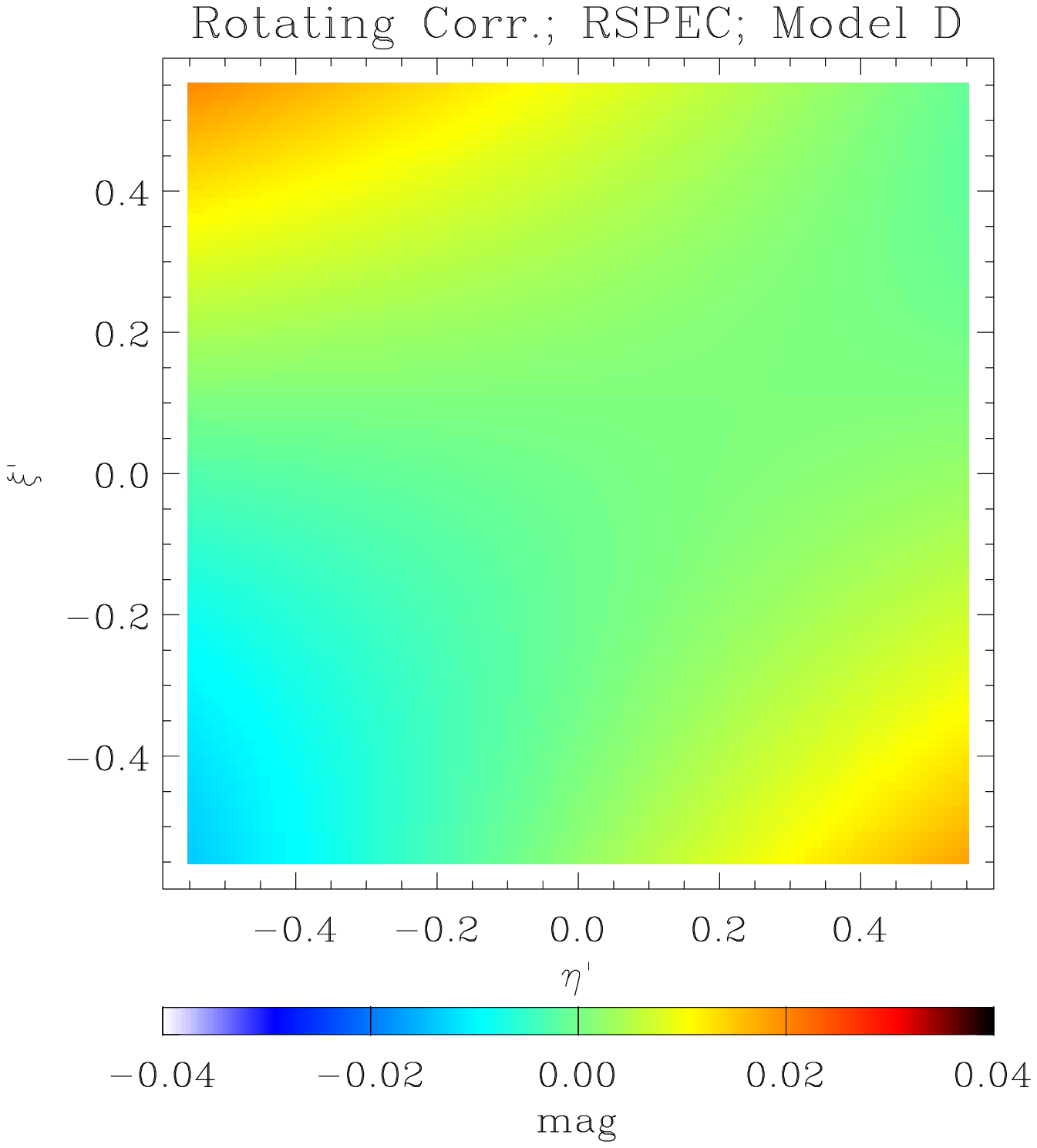,width=4.2cm,clip= ,bb=35 413 409 785}
     \psfig{file=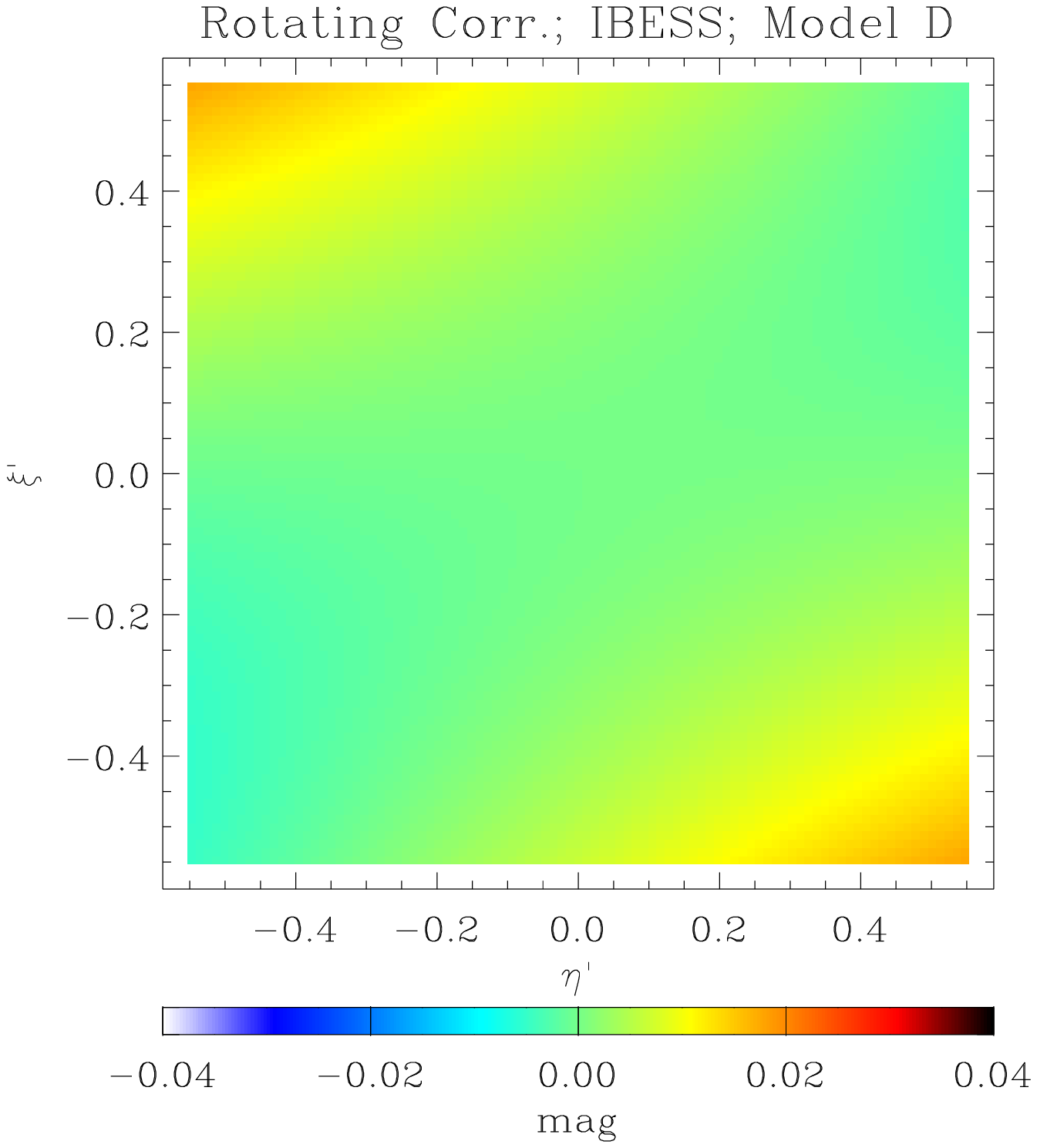,width=4.2cm,clip= ,bb=35 413 409 785}
     \psfig{file=rot_patt_2D_paperIBESS_2D_time_rangeB.ps,width=4.2cm,clip= ,bb=75 358 423 414,angle=90}
   }
  }
}
\caption{Best-fit polynomial surfaces (model D) representing the
  static illumination corrections $\static$ (upper panels) and the
  rotating illumination corrections $\rotating$ (lower panels) for
  different filters. Illumination correction surfaces in panels (a)
  refer to time range A, whereas surfaces in panels (b) refer to time
  range B.}
\label{fig:patterns_1}
\end{figure*}

Figure~\ref{fig:patterns_1} shows the best-fit two-dimensional surfaces
representing the static and rotating illumination corrections for the
different filters and for the two time ranges.

The shape of the illumination corrections for the two time ranges are
similar for all bands, except for the static correction in the $I$
filter. We are still investigating the source of this difference;
observations of standard stars after July 2013 will help to monitor
the variations of the illumination corrections with time.  At the
moment, we can exclude causes related to changes in the mechanical
components or the set-up of the rotator, because they would have
produced a large variation in the rotating patterns between the time
ranges for all bands. We can also exclude a degeneracy between the
static and rotating patterns, because it would have caused a variation
in the $I$ band rotating pattern coupled with the observed variation
in the static pattern.

\begin{table*}
\caption{Best-fit polynomial coefficients for each time range.}
\begin{center} 
 \begin{scriptsize}
\begin{tabular}{c c c c c c c c c c}
\hline
\noalign{\smallskip}
\noalign{\smallskip}
\multicolumn{10}{c}{\bf Time range A: 1st November 2011 -- 30th May 2012} \\
\noalign{\smallskip}
\multicolumn{2}{c}{} &  \multicolumn{2}{c}{\bf B filter:} & \multicolumn{2}{c}{\bf V filter:}  & \multicolumn{2}{c}{\bf R filter:} & \multicolumn{2}{c}{\bf I filter:}  \\
$m$  &  $n$&             $a_{mn}$          &            $b_{mn}$        &                $a_{mn}$          &          $b_{mn}$   &                  $a_{mn}$        &          $b_{mn}$         &                  $a_{mn}$          &          $b_{mn}$        \\
1    &  0  &     --0.043   $\pm$   0.001   &   --0.001  $\pm$   0.002  &     --0.054     $\pm$   0.001   &  0.001   $\pm$   0.001  &     --0.037  $\pm$   0.001   &   0.006   $\pm$   0.001  &    --0.055       $\pm$   0.001    &   0.004   $\pm$  0.002    \\
0    &  1  &     --0.011   $\pm$   0.003   &     0.031  $\pm$   0.001  &       0.016     $\pm$   0.002   &  0.001   $\pm$   0.001  &       0.015  $\pm$   0.002   &   0.003   $\pm$   0.001  &      0.009       $\pm$   0.003    & --0.006   $\pm$  0.001    \\
2    &  0  &     --0.089   $\pm$   0.008   &            --             &       0.023     $\pm$   0.007   &           --            &       0.081  $\pm$   0.007   &            --            &      0.059       $\pm$   0.008    &            --             \\
1    &  1  &       0.009   $\pm$   0.005   &   --0.052  $\pm$   0.008  &       0.011     $\pm$   0.005   & --0.027  $\pm$   0.007  &       0.023  $\pm$   0.004   &  --0.034  $\pm$   0.006  &      0.010       $\pm$   0.005    &   0.001   $\pm$  0.007    \\
0    &  2  &     --0.102   $\pm$   0.009   &     0.077  $\pm$   0.007  &       0.038     $\pm$   0.007   &   0.027  $\pm$   0.006  &       0.085  $\pm$   0.007   &    0.031  $\pm$   0.005  &      0.069       $\pm$   0.008    &   0.013   $\pm$  0.007    \\
\noalign{\smallskip}
\noalign{\smallskip}
\noalign{\smallskip}
\multicolumn{10}{c}{\bf Time range B: 12th June 2012 -- 7th July 2013} \\
\noalign{\smallskip}
\multicolumn{2}{c}{} &  \multicolumn{2}{c}{\bf B filter:} & \multicolumn{2}{c}{\bf V filter:}  & \multicolumn{2}{c}{\bf R filter:} & \multicolumn{2}{c}{\bf I filter:}  \\
$m$  &  $n$&             $a_{mn}$           &          $b_{mn}$          &                $a_{mn}$         &         $b_{mn}$         &            $a_{mn}$          &          $b_{mn}$        &                  $a_{mn}$          &          $b_{mn}$        \\
1    &  0  &    --0.0570   $\pm$  0.0005    &   0.0058   $\pm$  0.0006  &     --0.0574  $\pm$    0.0004  &  0.0063   $\pm$  0.0005  &    --0.0450  $\pm$    0.0004 &   0.0048  $\pm$   0.0005  &   --0.0544  $\pm$      0.0004   &     0.0014  $\pm$    0.0005    \\
0    &  1  &    --0.014    $\pm$  0.002     &   0.0154   $\pm$  0.0006  &       0.016   $\pm$     0.001  &  0.0109   $\pm$  0.0005  &    0.001     $\pm$     0.001 &   0.0065  $\pm$   0.0005  &     0.023   $\pm$      0.001    &     0.0013  $\pm$    0.0005    \\
2    &  0  &    --0.052    $\pm$  0.009     &              --           &       0.032   $\pm$     0.007  &         --               &      0.097   $\pm$     0.007 &            --             &   --0.045   $\pm$      0.008    &               --               \\
1    &  1  &    --0.013    $\pm$  0.002     & --0.075    $\pm$  0.002   &    --0.005    $\pm$     0.001  &  -0.052   $\pm$   0.002  &     0.008    $\pm$     0.001 &  --0.042  $\pm$    0.002  &     0.009    $\pm$     0.002    &   --0.034   $\pm$    0.002     \\
0    &  2  &    --0.082    $\pm$  0.009     &   0.050    $\pm$  0.002   &       0.033   $\pm$     0.007  &   0.031   $\pm$   0.002  &       0.113  $\pm$     0.007 &    0.019  $\pm$    0.002  &   --0.038   $\pm$      0.008    &     0.024   $\pm$    0.002     \\
\noalign{\smallskip}
\noalign{\smallskip}
\hline
\noalign{\smallskip}
\label{tab:coeff}
\end{tabular}
 \end{scriptsize}
\end{center}
\end{table*}

\section{Improving FORS2 photometric measurements}
\label{sec:improving}

We demonstrated in Section~\ref{sec:results} that the inclusion of static
and rotating illumination correction terms improves the photometric modelling
for FORS2 data. The best fit coefficients reported in Table~\ref{tab:coeff}
can be used to improve the calibration of photometric measurements
from any FORS2 observation.

The coefficients associated to time range B have on average smaller
errors than those of time range A, as a consequence of the larger
numbers of data points available in time range B (see Table
2). However, the different numbers of data points between the two time
ranges does not affect the improvement in photometry that can be
achieved: indeed we find later on that the systematic variations in
photometric data decrease by very similar amounts for both time ranges
when the correction coefficients are applied (see Section
\ref{sec:test_improvement}).

In this Section we test that photometric measurements that are
calibrated by the static and rotating illumination corrections
represent an improvement with respect to photometric
measurements that do not include these corrections.

Firstly, we define ``correct'' and ``incorrect'' photometric procedures to
be those that do and do not, respectively, take into
account the static and rotating large-scale spatial sensitivity
variations. ``Corrected'' magnitudes obtained by a correct photometric procedure
will be	denoted	by $m_i^C$. ``Uncorrected'' magnitudes obtained by an incorrect photometric procedure
will be denoted by $m_i^U$.

\subsection{``Uncorrected'' photometric measurements}
\label{sec:traditional}

``Uncorrected'' magnitudes can be obtained, for example, from a data
reduction procedure that follows the following steps:

\begin{enumerate}

\item Compute the master bias(es) and master flat(s), for example using {\tt
  fors\_bias} and {\tt fors\_img\_sky\_flat} recipes of the FORS2
  data reduction pipeline.\label{item:bias} Use them to calibrate the science image(s).

\item Perform the photometry of the stars on the science images, for
  example using the {\tt fors\_zeropoint} recipe of the FORS2 data
  reduction pipeline.

\item Using a set of standard star observations on photometric nights,
  determine the overall zero-point $Z$, detector zero-point offsets
  $\Delta Z_k^{\rm det}$, and nightly extinction coefficients $\lambda_{n}$
  where the index $n$ refers to the $n$th night. In other words, fit
  the following equation to the FORS2 standard star observations:

  \begin{equation}
    \overline{m}_i = M_j + Z + \Delta Z_k^{\rm det} + \lambda_{n} X_{i} 
    \label{eqn:apparent_mags}
  \end{equation}
  where $X_{i}$ is the airmass of the $i$th magnitude measurement.
  $\overline{m}_i$, $M_j$, $Z$, and $\Delta Z_k^{\rm det}$ are as in
  Equation \ref{eqn:phot_model_simple}.  This is a standard procedure
  for zero-point and atmospheric extinction determination for use in
  the calibration of photometric measurements of science objects.
 
\item Use the best-fit values of $Z$, $\Delta Z_k^{\rm det}$, and $\lambda_n$
  to calibrate the instrumental magnitudes of the stars in the science images to obtain the apparent
  ``uncorrected'' magnitudes $m_i^U$.

\end{enumerate}

\subsection{``Corrected'' photometric measurements}
\label{sec:improved}

A ``correct'' photometric data reduction scheme accounts for the
sensitivity variations $\static$ and $\rotating$. The most efficient
way to correct for $\static$ and $\rotating$ is to follow the points
listed in Section \ref{sec:traditional} above, but using a different
master flat frame than the one determined in the 1st point of the
list.
This ``improved'' master flat frame to be used to calibrate the
science images can be obtained as follows:

\begin{itemize}

\item Compute master bias(es) and master flat(s), for example using the {\tt
  fors\_bias} and {\tt fors\_img\_sky\_flat} recipes of the FORS2
  data reduction pipeline.

\item Remove any large-scale variations from the master flat(s) to leave only
  the pixel-to-pixel variations. This can be done for example by dividing
  the master flat by a smoothed version of itself, or by a best fitting
  polynomial surface (see Section~\ref{sec:obs_and_red}).

\item Multiply each pixel-to-pixel master flat determined above by the
  corresponding correcting surface $C(x,y,\theta)$ to obtain the
  improved master flat. The correcting surface is given by:

\begin{equation}
 C(x,y,\theta) =10^{-0.4 \, \left[ \static(\eta,\xi) \, + \, \rotating(\eta',\xi')\right]}
\end{equation}

$\static(\eta,\xi)$ and $\rotating(\eta',\xi')$ are the best-fit
polynomial-surface static and rotating illumination corrections,
respectively, whose coefficients are specified in Table
\ref{tab:coeff}.  We recommend to use the $a_{mn}$ and $b_{mn}$
coefficients determined for time range A if observations are prior to
31st May 2012, and those for time range B if observations are obtained
after 1st June 2012\footnote{As consistency check, we tested that if
  coefficients are applied to observations of the wrong time range,
  the systematic variations in photometric data can increase up to
  $\sim 1.7$\%, as measured following the prescriptions detailed in Section
  \ref{sec:test_improvement}.}.  Equations \ref{eqn:eta_rot},
\ref{eqn:xi_rot}, \ref{eqn:coords1a}, and \ref{eqn:coords1b} specify
the transformations to convert the detector pixel coordinates ($x$,
$y$) and rotator angle $\theta$ into the coordinates $(\eta,\xi)$ and
$(\eta',\xi')$ that enable the calculation of $\static(\eta,\xi)$ and
$\rotating(\eta',\xi')$ via Equations \ref{eqn:stat_poly} and
\ref{eqn:rot_poly}.  The rotator angle $\theta$ is read from the
header keywords of the master flat, as specified in Section
\ref{sec:reference_system}. The use of the improved master flats
automatically corrects for the large-scale spatial sensitivity
variations across the focal plane.

\end{itemize}

``Corrected'' instrumental magnitudes $m_i^C$ will be obtained
following the steps 2-4 in Section \ref{sec:traditional} and using the
improved master flats determined above.

\subsection{Improvement in photometry}
\label{sec:test_improvement}
We now want to test if, and quantify by how much, the application of
the static and rotating illumination corrections determined in Section
\ref{sec:results} to the master flats improves the photometry that is
performed on FORS2 images.

We consider now the uncorrected (Section \ref{sec:traditional}) and
corrected apparent magnitudes (Section \ref{sec:improved}) determined
for all of the standard stars in our data set (see Equation
\ref{eqn:apparent_mags}) observed at least 10 times under photometric
conditions in each filter.

\begin{figure*}
\vbox{
  \hbox{
     \psfig{file=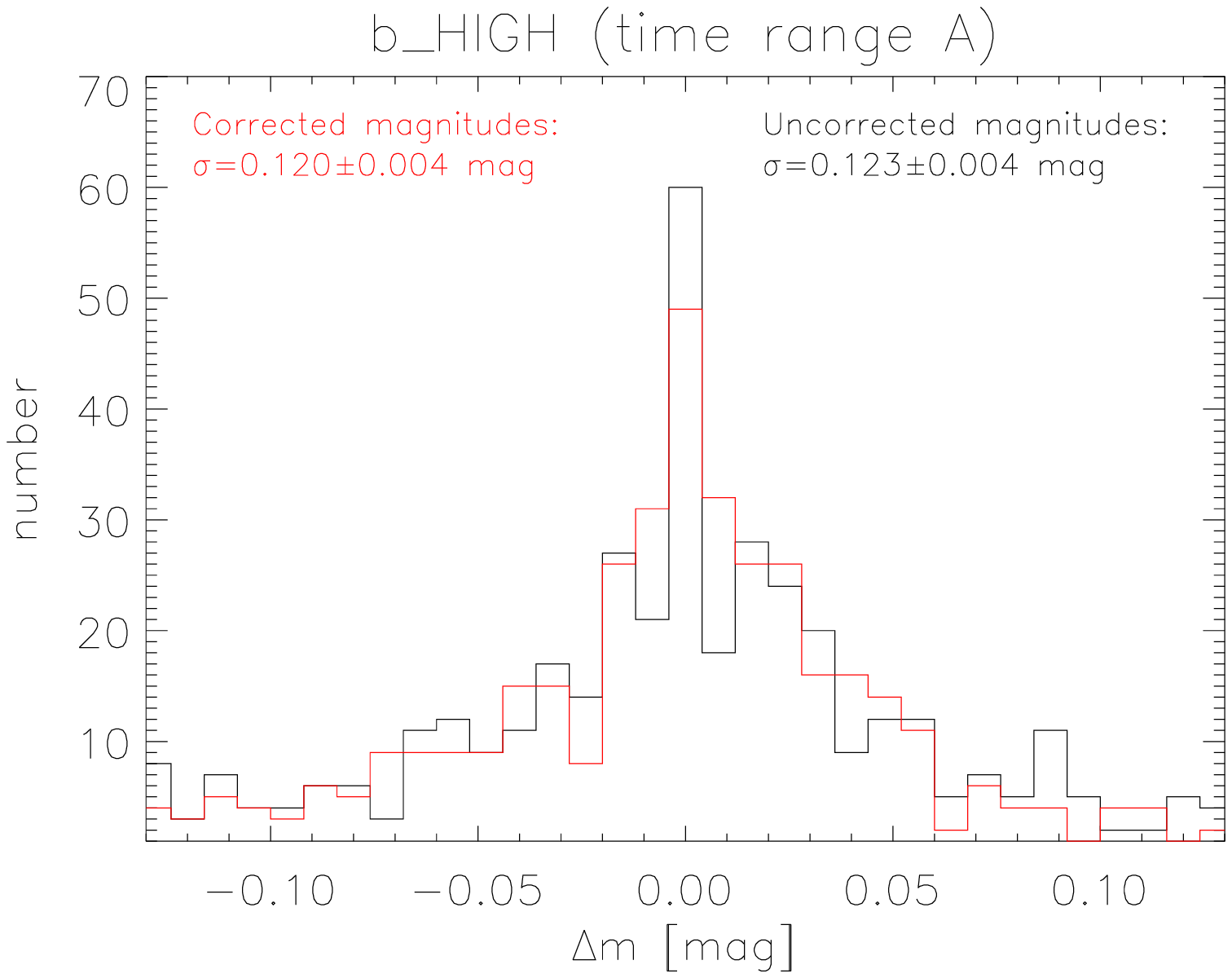, width=8.0cm,clip= }
     \psfig{file=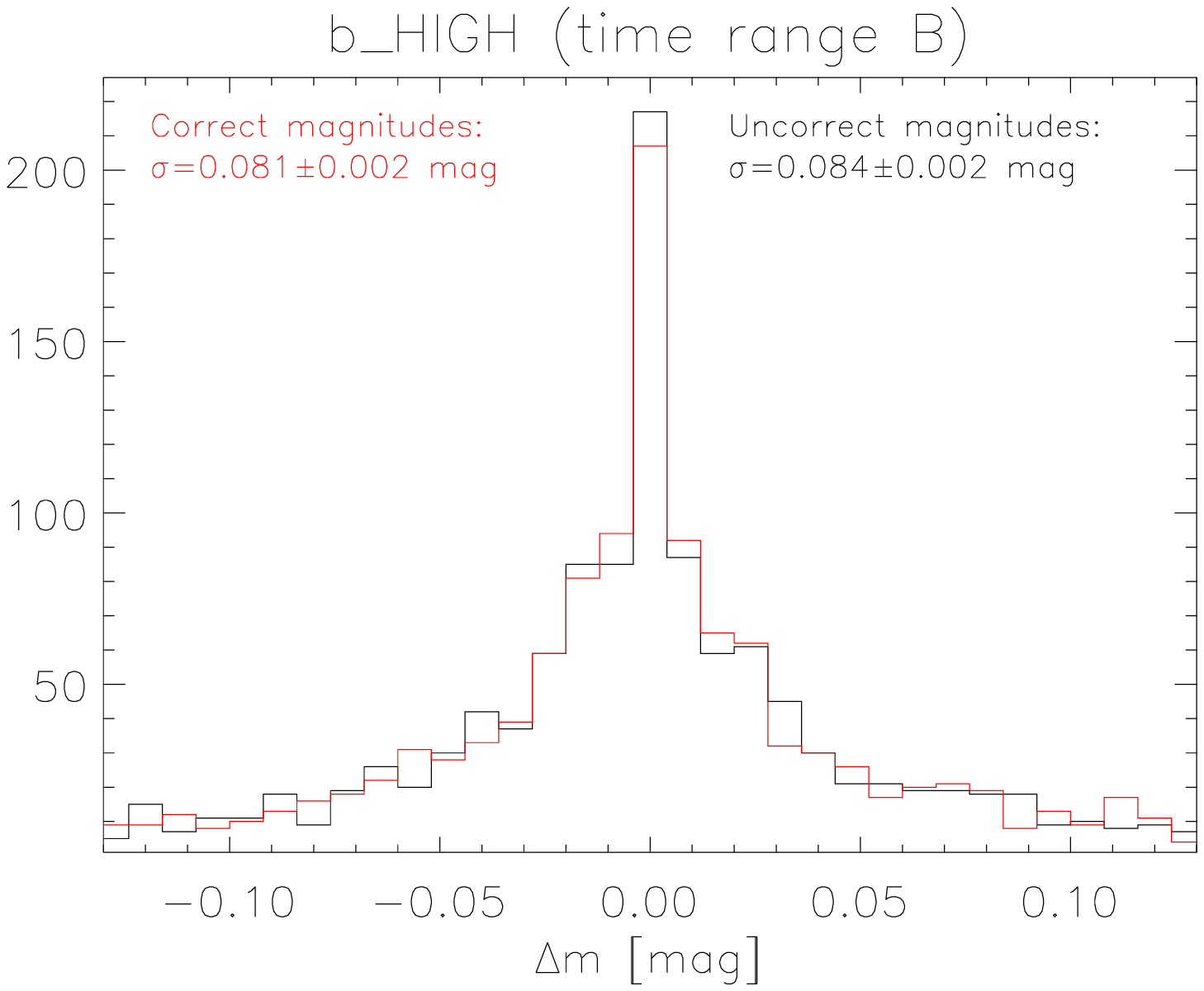, width=8.0cm,clip= }
}
  \hbox{
     \psfig{file=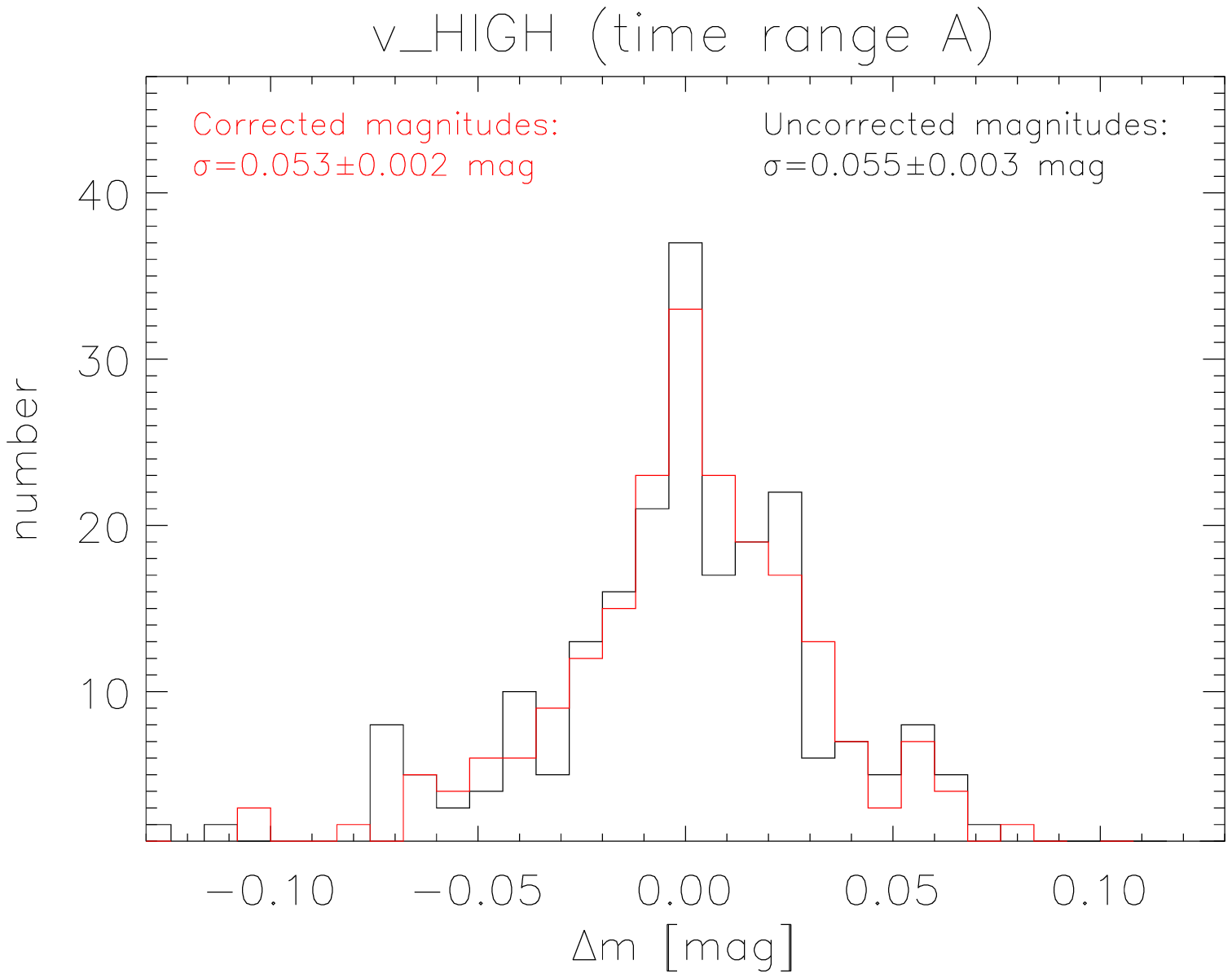, width=8.0cm,clip= }
     \psfig{file=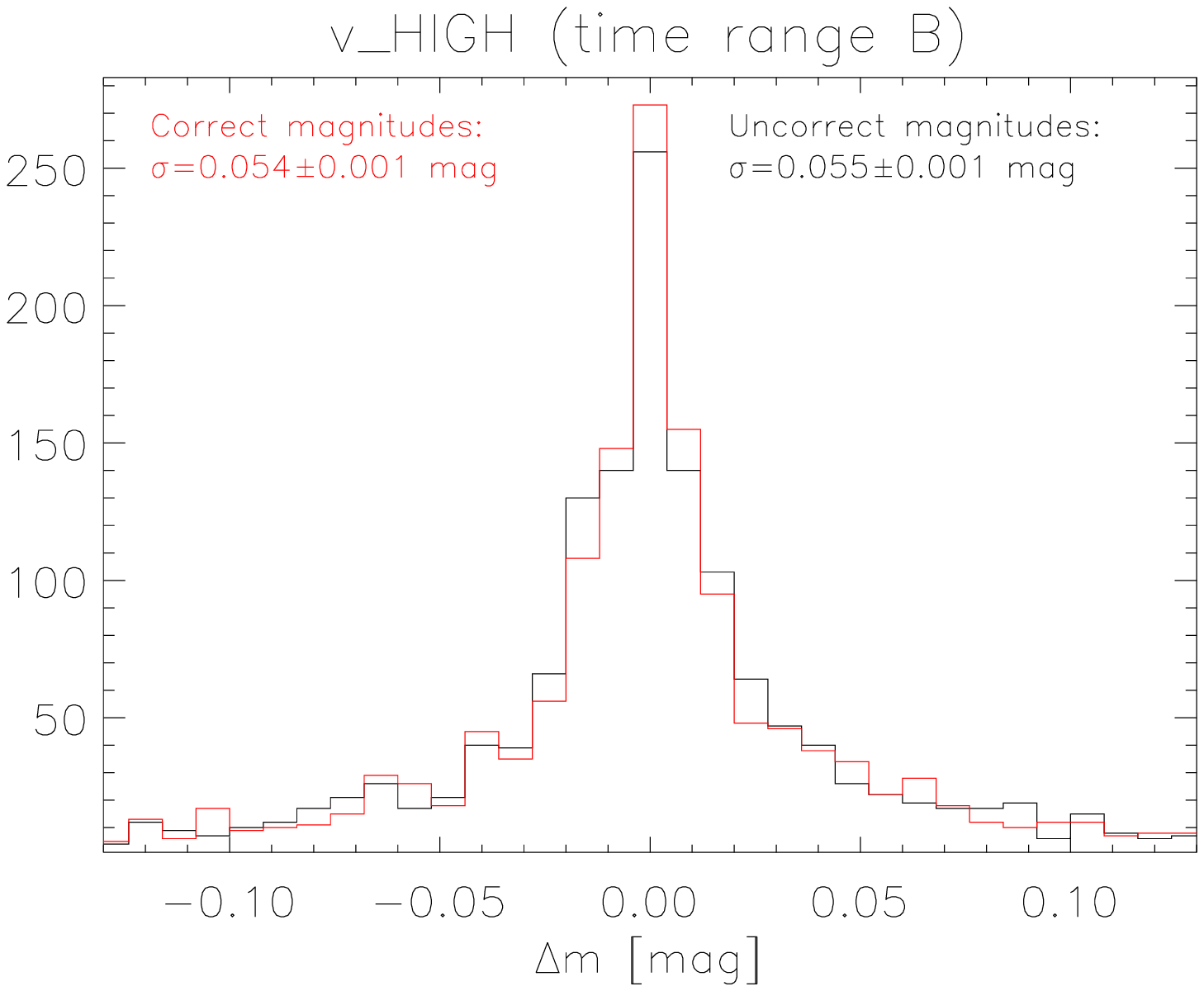, width=8.0cm,clip= }
}
\hbox{
     \psfig{file=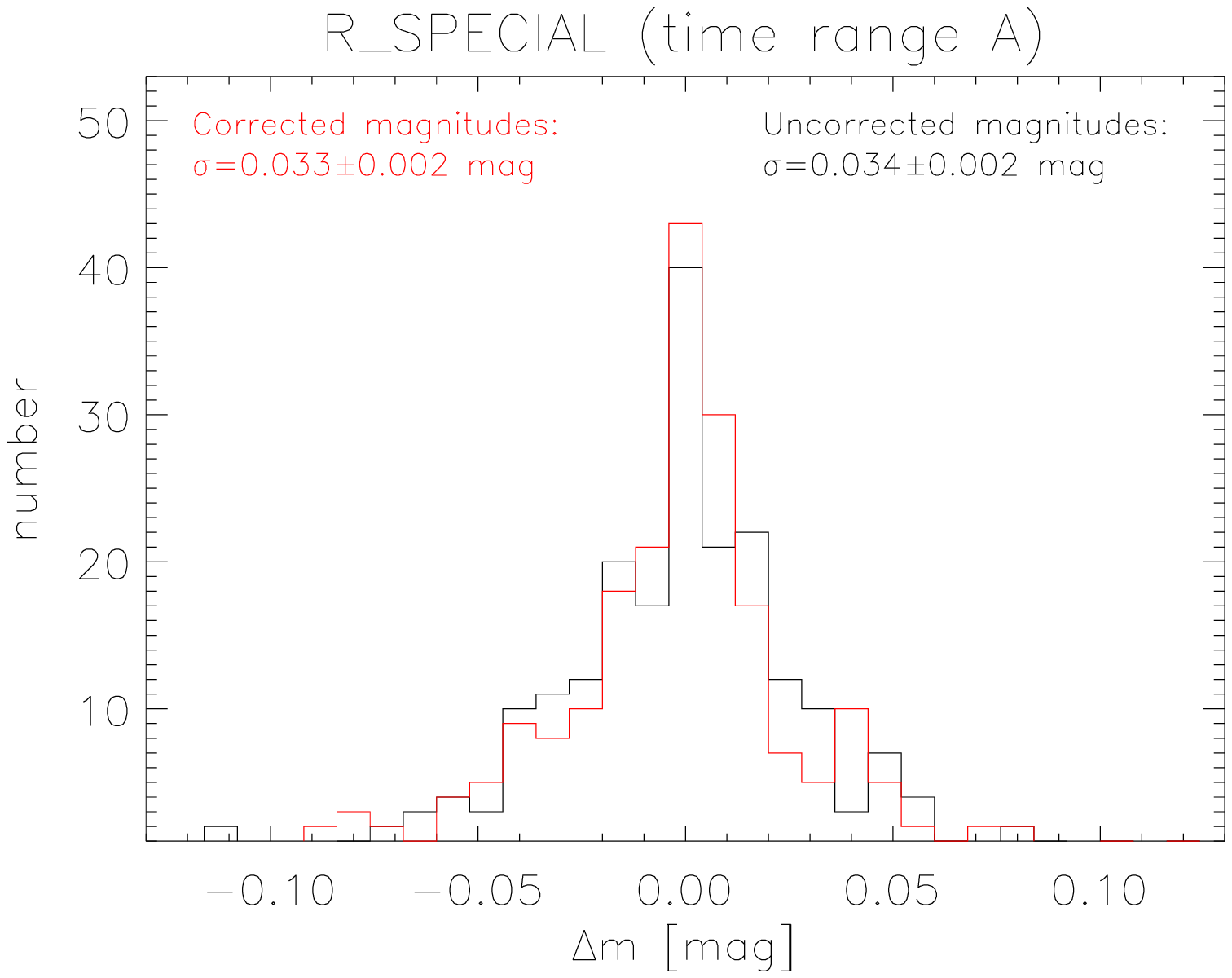, width=8.0cm,clip= }
     \psfig{file=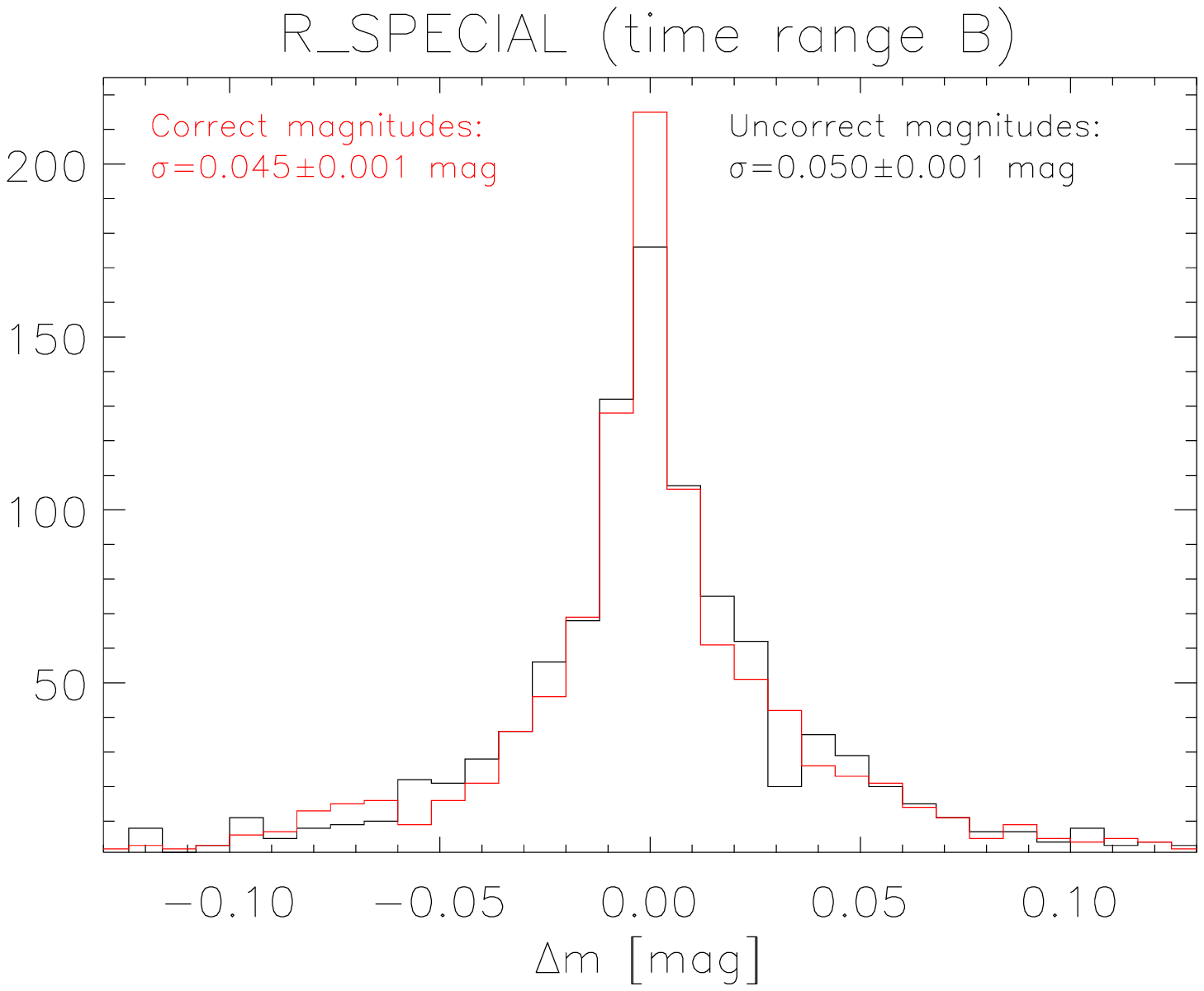 , width=8.0cm,clip= }
}
\hbox{
     \psfig{file=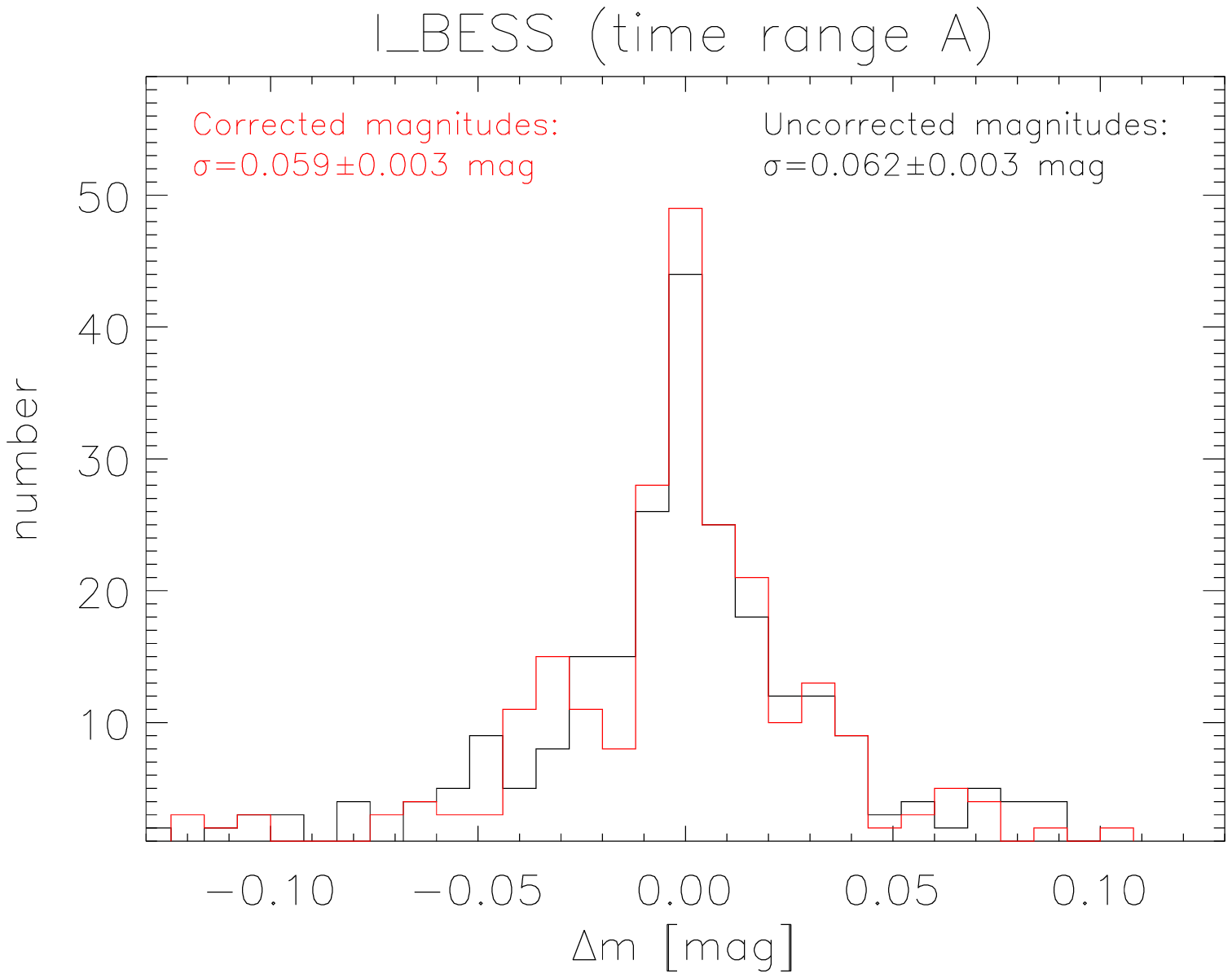, width=8.0cm,clip= }
     \psfig{file=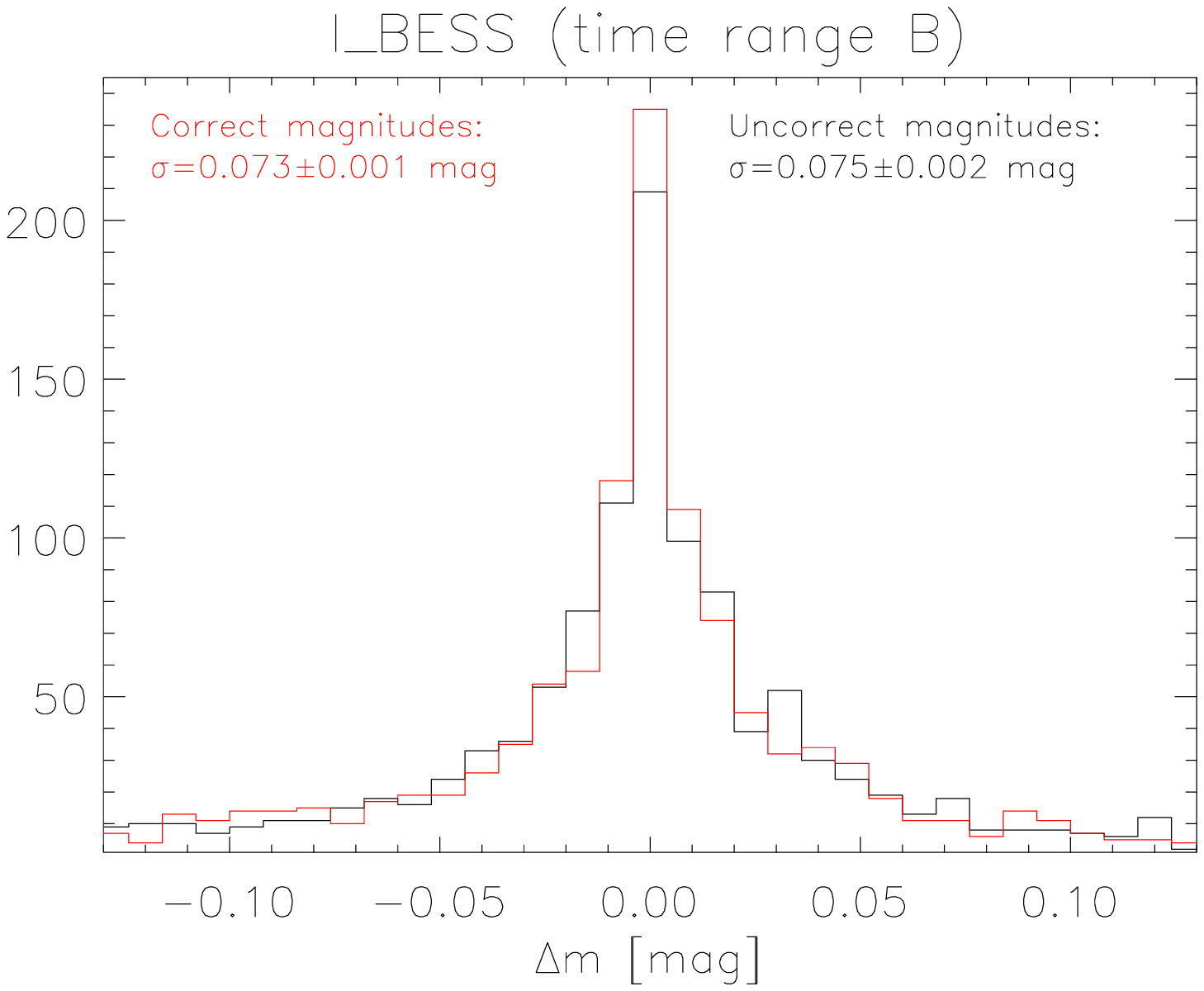 , width=8.0cm,clip= }
}
}
\caption{Comparison between the magnitude scatter of uncorrected
  magnitudes (black) and corrected magnitudes (red). 
 Histograms include only
  standard stars observed at least 10 times in our dataset, and with
  the standard deviation of $\Delta m_i^{U,C}$ less than 0.3
  magnitudes.}
\label{fig:improvement}
\end{figure*}

We compare the histograms of the residuals of the uncorrected and
corrected magnitudes, $\Delta m_i^U$ and $\Delta m_i^C$, respectively
in Figure \ref{fig:improvement}. Residuals are computed as the
difference between the apparent magnitude $m_i$ and the median
apparent magnitude for the relevant star.  In detail, for each
standard star we compute:

\begin{eqnarray}
   \Delta m_{i}^{U} &=& m_{i}^U - \langle m_{i}^U \rangle \nonumber \\
   \Delta m_{i}^{C} &=& m_{i}^C - \langle m_{i}^C \rangle
\end{eqnarray}

where $\langle \ \rangle$ indicates the median. We find that the use
of the corrected magnitudes reduces the scatter in the histograms by
$c = \sqrt{\sigma(\Delta m_i^U)^2 - \sigma(\Delta m_i^C)^2} =$ 0.027,
0.015, 0.008, and 0.019 magnitudes for filters $B$, $V$, $R$, and $I$,
respectively, in the time range A. Results for time range B are: $c=$
0.022, 0.010, 0.022, 0.017 magnitudes for filters $B$, $V$, $R$, and
$I$, respectively. They represent an improvement of $(1 - 10^{-0.4 \,
  c}) \times 100$ \% in the photometry. This ranges from 2.5\% (filter
$B$) to 0.75\% (filter $R$) for time range A, and from 2.0\% (filters
$B$ and $R$) to 0.96\% (filter $V$) for time range B. Standard
deviations are calculated removing the $5\sigma$ outliers.

We tested that the use of higher polynomial degrees in the parametrisation of
$\static$ and $\rotating$ does not significantly improve the results.

\section {Summary}
\label{sec:summary}

In Paper~I we found a rotating illumination pattern in imaging data
from ESO's FORS1 and FORS2 instruments and attributed it to the
rotating Field Rotator Unit. The photometric zero point variations of
up to $\sim$4\% across the detectors are big enough to be a concern
for precision photometry with these instruments.  We also found that
this pattern is stable between instrument interventions.  The usual
strategy of obtaining flat frames that match the position angle of the
science observations is cumbersome and suffers from a static
illumination pattern that has to be corrected for. On the other hand,
the stability of the pattern suggests that it is possible to correct
for the rotating illumination pattern and significantly improve the
photometric accuracy that can be obtained with these instruments.  In
this paper, we therefore aimed to derive an analytic correction to the
flat frames that can be applied to improve the photometric accuracy.

We used data from the FORS2 nightly standard star observations to
derive a photometric model that includes terms for both a static and
rotating illumination correction. For this purpose, we generalised the
fitting method from \cite{bra2012}.  We then derived the corrections
to the flat frames, and investigated how the photometric scatter of
individual stars improves depending on whether our corrections are
used or not. We found that using the analytic corrections, the
systematic variations in photometric data taken from 1st November 2011
to 30th May 2012 decreases by $\sim$2.5\%, 1.3\%, 0.75\%, and 1.7\% in
the $BVRI$ filters, respectively.  Similarly, for data taken from 12th
June 2012 to 7th July 2013, the improvements are $\sim$2.0\%, 0.96\%,
2.0\%, and 1.6\% in the $BVRI$ filters, respectively.

The amplitude of the improvements are consistent with the pattern
amplitudes found in Paper~I for FORS2. We therefore suggest that
flat frames for the FORS2 instrument should be corrected by the
applicable analytic expressions of
Equations~\ref{eqn:stat_poly}~and~\ref{eqn:rot_poly}, whose
coefficients are given in Table~\ref{tab:coeff}, before they are
applied to science data.

\section*{Acknowledgments}

L.C. acknowledges funding from the European Community Seventh
Framework Programme (FP7/2007-2013/) under grant agreement No 229517.
This research has benefited from the use of the following IDL libraries:
{\tt DanIDL} ({\tt http://www.danidl.co.uk}) and {\tt Markwardt} ({\tt
  http://cow.physics.wisc.edu/\textasciitilde
  craigm/idl/idl.html}). This research is based on data obtained from
the ESO Science Archive Facility under request ID: LCOCCATO-77441.

\bibliography{FORS}

\begin{thebibliography}{}

\bibitem[\protect\citeauthoryear{{Appenzeller}, {Fricke}, {F{\"u}rtig},
  {G{\"a}ssler}, {H{\"a}fner} \& et al.}{{Appenzeller}
  et~al.}{1998}]{Appenzeller+98}
{Appenzeller} I.,  {Fricke} K.,  {F{\"u}rtig} W.,  {G{\"a}ssler} W.,
  {H{\"a}fner} R.,    et al. 1998, The Messenger, 94, 1

\bibitem[\protect\citeauthoryear{{Bertin} \& {Arnouts}}{{Bertin} \&
  {Arnouts}}{1996}]{Bertin+96}
{Bertin} E.,  {Arnouts} S.,  1996, \aaps, 117, 393

\bibitem[\protect\citeauthoryear{{Betoule}, {Marriner}, {Regnault},
  {Cuillandre}, {Astier}, {Guy}, {Balland}, {El Hage}, {Hardin}, {Kessler}, {Le
  Guillou}, {Mosher}, {Pain}, {Rocci}, {Sako} \& {Schahmaneche}}{{Betoule}
  et~al.}{2013}]{bet2013}
{Betoule} M.,  {Marriner} J.,  {Regnault} N.,  {Cuillandre} J.-C.,  {Astier}
  P.,  {Guy} J.,  {Balland} C.,  {El Hage} P.,  {Hardin} D.,  {Kessler} R.,
  {Le Guillou} L.,  {Mosher} J.,  {Pain} R.,  {Rocci} P.-F.,  {Sako} M.,
  {Schahmaneche} K.,  2013, \aap, 552, A124

\bibitem[\protect\citeauthoryear{{Bramich}, {Moehler}, {Coccato}, {Freudling},
  {Garcia-Dab{\'o}}, {M{\"u}ller} \& {Saviane}}{{Bramich}
  et~al.}{2012}]{Bramich+12}
{Bramich} D.,  {Moehler} S.,  {Coccato} L.,  {Freudling} W.,  {Garcia-Dab{\'o}}
  C.~E.,  {M{\"u}ller} P.,    {Saviane} I.,  2012, The Messenger, 149, 12

\bibitem[\protect\citeauthoryear{{Bramich} \& {Freudling}}{{Bramich} \&
  {Freudling}}{2012}]{bra2012}
{Bramich} D.~M.,  {Freudling} W.,  2012, \mnras, 424, 1584

\bibitem[\protect\citeauthoryear{{Harris}, {Fitzgerald} \& {Reed}}{{Harris}
  et~al.}{1981}]{har1981}
{Harris} W.~E.,  {Fitzgerald} M.~P.,    {Reed} B.~C.,  1981, \pasp, 93, 507

\bibitem[\protect\citeauthoryear{{Honeycutt}}{{Honeycutt}}{1992}]{hon1992}
{Honeycutt} R.~K.,  1992, \pasp, 104, 435

\bibitem[\protect\citeauthoryear{{Hrudkov{\'a}}, {Skillen}, {Benn}, {Gibson},
  {Pollacco}, {Nesvorn{\'y}}, {Augusteijn}, {Tulloch} \&
  {Joshi}}{{Hrudkov{\'a}} et~al.}{2010}]{hru2010}
{Hrudkov{\'a}} M.,  {Skillen} I.,  {Benn} C.~R.,  {Gibson} N.~P.,  {Pollacco}
  D.,  {Nesvorn{\'y}} D.,  {Augusteijn} T.,  {Tulloch} S.~M.,    {Joshi} Y.~C.,
   2010, \mnras, 403, 2111

\bibitem[\protect\citeauthoryear{{Manfroid}}{{Manfroid}}{1995}]{man1995}
{Manfroid} J.,  1995, \aaps, 113, 587

\bibitem[\protect\citeauthoryear{{Manfroid} \& {Heck}}{{Manfroid} \&
  {Heck}}{1983}]{man1983}
{Manfroid} J.,  {Heck} A.,  1983, \aap, 120, 302

\bibitem[\protect\citeauthoryear{{Moehler}, {Freudling}, {M{\o}ller}, {Patat},
  {Rupprecht} \& {O'Brien}}{{Moehler} et~al.}{2010}]{moe2010}
{Moehler} S.,  {Freudling} W.,  {M{\o}ller} P.,  {Patat} F.,  {Rupprecht} G.,
   {O'Brien} K.,  2010, \pasp, 122, 93~(Paper~I)

\bibitem[\protect\citeauthoryear{{Padmanabhan}, {Schlegel}, {Finkbeiner},
  {Barentine}, {Blanton} \& et al.}{{Padmanabhan} et~al.}{2008}]{pad2008}
{Padmanabhan} N.,  {Schlegel} D.~J.,  {Finkbeiner} D.~P.,  {Barentine} J.~C.,
  {Blanton} M.~R.,    et al. 2008, \apj, 674, 1217

\bibitem[\protect\citeauthoryear{{Piterman} \& {Ninkov}}{{Piterman} \&
  {Ninkov}}{2002}]{pit2002}
{Piterman} A.,  {Ninkov} Z.,  2002, Optical Engineering, 41, 1192

\bibitem[\protect\citeauthoryear{{Popper}}{{Popper}}{1982}]{pop1982}
{Popper} D.~M.,  1982, \pasp, 94, 204

\bibitem[\protect\citeauthoryear{{Press}, {Teukolsky}, {Vetterling} \&
  {Flannery}}{{Press} et~al.}{2007}]{Press+07}
{Press} W.~H.,  {Teukolsky} S.~A.,  {Vetterling} W.~T.,    {Flannery} B.~P.,
  2007, {Numerical Recipes: The Art Of Scientific Computing}.
New York: Cambridge University Press

\bibitem[\protect\citeauthoryear{{Reed} \& {Fitzgerald}}{{Reed} \&
  {Fitzgerald}}{1982}]{ree1982}
{Reed} B.~D.,  {Fitzgerald} M.~P.,  1982, \aap, 111, 81

\bibitem[\protect\citeauthoryear{{Regnault}, {Conley}, {Guy}, {Sullivan},
  {Cuillandre}, {Astier}, {Balland}, {Basa}, {Carlberg}, {Fouchez}, {Hardin},
  {Hook}, {Howell}, {Pain}, {Perrett} \& {Pritchet}}{{Regnault}
  et~al.}{2009}]{reg2009}
{Regnault} N.,  {Conley} A.,  {Guy} J.,  {Sullivan} M.,  {Cuillandre} J.-C.,
  {Astier} P.,  {Balland} C.,  {Basa} S.,  {Carlberg} R.~G.,  {Fouchez} D.,
  {Hardin} D.,  {Hook} I.~M.,  {Howell} D.~A.,  {Pain} R.,  {Perrett} K.,
  {Pritchet} C.~J.,  2009, \aap, 506, 999

\bibitem[\protect\citeauthoryear{{Schlafly}, {Finkbeiner}, {Juri{\'c}},
  {Magnier}, {Burgett}, {Chambers}, {Grav}, {Hodapp}, {Kaiser}, {Kudritzki},
  {Martin}, {Morgan}, {Price}, {Rix}, {Stubbs}, {Tonry} \&
  {Wainscoat}}{{Schlafly} et~al.}{2012}]{sch2012}
{Schlafly} E.~F.,  {Finkbeiner} D.~P.,  {Juri{\'c}} M.,  {Magnier} E.~A.,
  {Burgett} W.~S.,  {Chambers} K.~C.,  {Grav} T.,  {Hodapp} K.~W.,  {Kaiser}
  N.,  {Kudritzki} R.-P.,  {Martin} N.~F.,  {Morgan} J.~S.,  {Price} P.~A.,
  {Rix} H.-W.,  {Stubbs} C.~W.,  {Tonry} J.~L.,    {Wainscoat} R.~J.,  2012,
  \apj, 756, 158

\bibitem[\protect\citeauthoryear{{Schwarz}}{{Schwarz}}{1978}]{sch78}
{Schwarz} G.,  1978, Ann. Statist., 6, 461

\end{thebibliography}

\appendix

\section{The dataset}

We list in Table \ref{tab:log} the images obtained during photometric
conditions that are used in the fit of the photometric model (Section \ref{sec:phot_models}).

\begin{onecolumn}
\setlongtables
\begin{small}
\begin{longtable}{l c c c c c}
\caption{List of the images obtained during photometric nights as part of the FORS2 photometric calibration plan.} \\ 
\endfirsthead
\multicolumn{6}{c}{Tab. A1: {\tiny (continue).}} \\
\noalign{\smallskip}
\hline
\noalign{\smallskip}
\endhead
\noalign{\smallskip}
\hline
\endfoot
\noalign{\smallskip}
\hline
\endlastfoot
\hline
\noalign{\smallskip}
\multicolumn{1}{l}{NIGHT} &
\multicolumn{1}{c}{FIELD} &
\multicolumn{1}{c}{File name ($B$-filter)} &
\multicolumn{1}{c}{File name ($V$-filter)} &
\multicolumn{1}{c}{File name ($R$-filter)} &
\multicolumn{1}{c}{File name ($I$-filter)} \\
\multicolumn{1}{l}{(yyyy-mm-dd)} &
\multicolumn{1}{c}{} &
\multicolumn{1}{c}{(FORS2.xxx)} &
\multicolumn{1}{c}{(FORS2.xxx)} &
\multicolumn{1}{c}{(FORS2.xxx)} &
\multicolumn{1}{c}{(FORS2.xxx)} \\
\noalign{\smallskip}
\hline
\noalign{\smallskip}
2011-11-07 & MarkA   &  2011-11-07T23:59:08.771 & 2011-11-08T00:00:06.445 &  2011-11-08T00:00:59.179 &  2011-11-08T00:01:46.523 \\ 
2011-11-07 & N6940   &  2011-11-08T00:09:59.751 & 2011-11-08T00:10:56.445 &  2011-11-08T00:11:47.959 &  2011-11-08T00:12:34.083 \\ 
2011-11-07 & PG0231  &         --               &     --                  &  2011-11-08T03:51:20.551 &  2011-11-08T03:52:06.794 \\ 
2011-11-21 & T\_Phe  &  2011-11-21T23:54:34.442 & 2011-11-21T23:55:33.357 &  2011-11-21T23:56:25.431 &  2011-11-21T23:57:12.034 \\ 
2011-11-21 & PG0231  &  2011-11-22T00:01:31.814 & 2011-11-22T00:02:29.408 &  2011-11-22T00:03:21.012 &  2011-11-22T00:04:07.185 \\  
2011-11-21 & N2298   &  2011-11-22T05:29:38.410 & 2011-11-22T05:30:36.154 &  2011-11-22T05:31:28.428 &  2011-11-22T05:32:15.101 \\ 
2011-11-23 & T\_Phe  &  2011-11-24T02:48:58.442 & 2011-11-24T02:49:56.677 &  2011-11-24T02:50:48.790 &  2011-11-24T02:51:34.924 \\  
2011-12-01 & L95     &  2011-12-02T03:15:25.425 & 2011-12-02T03:16:22.810 &  2011-12-02T03:17:14.523 &  2011-12-02T03:18:00.877 \\  
2011-12-01 & L98     &  2011-12-02T03:23:06.041 & 2011-12-02T03:24:03.195 &  2011-12-02T03:24:54.609 &  2011-12-02T03:25:40.893 \\  
2011-12-29 & T\_Phe  &  2011-12-30T00:28:54.179 & 2011-12-30T00:29:52.444 &  2011-12-30T00:30:44.908 &  2011-12-30T00:31:30.972 \\  
2011-12-29 & E3      &  2011-12-30T00:37:31.221 & 2011-12-30T00:38:28.485 &  2011-12-30T00:39:20.360 &  2011-12-30T00:40:06.543 \\   
2012-01-11 & N2298   &  2012-01-12T04:59:37.902 & 2012-01-12T05:00:35.916 &  2012-01-12T05:01:28.129 &  2012-01-12T05:02:14.233 \\ 
2012-01-11 & L95     &  2012-01-12T05:21:10.742 & 2012-01-12T05:22:07.876 &  2012-01-12T05:22:59.409 &  2012-01-12T05:23:44.953 \\ 
2012-01-11 & N5139   &  2012-01-12T07:53:30.320 & 2012-01-12T07:54:30.135 &  2012-01-12T07:55:22.319 &  2012-01-12T07:56:10.662 \\ 
2012-01-13 & L95     &  2012-01-14T00:35:12.033 & 2012-01-14T00:36:09.168 &  2012-01-14T00:37:01.512 &  2012-01-14T00:37:47.476 \\ 
2012-01-13 & N2298   &  2012-01-14T00:46:20.345 & 2012-01-14T00:47:18.880 &  2012-01-14T00:48:11.324 &  2012-01-14T00:48:58.878 \\ 
2012-01-13 & L95     &  2012-01-14T01:26:03.550 & 2012-01-14T01:27:01.244 &  2012-01-14T01:27:53.569 &  2012-01-14T01:28:39.532 \\ 
2012-01-16 & L95     &  2012-01-17T00:45:25.051 & 2012-01-17T00:46:21.405 &  2012-01-17T00:47:12.930 &  2012-01-17T00:47:59.193 \\ 
2012-01-16 & Ru149   &  2012-01-17T00:53:16.698 & 2012-01-17T00:54:13.713 &  2012-01-17T00:55:04.686 &  2012-01-17T00:55:50.309 \\ 
2012-01-16 & LeoI    &  2012-01-17T08:08:17.034 & 2012-01-17T08:09:14.629 &  2012-01-17T08:10:06.673 &  2012-01-17T08:10:52.636 \\ 
2012-01-16 & L101    &  2012-01-17T08:51:22.091 & 2012-01-17T08:52:20.935 &  2012-01-17T08:53:14.429 &  2012-01-17T08:54:01.103 \\ 
2012-01-18 & L95     &  2012-01-19T00:31:55.149 & 2012-01-19T00:32:51.654 &  2012-01-19T00:33:43.158 &  2012-01-19T00:34:28.611 \\ 
2012-01-18 & Ru149   &  2012-01-19T00:38:28.170 & 2012-01-19T00:39:25.604 &  2012-01-19T00:40:16.918 &  2012-01-19T00:41:02.212 \\ 
2012-01-18 & N2298   &  2012-01-19T07:14:08.051 & 2012-01-19T07:15:05.665 &  2012-01-19T07:15:57.149 &  2012-01-19T07:16:43.382 \\ 
2012-01-19 & L95     &  2012-01-20T00:32:14.409 & 2012-01-20T00:33:11.393 &       --                 &  2012-01-20T00:34:48.621 \\ 
2012-01-19 & N2298   &  2012-01-20T05:49:22.904 &     --                  &       --                 &         --               \\   
2012-01-22 & L95     &  2012-01-23T00:31:16.761 & 2012-01-23T00:32:12.795 &  2012-01-23T00:33:03.769 &  2012-01-23T00:33:49.572 \\ 
2012-01-22 & Ru149   &  2012-01-23T00:38:25.763 & 2012-01-23T00:39:23.118 &  2012-01-23T00:40:14.281 &  2012-01-23T00:40:59.655 \\ 
2012-01-22 & LeoI    &  2012-01-23T08:36:29.417 & 2012-01-23T08:37:25.992 &  2012-01-23T08:38:17.545 &  2012-01-23T08:39:03.749 \\ 
2012-01-23 & L95     &  2012-01-24T00:19:46.592 & 2012-01-24T00:20:42.986 &  2012-01-24T00:21:34.720 &  2012-01-24T00:22:20.183 \\ 
2012-01-23 & Ru149   &  2012-01-24T00:47:29.869 & 2012-01-24T00:48:41.514 &  2012-01-24T00:49:44.739 &  2012-01-24T00:50:42.763 \\ 
2012-01-23 & Ru149   &  2012-01-24T00:52:19.941 & 2012-01-24T00:53:24.245 &  2012-01-24T00:54:27.441 &  2012-01-24T00:55:16.404 \\ 
2012-01-23 & N2298   &  2012-01-24T05:05:30.650 & 2012-01-24T05:06:28.274 &  2012-01-24T05:07:20.318 &  2012-01-24T05:08:06.662 \\ 
2012-01-23 & LeoI    &  2012-01-24T08:57:47.595 & 2012-01-24T08:58:44.569 &  2012-01-24T08:59:35.593 &  2012-01-24T09:00:21.197 \\ 
2012-01-24 & L95     &  2012-01-25T00:20:08.189 & 2012-01-25T00:21:04.723 &  2012-01-25T00:21:56.217 &  2012-01-25T00:22:42.070 \\ 
2012-01-24 & Ru149   &  2012-01-25T00:44:56.535 & 2012-01-25T00:45:56.030 &  2012-01-25T00:46:54.054 &  2012-01-25T00:47:41.748 \\ 
2012-01-24 & N2298   &  2012-01-25T06:19:24.537 & 2012-01-25T06:20:22.231 &  2012-01-25T06:21:15.265 &  2012-01-25T06:22:01.819 \\ 
2012-01-24 & LeoI    &  2012-01-25T08:54:39.088 & 2012-01-25T08:55:36.983 &  2012-01-25T08:56:28.686 &  2012-01-25T08:57:14.590 \\ 
2012-01-27 & N2818   &  2012-01-28T00:59:38.545 & 2012-01-28T01:00:36.219 &  2012-01-28T01:01:27.923 &  2012-01-28T01:02:13.517 \\ 
2012-01-27 & L95     &  2012-01-28T01:05:57.963 & 2012-01-28T01:06:56.187 &  2012-01-28T01:07:48.832 &  2012-01-28T01:08:35.555 \\ 
2012-01-27 & N2298   &  2012-01-28T04:06:06.952 & 2012-01-28T04:07:05.847 &  2012-01-28T04:07:58.111 &  2012-01-28T04:08:44.294 \\ 
2012-01-31 & N2298   &  2012-02-01T01:52:04.408 & 2012-02-01T01:53:01.852 &  2012-02-01T01:53:54.606 &  2012-02-01T01:54:40.920 \\ 
2012-01-31 & N2818   &  2012-02-01T02:07:33.518 & 2012-02-01T02:08:31.902 &  2012-02-01T02:09:25.276 &  2012-02-01T02:10:12.780 \\ 
2012-01-31 & N2818   &  2012-02-01T03:39:16.588 & 2012-02-01T03:40:15.602 &  2012-02-01T03:41:07.377 &  2012-02-01T03:41:53.990 \\ 
2012-01-31 & N2818   &  2012-02-01T08:17:25.810 & 2012-02-01T08:18:24.925 &  2012-02-01T08:19:17.859 &  2012-02-01T08:20:04.973 \\ 
2012-01-31 & I4499   &  2012-02-01T08:25:50.690 & 2012-02-01T08:26:47.894 &  2012-02-01T08:27:39.248 &  2012-02-01T08:28:25.482 \\ 
2012-02-22 & N2298   &  2012-02-23T00:20:33.355 & 2012-02-23T00:21:33.001 &  2012-02-23T00:22:25.147 &  2012-02-23T00:23:11.432 \\ 
2012-02-22 & N2420   &  2012-02-23T00:26:50.027 & 2012-02-23T00:27:47.443 &  2012-02-23T00:28:38.359 &  2012-02-23T00:29:23.904 \\ 
2012-02-22 & Ru149   &  2012-02-23T04:35:51.680 & 2012-02-23T04:36:50.616 &  2012-02-23T04:37:41.612 &  2012-02-23T04:38:28.007 \\ 
2012-02-22 & E7      &  2012-02-23T08:33:31.416 & 2012-02-23T08:34:27.752 &  2012-02-23T08:35:19.667 &  2012-02-23T08:36:05.022 \\ 
2012-03-03 & N2298   &  2012-03-04T00:31:10.106 & 2012-03-04T00:32:09.943 &  2012-03-04T00:33:03.038 &  2012-03-04T00:33:50.964 \\ 
2012-06-15 & E5      &  2012-06-15T22:49:41.739 & 2012-06-15T22:50:41.624 &  2012-06-15T22:51:36.009 &  2012-06-15T22:52:25.053 \\           
2012-06-15 & M4      &  2012-06-15T22:57:48.980 & 2012-06-15T22:58:48.466 &  2012-06-15T22:59:42.510 &  2012-06-15T23:00:30.314 \\         
2012-06-15 & M4      &  2012-06-16T06:48:20.959 & 2012-06-16T06:51:00.832 &  2012-06-16T06:51:54.617 &  2012-06-16T06:52:43.051 \\         
2012-06-15 & MarkA   &  2012-06-16T10:24:33.063 & 2012-06-16T10:25:33.708 &  2012-06-16T10:26:28.882 &  2012-06-16T10:27:19.397 \\         
2012-06-23 & E5      &  2012-06-23T22:47:45.852 & 2012-06-23T22:48:44.597 &  2012-06-23T22:49:39.008 &  2012-06-23T22:50:27.256 \\         
2012-06-23 & PG2213  &  2012-06-24T08:54:04.173 & 2012-06-24T08:55:04.358 &  2012-06-24T08:55:58.312 &  2012-06-24T08:56:46.187 \\         
2012-06-23 & N6940   &  2012-06-24T09:01:14.448 & 2012-06-24T09:02:13.694 &  2012-06-24T09:03:07.658 &  2012-06-24T09:03:56.352 \\         
2012-06-23 & MarkA   &  2012-06-24T10:10:28.690 & 2012-06-24T10:11:28.925 &  2012-06-24T10:12:22.940 &  2012-06-24T10:13:10.724 \\         
2012-06-30 & E5      &  2012-06-30T23:30:42.323 & 2012-06-30T23:31:42.878 &          --              &       --                 \\      
2012-06-30 & L110    &  2012-07-01T04:56:29.450 & 2012-07-01T04:57:28.735 &  2012-07-01T04:58:22.769 &  2012-07-01T04:59:11.454 \\         
2012-06-30 & N6940   &  2012-07-01T05:04:01.769 & 2012-07-01T05:05:01.725 &  2012-07-01T05:05:55.820 &  2012-07-01T05:06:43.754 \\         
2012-06-30 & MarkA   &  2012-07-01T08:12:24.896 & 2012-07-01T08:13:25.532 &  2012-07-01T08:14:20.507 &  2012-07-01T08:15:10.331 \\         
2012-06-30 & L92     &  2012-07-01T08:26:16.368 & 2012-07-01T08:27:15.843 &  2012-07-01T08:28:09.847 &  2012-07-01T08:28:58.172 \\         
2012-06-30 & MarkA   &        --                & 2012-07-01T09:57:56.388 &  2012-07-01T09:58:50.542 &  2012-07-01T09:59:39.967 \\         
2012-07-02 & N5139   &  2012-07-02T22:58:59.547 & 2012-07-02T22:59:58.852 &  2012-07-02T23:00:52.767 &  2012-07-02T23:01:40.851 \\         
2012-07-02 & L101    &  2012-07-02T23:06:32.037 & 2012-07-02T23:07:31.302 &  2012-07-02T23:08:24.297 &  2012-07-02T23:09:11.741 \\         
2012-07-03 & N5139   &  2012-07-03T22:50:43.488 & 2012-07-03T22:51:42.983 &  2012-07-03T22:52:37.498 &  2012-07-03T22:53:25.611 \\         
2012-07-03 & L101    &  2012-07-03T22:56:53.039 & 2012-07-03T22:57:52.664 &  2012-07-03T22:58:46.039 &  2012-07-03T22:59:34.023 \\         
2012-08-18 & E7      &  2012-08-18T23:11:12.316 & 2012-08-18T23:12:13.220 &  2012-08-18T23:13:07.824 &  2012-08-18T23:14:01.028 \\         
2012-08-18 & E5      &  2012-08-18T23:17:41.753 & 2012-08-18T23:18:39.037 &  2012-08-18T23:19:30.080 &  2012-08-18T23:20:16.064 \\         
2012-08-18 & PG2213  &  2012-08-19T05:37:53.828 & 2012-08-19T05:38:51.552 &  2012-08-19T05:39:43.376 &  2012-08-19T05:40:30.090 \\         
2012-08-18 & PG2213  &  2012-08-19T06:15:07.326 & 2012-08-19T06:16:05.491 &  2012-08-19T06:16:57.784 &  2012-08-19T06:17:44.497 \\         
2012-08-18 & N2298   &  2012-08-19T10:01:12.931 & 2012-08-19T10:02:10.266 &  2012-08-19T10:03:01.499 &  2012-08-19T10:03:47.592 \\         
2012-08-20 & E7      &  2012-08-20T23:16:31.628 & 2012-08-20T23:17:28.583 &  2012-08-20T23:18:21.027 &  2012-08-20T23:19:07.221 \\         
2012-08-20 & E5      &  2012-08-20T23:23:25.190 & 2012-08-20T23:24:22.334 &  2012-08-20T23:25:13.359 &  2012-08-20T23:25:59.602 \\         
2012-08-20 & T\_Phe  &  2012-08-21T06:58:46.675 & 2012-08-21T06:59:43.949 &  2012-08-21T07:00:35.874 &  2012-08-21T07:01:21.876 \\         
2012-08-20 & N2298   &  2012-08-21T09:41:59.072 & 2012-08-21T09:42:56.807 &  2012-08-21T09:43:48.221 &  2012-08-21T09:44:34.004 \\         
2012-08-21 & E7      &  2012-08-21T23:15:15.667 & 2012-08-21T23:16:13.891 &  2012-08-21T23:17:05.225 &  2012-08-21T23:17:51.499 \\         
2012-08-21 & E5      &  2012-08-21T23:29:20.571 & 2012-08-21T23:30:16.925 &  2012-08-21T23:31:08.289 &  2012-08-21T23:31:54.212 \\         
2012-08-21 & L110    &  2012-08-22T05:09:25.499 & 2012-08-22T05:10:22.543 &  2012-08-22T05:11:14.106 &  2012-08-22T05:11:59.430 \\         
2012-08-21 & N2298   &  2012-08-22T09:47:13.700 & 2012-08-22T09:48:11.354 &  2012-08-22T09:49:02.017 &  2012-08-22T09:49:47.650 \\         
2012-08-22 & E7      &  2012-08-22T23:20:24.211 & 2012-08-22T23:21:25.345 &  2012-08-22T23:22:20.849 &  2012-08-22T23:23:12.662 \\         
2012-08-22 & E5      &  2012-08-22T23:28:27.884 & 2012-08-22T23:29:25.679 &  2012-08-22T23:30:16.892 &  2012-08-22T23:31:02.676 \\         
2012-08-22 & T\_Phe  &  2012-08-23T04:35:47.357 & 2012-08-23T04:36:48.182 &  2012-08-23T04:37:40.695 &  2012-08-23T04:38:27.379 \\         
2012-08-23 & E7      &  2012-08-23T23:07:32.444 & 2012-08-23T23:08:30.117 &  2012-08-23T23:09:21.701 &  2012-08-23T23:10:07.754 \\         
2012-08-23 & E5      &  2012-08-23T23:16:56.982 & 2012-08-23T23:17:54.506 &  2012-08-23T23:18:45.730 &  2012-08-23T23:19:31.353 \\         
2012-08-24 & E7      &  2012-08-24T23:06:56.080 & 2012-08-24T23:07:53.744 &  2012-08-24T23:08:45.727 &  2012-08-24T23:09:31.271 \\         
2012-08-24 & E5      &  2012-08-24T23:16:52.372 & 2012-08-24T23:17:49.686 &  2012-08-24T23:18:41.200 &  2012-08-24T23:19:26.983 \\         
2012-08-24 & L110    &  2012-08-25T05:38:20.856 & 2012-08-25T05:39:17.980 &  2012-08-25T05:40:09.524 &  2012-08-25T05:40:55.018 \\         
2012-08-26 & E7      &  2012-08-26T23:12:31.441 & 2012-08-26T23:13:29.396 &  2012-08-26T23:14:20.799 &  2012-08-26T23:15:07.013 \\         
2012-08-26 & E5      &  2012-08-26T23:19:55.635 & 2012-08-26T23:20:52.599 &  2012-08-26T23:21:43.653 &  2012-08-26T23:22:29.166 \\         
2012-08-26 & N2298   &  2012-08-27T09:40:03.038 & 2012-08-27T09:40:59.902 &  2012-08-27T09:41:51.146 &  2012-08-27T09:42:36.700 \\         
2012-08-28 & E7      &  2012-08-28T23:13:12.597 & 2012-08-28T23:14:10.222 &  2012-08-28T23:15:01.876 &  2012-08-28T23:15:48.119 \\         
2012-08-28 & N5139   &  2012-08-28T23:19:33.896 & 2012-08-28T23:20:31.461 &  2012-08-28T23:21:23.174 &  2012-08-28T23:22:09.639 \\         
2012-09-07 & PG2213  &  2012-09-08T05:28:29.791 & 2012-09-08T05:29:28.615 &  2012-09-08T05:30:22.058 &  2012-09-08T05:31:09.332 \\         
2012-09-07 & N6822   &  2012-09-08T05:36:19.503 & 2012-09-08T05:37:16.617 &  2012-09-08T05:38:08.811 &  2012-09-08T05:38:53.594 \\         
2012-09-07 & N2298   &  2012-09-08T09:39:07.269 & 2012-09-08T09:40:07.193 &  2012-09-08T09:40:59.967 &  2012-09-08T09:41:47.490 \\         
2012-09-09 & MarkA   &  2012-09-10T03:27:26.592 & 2012-09-10T03:28:24.637 &  2012-09-10T03:29:18.031 &  2012-09-10T03:30:05.494 \\         
2012-09-09 & L92     &  2012-09-10T09:44:00.956 & 2012-09-10T09:44:58.401 &  2012-09-10T09:45:50.904 &  2012-09-10T09:46:37.658 \\         
2012-09-12 & E7      &  2012-09-12T23:24:48.260 & 2012-09-12T23:25:45.914 &  2012-09-12T23:26:36.978 &  2012-09-12T23:27:23.672 \\         
2012-09-12 & N5139   &  2012-09-12T23:31:07.080 & 2012-09-12T23:32:04.815 &  2012-09-12T23:32:56.690 &  2012-09-12T23:33:43.063 \\         
2012-09-12 & MarkA   &  2012-09-13T05:14:10.985 & 2012-09-13T05:15:08.560 &  2012-09-13T05:16:00.204 &  2012-09-13T05:16:46.228 \\         
2012-09-13 & E7      &  2012-09-13T23:29:00.110 & 2012-09-13T23:29:58.645 &  2012-09-13T23:30:49.519 &  2012-09-13T23:31:35.312 \\         
2012-09-13 & MarkA   &  2012-09-14T03:38:58.613 & 2012-09-14T03:39:57.158 &  2012-09-14T03:40:49.072 &  2012-09-14T03:41:35.485 \\         
2012-09-13 & I4499   &  2012-09-14T03:45:48.855 & 2012-09-14T03:46:46.370 &  2012-09-14T03:47:38.593 &  2012-09-14T03:48:24.637 \\         
2012-09-23 & MarkA   &  2012-09-24T01:42:00.729 & 2012-09-24T01:42:58.714 &  2012-09-24T01:43:50.988 &  2012-09-24T01:44:37.302 \\         
2012-09-23 & N6940   &  2012-09-24T01:51:51.821 & 2012-09-24T01:52:49.106 &  2012-09-24T01:53:40.831 &  2012-09-24T01:54:27.174 \\         
2012-09-23 & N6940   &  2012-09-24T01:56:17.544 & 2012-09-24T01:57:13.189 &  2012-09-24T01:58:02.483 &  2012-09-24T01:58:46.897 \\         
2012-09-29 & E7      &  2012-09-30T00:38:48.763 & 2012-09-30T00:39:47.438 &  2012-09-30T00:40:39.543 &  2012-09-30T00:41:25.777 \\         
2012-09-29 & N6940   &  2012-09-30T00:45:23.839 & 2012-09-30T00:46:21.044 &  2012-09-30T00:47:12.519 &  2012-09-30T00:47:58.283 \\         
2012-09-29 & L98     &  2012-09-30T08:49:32.983 & 2012-09-30T08:50:31.468 &  2012-09-30T08:51:23.533 &  2012-09-30T08:52:10.997 \\         
2012-09-29 & N2420   &  2012-09-30T09:00:14.529 & 2012-09-30T09:01:11.174 &  2012-09-30T09:02:11.149 &  2012-09-30T09:02:57.203 \\         
2012-10-01 & MarkA   &  2012-10-02T03:10:16.368 & 2012-10-02T03:11:16.943 &  2012-10-02T03:12:10.288 &  2012-10-02T03:12:58.003 \\         
2012-10-01 & PG2213  &  2012-10-02T03:18:02.759 & 2012-10-02T03:19:00.754 &  2012-10-02T03:19:53.219 &  2012-10-02T03:20:40.612 \\         
2012-10-01 & PG221   &           --             & 2012-10-02T03:25:53.659 &  2012-10-02T03:26:45.614 &  2012-10-02T03:27:31.798 \\               
2012-10-01 & T\_Phe  &  2012-10-02T05:02:16.853 & 2012-10-02T05:03:14.908 &  2012-10-02T05:04:07.092 &  2012-10-02T05:04:53.417 \\         
2012-10-01 & T\_Phe  &  2012-10-02T05:05:52.792 &     --                  &          --              &             --           \\ 
2012-10-05 & MarkA   &  2012-10-06T00:21:58.765 & 2012-10-06T00:22:59.189 &  2012-10-06T00:23:53.954 &  2012-10-06T00:24:42.268 \\        
2012-10-05 & M4      &  2012-10-06T00:29:51.153 & 2012-10-06T00:30:51.929 &  2012-10-06T00:31:46.113 &  2012-10-06T00:32:33.357 \\        
2012-10-05 & PG2213  &  2012-10-06T05:11:23.869 & 2012-10-06T05:12:23.664 &  2012-10-06T05:13:17.278 &  2012-10-06T05:14:05.392 \\        
2012-10-05 & L95     &  2012-10-06T06:38:12.087 & 2012-10-06T06:39:12.101 &  2012-10-06T06:40:05.856 &  2012-10-06T06:40:53.979 \\        
2012-10-06 & M4      &  2012-10-06T23:40:43.016 & 2012-10-06T23:41:42.791 &  2012-10-06T23:42:36.565 &  2012-10-06T23:43:25.119 \\        
2012-10-06 & T\_Phe  &  2012-10-06T23:52:50.745 & 2012-10-06T23:53:55.351 &  2012-10-06T23:54:51.835 &  2012-10-06T23:55:41.890 \\        
2012-10-06 & PG2213  &  2012-10-07T04:11:58.711 & 2012-10-07T04:12:59.686 &  2012-10-07T04:13:54.830 &  2012-10-07T04:14:43.824 \\        
2012-10-06 & E3      &  2012-10-07T09:19:37.114 & 2012-10-07T09:20:37.188 &  2012-10-07T09:21:31.443 &  2012-10-07T09:22:20.647 \\        
2012-10-07 & MarkA   &  2012-10-07T23:30:12.587 & 2012-10-07T23:31:12.642 &  2012-10-07T23:32:06.596 &  2012-10-07T23:32:54.630 \\        
2012-10-07 & PG1633  &  2012-10-07T23:40:03.745 & 2012-10-07T23:41:03.100 &  2012-10-07T23:41:57.135 &  2012-10-07T23:42:45.119 \\        
2012-10-07 & PG2213  &  2012-10-08T05:36:17.762 & 2012-10-08T05:37:17.707 &  2012-10-08T05:38:12.352 &  2012-10-08T05:39:00.205 \\        
2012-10-07 & Ru149   &  2012-10-08T09:19:27.999 & 2012-10-08T09:20:31.744 &  2012-10-08T09:21:26.678 &  2012-10-08T09:22:14.902 \\        
2012-10-09 & MarkA   &  2012-10-09T23:30:44.029 & 2012-10-09T23:31:45.544 &  2012-10-09T23:32:39.509 &  2012-10-09T23:33:28.303 \\        
2012-10-09 & PG1633  &  2012-10-09T23:40:43.640 & 2012-10-09T23:41:43.285 &  2012-10-09T23:42:37.379 &  2012-10-09T23:43:25.834 \\        
2012-10-09 & PG2213  &  2012-10-10T04:29:07.922 & 2012-10-10T04:30:08.016 &  2012-10-10T04:31:01.931 &  2012-10-10T04:31:49.425 \\        
2012-10-09 & Ru149   &  2012-10-10T09:32:34.528 & 2012-10-10T09:33:35.582 &  2012-10-10T09:34:30.067 &  2012-10-10T09:35:19.421 \\        
2012-10-10 & MarkA   &  2012-10-10T23:24:41.464 & 2012-10-10T23:25:41.899 &  2012-10-10T23:26:37.124 &  2012-10-10T23:27:25.808 \\        
2012-10-10 & PG1633  &  2012-10-10T23:32:50.835 & 2012-10-10T23:33:51.010 &  2012-10-10T23:34:44.994 &  2012-10-10T23:35:33.748 \\        
2012-10-10 & PG0231  &  2012-10-11T05:45:07.768 & 2012-10-11T05:46:06.853 &  2012-10-11T05:47:01.047 &  2012-10-11T05:47:49.061 \\        
2012-10-10 & N2298   &  2012-10-11T09:08:33.054 & 2012-10-11T09:09:33.690 &  2012-10-11T09:10:27.894 &  2012-10-11T09:11:17.078 \\        
2012-10-12 & MarkA   &  2012-10-12T23:27:21.400 & 2012-10-12T23:28:21.795 &  2012-10-12T23:29:16.469 &  2012-10-12T23:30:05.164 \\        
2012-10-12 & PG1633  &  2012-10-12T23:39:35.042 & 2012-10-12T23:40:34.477 &  2012-10-12T23:41:28.481 &  2012-10-12T23:42:16.315 \\        
2012-10-12 & PG2213  &  2012-10-13T03:36:38.696 & 2012-10-13T03:37:39.421 &  2012-10-13T03:38:34.526 &  2012-10-13T03:39:23.830 \\        
2012-10-12 & N2298   &  2012-10-13T07:21:11.395 & 2012-10-13T07:22:12.571 &  2012-10-13T07:23:07.705 &  2012-10-13T07:23:56.719 \\        
2012-10-14 & MarkA   &  2012-10-14T23:28:41.392 & 2012-10-14T23:29:42.128 &  2012-10-14T23:30:37.222 &  2012-10-14T23:31:25.206 \\        
2012-10-14 & PG1633  &  2012-10-14T23:36:31.762 & 2012-10-14T23:37:31.207 &  2012-10-14T23:38:24.721 &  2012-10-14T23:39:12.596 \\        
2012-10-14 & PG2213  &  2012-10-15T04:03:28.034 & 2012-10-15T04:04:28.449 &  2012-10-15T04:05:22.434 &  2012-10-15T04:06:11.208 \\        
2012-10-14 & Ru149   &  2012-10-15T09:24:14.038 & 2012-10-15T09:25:15.282 &  2012-10-15T09:26:09.307 &  2012-10-15T09:26:58.631 \\        
2012-10-15 & MarkA   &  2012-10-16T00:02:09.259 & 2012-10-16T00:03:10.534 &  2012-10-16T00:04:04.878 &  2012-10-16T00:04:52.713 \\        
2012-10-15 & M4      &  2012-10-16T00:13:29.308 & 2012-10-16T00:14:34.004 &  2012-10-16T00:15:28.619 &  2012-10-16T00:16:16.563 \\        
2012-10-15 & L95     &  2012-10-16T04:43:46.788 & 2012-10-16T04:44:47.323 &  2012-10-16T04:45:41.407 &  2012-10-16T04:46:30.662 \\        
2012-10-15 & N2818   &  2012-10-16T08:09:45.079 & 2012-10-16T08:10:45.104 &  2012-10-16T08:11:38.639 &  2012-10-16T08:12:26.423 \\        
2012-10-20 & L92     &  2012-10-21T02:51:50.682 & 2012-10-21T02:52:50.157 &  2012-10-21T02:53:43.932 &  2012-10-21T02:54:32.656 \\        
2012-10-20 & MarkA   &  2012-10-21T02:59:36.393 & 2012-10-21T03:00:35.968 &  2012-10-21T03:01:29.732 &  2012-10-21T03:02:18.047 \\        
2012-10-20 & Ru149   &  2012-10-21T08:12:42.569 & 2012-10-21T08:13:43.514 &  2012-10-21T08:14:39.268 &  2012-10-21T08:15:27.982 \\        
2012-10-22 & L95     &  2012-10-23T04:33:47.062 & 2012-10-23T04:34:49.567 &  2012-10-23T04:35:44.561 &  2012-10-23T04:36:33.785 \\        
2012-10-22 & E3      &  2012-10-23T04:41:14.798 & 2012-10-23T04:42:15.123 &  2012-10-23T04:43:09.327 &  2012-10-23T04:43:59.142 \\        
2012-10-22 & N2818   &  2012-10-23T08:50:39.479 & 2012-10-23T08:51:40.794 &  2012-10-23T08:52:36.768 &  2012-10-23T08:53:26.192 \\        
2012-10-31 & MarkA   &  2012-11-01T00:08:34.246 & 2012-11-01T00:09:36.491 &  2012-11-01T00:10:28.326 &  2012-11-01T00:11:14.470 \\        
2012-10-31 & N6940   &  2012-11-01T00:15:17.071 & 2012-11-01T00:16:14.176 &  2012-11-01T00:17:05.630 &  2012-11-01T00:17:51.125 \\        
2012-10-31 & L95     &  2012-11-01T05:07:20.400 & 2012-11-01T05:08:18.734 &  2012-11-01T05:09:09.759 &  2012-11-01T05:09:55.682 \\        
2012-11-03 & MarkA   &  2012-11-03T23:54:27.393 & 2012-11-03T23:55:24.938 &  2012-11-03T23:56:17.283 &  2012-11-03T23:57:03.947 \\        
2012-11-03 & N6940   &  2012-11-04T00:01:27.101 & 2012-11-04T00:02:24.406 &  2012-11-04T00:03:15.250 &  2012-11-04T00:04:01.145 \\        
2012-11-07 & MarkA   &  2012-11-08T00:05:52.138 & 2012-11-08T00:06:49.782 &  2012-11-08T00:07:41.707 &  2012-11-08T00:08:28.081 \\        
2012-11-07 & N6940   &  2012-11-08T00:14:36.905 & 2012-11-08T00:15:33.359 &  2012-11-08T00:16:23.924 &  2012-11-08T00:17:09.508 \\        
2012-11-07 & E3      &  2012-11-08T04:32:04.172 & 2012-11-08T04:33:01.767 &  2012-11-08T04:33:52.652 &  2012-11-08T04:34:38.456 \\        
2012-11-08 & MarkA   &  2012-11-09T01:50:02.955 & 2012-11-09T01:50:59.880 &  2012-11-09T01:51:50.164 &  2012-11-09T01:52:35.898 \\        
2012-11-08 & T\_Phe  &  2012-11-09T02:02:42.547 & 2012-11-09T02:03:42.672 &  2012-11-09T02:04:39.137 &  2012-11-09T02:05:29.351 \\        
2012-11-08 & E3      &  2012-11-09T04:38:35.443 & 2012-11-09T04:39:32.058 &  2012-11-09T04:40:23.443 &  2012-11-09T04:41:08.836 \\        
2012-11-09 & MarkA   &  2012-11-10T01:26:42.428 & 2012-11-10T01:27:39.523 &  2012-11-10T01:28:30.377 &  2012-11-10T01:29:16.181 \\        
2012-11-09 & T\_Phe  &  2012-11-10T01:34:09.336 & 2012-11-10T01:35:10.421 &  2012-11-10T01:36:05.725 &  2012-11-10T01:37:01.419 \\        
2012-12-08 & L92     &  2012-12-09T00:06:31.997 & 2012-12-09T00:07:29.472 &  2012-12-09T00:08:20.635 &  2012-12-09T00:09:06.149 \\        
2012-12-08 & PG2213  &  2012-12-09T00:42:51.045 & 2012-12-09T00:43:48.410 &  2012-12-09T00:44:39.624 &  2012-12-09T00:45:24.738 \\        
2012-12-08 & PG2213  &  2012-12-09T00:47:57.790 & 2012-12-09T00:48:54.924 &  2012-12-09T00:49:46.238 &  2012-12-09T00:50:33.302 \\        
2012-12-08 & N2298   &  2012-12-09T04:07:41.094 & 2012-12-09T04:08:39.238 &  2012-12-09T04:09:31.273 &  2012-12-09T04:10:17.397 \\        
2012-12-09 & PG0231  &          --              & 2012-12-10T00:15:21.796 &  2012-12-10T00:16:14.629 &  2012-12-10T00:17:00.184 \\        
2012-12-09 & L95     &  2012-12-10T00:47:35.762 & 2012-12-10T00:48:32.496 &  2012-12-10T00:49:24.921 &  2012-12-10T00:50:11.554 \\        
2012-12-09 & E3      &  2012-12-10T02:17:13.951 & 2012-12-10T02:18:11.056 &  2012-12-10T02:19:01.980 &  2012-12-10T02:19:47.753 \\        
2012-12-09 & N2298   &  2012-12-10T08:11:22.014 & 2012-12-10T08:12:19.629 &  2012-12-10T08:13:11.773 &  2012-12-10T08:14:00.087 \\        
2012-12-12 & L92     &  2012-12-13T00:51:47.717 & 2012-12-13T00:52:45.471 &  2012-12-13T00:53:36.835 &  2012-12-13T00:54:23.199 \\        
2012-12-12 & E3      &  2012-12-13T01:02:04.155 & 2012-12-13T01:03:01.160 &  2012-12-13T01:03:52.134 &  2012-12-13T01:04:37.348 \\        
2012-12-12 & L95     &  2012-12-13T03:38:08.916 & 2012-12-13T03:39:05.860 &  2012-12-13T03:39:57.584 &  2012-12-13T03:40:44.228 \\        
2012-12-12 & N2298   &  2012-12-13T07:29:16.763 & 2012-12-13T07:30:14.197 &  2012-12-13T07:31:05.831 &  2012-12-13T07:31:51.874 \\        
2012-12-13 & L92     &  2012-12-14T00:14:33.080 & 2012-12-14T00:15:30.715 &  2012-12-14T00:16:21.679 &  2012-12-14T00:17:08.572 \\        
2012-12-13 & MarkA   &  2012-12-14T00:29:14.450 & 2012-12-14T00:30:11.255 &  2012-12-14T00:31:02.819 &  2012-12-14T00:31:48.123 \\        
2012-12-13 & Ru149   &  2012-12-14T04:41:41.952 & 2012-12-14T04:42:39.806 &  2012-12-14T04:43:32.080 &  2012-12-14T04:44:18.824 \\        
2012-12-14 & L92     &  2012-12-15T00:22:01.616 & 2012-12-15T00:22:58.860 &  2012-12-15T00:23:50.464 &  2012-12-15T00:24:36.508 \\        
2012-12-14 & L92     &  2012-12-15T00:25:30.182 & 2012-12-15T00:26:28.277 &  2012-12-15T00:27:20.240 &  2012-12-15T00:28:06.464 \\        
2012-12-14 & E3      &  2012-12-15T00:51:18.231 & 2012-12-15T00:52:15.805 &  2012-12-15T00:53:07.259 &  2012-12-15T00:53:53.652 \\        
2012-12-15 & L92     &  2012-12-16T00:15:41.145 & 2012-12-16T00:16:37.869 &  2012-12-16T00:17:29.873 &  2012-12-16T00:18:15.727 \\        
2012-12-15 & E3      &  2012-12-16T00:31:48.511 & 2012-12-16T00:32:45.515 &  2012-12-16T00:33:37.430 &  2012-12-16T00:34:22.773 \\        
2012-12-15 & L95     &  2012-12-16T03:33:18.599 & 2012-12-16T03:34:17.434 &  2012-12-16T03:35:09.727 &  2012-12-16T03:35:56.041 \\        
2012-12-16 & L92     &  2012-12-17T00:59:48.107 & 2012-12-17T01:00:45.741 &  2012-12-17T01:01:37.975 &  2012-12-17T01:02:25.500 \\        
2012-12-16 & N2298   &  2012-12-17T01:07:33.364 & 2012-12-17T01:08:29.899 &  2012-12-17T01:09:21.783 &  2012-12-17T01:10:07.926 \\        
2012-12-18 & N2298   &  2012-12-19T03:17:56.649 & 2012-12-19T03:18:54.884 &  2012-12-19T03:19:46.978 &  2012-12-19T03:20:33.321 \\        
2012-12-18 & L101    &  2012-12-19T05:38:32.914 & 2012-12-19T05:39:31.629 &  2012-12-19T05:40:22.982 &  2012-12-19T05:41:08.456 \\        
2012-12-21 & L98     &  2012-12-22T08:52:41.712 & 2012-12-22T08:53:38.376 &  2012-12-22T08:54:30.159 &  2012-12-22T08:55:15.983 \\        
2012-12-21 & N2818   &  2012-12-22T08:59:51.024 & 2012-12-22T09:00:49.208 &  2012-12-22T09:01:42.193 &  2012-12-22T09:02:28.687 \\        
2012-12-28 & T\_Phe  &  2012-12-29T00:47:20.607 & 2012-12-29T00:48:19.862 &  2012-12-29T00:49:12.047 &  2012-12-29T00:49:57.870 \\        
2012-12-28 & PG2213  &  2012-12-29T00:54:11.400 & 2012-12-29T00:55:08.486 &  2012-12-29T00:56:00.340 &  2012-12-29T00:56:46.564 \\        
2012-12-28 & N2298   &  2012-12-29T03:26:04.965 & 2012-12-29T03:27:03.169 &  2012-12-29T03:27:55.913 &  2012-12-29T03:28:43.596 \\        
2013-01-01 & L95     &  2013-01-02T00:35:59.578 & 2013-01-02T00:36:58.163 &  2013-01-02T00:37:50.357 &  2013-01-02T00:38:36.360 \\        
2013-01-01 & N2298   &  2013-01-02T00:42:36.840 & 2013-01-02T00:43:33.934 &  2013-01-02T00:44:24.828 &  2013-01-02T00:45:10.431 \\        
2013-01-01 & N2298   &        --                & 2013-01-02T00:49:04.120 &     --                   &    --                    \\ 
2013-01-02 & L95     &  2013-01-03T00:18:52.444 & 2013-01-03T00:19:50.659 &  2013-01-03T00:20:42.853 &  2013-01-03T00:21:29.157 \\        
2013-01-02 & N2298   &  2013-01-03T00:25:06.974 & 2013-01-03T00:26:03.768 &  2013-01-03T00:26:55.132 &  2013-01-03T00:27:41.126 \\        
2013-01-08 & L95     &  2013-01-09T00:26:16.079 & 2013-01-09T00:27:12.943 &  2013-01-09T00:28:04.207 &  2013-01-09T00:28:50.141 \\        
2013-01-08 & N2298   &  2013-01-09T00:32:44.530 & 2013-01-09T00:33:41.263 &  2013-01-09T00:34:31.597 &  2013-01-09T00:35:17.361 \\        
2013-01-08 & N2298   &  2013-01-09T05:08:28.817 & 2013-01-09T05:09:25.872 &  2013-01-09T05:10:17.256 &  2013-01-09T05:11:03.580 \\        
2013-01-31 & L95     &  2013-02-01T01:07:40.436 & 2013-02-01T01:08:38.491 &  2013-02-01T01:09:30.205 &  2013-02-01T01:10:16.558 \\        
2013-01-31 & N2818   &  2013-02-01T01:15:06.932 & 2013-02-01T01:16:02.867 &  2013-02-01T01:16:54.471 &  2013-02-01T01:17:39.915 \\        
2013-01-31 & N2298   &  2013-02-01T03:37:56.295 & 2013-02-01T03:38:55.150 &  2013-02-01T03:39:47.214 &  2013-02-01T03:40:33.598 \\        
2013-01-31 & LeoI    &  2013-02-01T03:44:38.168 & 2013-02-01T03:45:35.412 &  2013-02-01T03:46:26.276 &  2013-02-01T03:47:11.990 \\        
2013-01-31 & N2818   &  2013-02-01T05:56:04.254 & 2013-02-01T05:57:01.789 &  2013-02-01T05:57:54.393 &  2013-02-01T05:58:41.266 \\        
2013-02-01 & N2298   &  2013-02-02T00:31:07.088 & 2013-02-02T00:32:05.302 &  2013-02-02T00:32:57.397 &  2013-02-02T00:33:43.821 \\        
2013-02-01 & N2818   &  2013-02-02T00:38:42.045 & 2013-02-02T00:39:38.290 &  2013-02-02T00:40:29.874 &  2013-02-02T00:41:14.737 \\        
2013-02-01 & N2818   &  2013-02-02T03:18:02.909 & 2013-02-02T03:19:00.153 &  2013-02-02T03:19:52.628 &  2013-02-02T03:20:38.931 \\        
2013-02-03 & N2437   &  2013-02-04T04:32:33.262 & 2013-02-04T04:33:32.397 &  2013-02-04T04:34:24.801 &  2013-02-04T04:35:10.395 \\        
2013-02-03 & N5139   &  2013-02-04T04:40:46.353 & 2013-02-04T04:41:43.397 &  2013-02-04T04:42:34.371 &  2013-02-04T04:43:20.795 \\        
2013-02-07 & N2818   &  2013-02-08T05:50:34.518 & 2013-02-08T05:51:31.743 &  2013-02-08T05:52:23.717 &  2013-02-08T05:53:10.431 \\        
2013-02-07 & L98     &  2013-02-08T05:57:42.312 & 2013-02-08T05:58:40.697 &  2013-02-08T05:59:32.101 &  2013-02-08T06:00:17.045 \\        
2013-02-08 & N2298   &  2013-02-09T00:19:47.717 & 2013-02-09T00:20:45.391 &  2013-02-09T00:21:37.325 &  2013-02-09T00:22:23.739 \\        
2013-02-08 & N2818   &  2013-02-09T00:26:54.841 & 2013-02-09T00:27:51.866 &  2013-02-09T00:28:43.550 &  2013-02-09T00:29:29.074 \\        
2013-02-08 & N2818   &  2013-02-09T03:37:26.302 & 2013-02-09T03:38:23.706 &  2013-02-09T03:39:17.100 &  2013-02-09T03:40:03.403 \\        
2013-02-09 & N2298   &  2013-02-10T00:22:20.949 & 2013-02-10T00:23:18.873 &  2013-02-10T00:24:11.248 &  2013-02-10T00:24:57.521 \\        
2013-02-09 & N2818   &  2013-02-10T00:28:54.759 & 2013-02-10T00:29:51.894 &  2013-02-10T00:30:43.588 &  2013-02-10T00:31:29.711 \\        
2013-02-09 & N2818   &  2013-02-10T05:11:00.365 & 2013-02-10T05:11:57.430 &  2013-02-10T05:12:49.734 &  2013-02-10T05:13:36.447 \\        
2013-02-09 & N2818   &  2013-02-10T09:23:57.685 & 2013-02-10T09:24:53.929 &  2013-02-10T09:25:45.142 &  2013-02-10T09:26:30.626 \\        
2013-03-12 & N2298   &  2013-03-12T23:49:45.916 & 2013-03-12T23:50:42.973 &  2013-03-12T23:51:33.769 &  2013-03-12T23:52:19.644 \\        
2013-03-12 & L101    &  2013-03-12T23:57:41.840 & 2013-03-12T23:58:38.927 &  2013-03-12T23:59:29.632 &  2013-03-13T00:00:15.698 \\        
2013-03-12 & E5      &  2013-03-13T05:35:18.912 & 2013-03-13T05:36:15.679 &  2013-03-13T05:37:07.125 &  2013-03-13T05:37:53.360 \\        
2013-03-14 & N2298   &  2013-03-14T23:55:19.137 & 2013-03-14T23:56:15.973 &  2013-03-14T23:57:07.419 &  2013-03-14T23:57:52.614 \\        
2013-03-14 & L101    &  2013-03-15T00:02:10.080 & 2013-03-15T00:03:07.126 &  2013-03-15T00:03:59.182 &  2013-03-15T00:04:45.037 \\        
2013-03-14 & E5      &  2013-03-15T03:48:44.281 & 2013-03-15T03:49:43.268 &  2013-03-15T03:50:36.274 &  2013-03-15T03:51:22.069 \\        
2013-03-14 & L104    &  2013-03-15T05:30:22.318 & 2013-03-15T05:31:19.575 &  2013-03-15T05:32:10.820 &  2013-03-15T05:32:56.736 \\        
2013-03-16 & N2298   &  2013-03-16T23:47:27.849 & 2013-03-16T23:48:28.216 &  2013-03-16T23:49:23.332 &  2013-03-16T23:50:14.019 \\        
2013-03-16 & LeoI    &  2013-03-16T23:54:44.380 & 2013-03-16T23:55:41.327 &  2013-03-16T23:56:32.253 &  2013-03-16T23:57:17.878 \\        
2013-03-16 & N2818   &  2013-03-17T02:59:56.187 & 2013-03-17T03:00:54.464 &  2013-03-17T03:01:45.449 &  2013-03-17T03:02:32.315 \\        
2013-03-16 & L104    &  2013-03-17T06:20:15.463 & 2013-03-17T06:21:13.550 &  2013-03-17T06:22:05.026 &  2013-03-17T06:22:51.012 \\        
2013-03-16 & E7      &  2013-03-17T09:54:49.166 & 2013-03-17T09:55:46.453 &  2013-03-17T09:56:38.548 &  2013-03-17T09:57:25.374 \\        
2013-03-17 & N2298   &  2013-03-17T23:37:51.060 & 2013-03-17T23:38:51.658 &  2013-03-17T23:39:45.184 &  2013-03-17T23:40:31.979 \\        
2013-03-17 & LeoI    &  2013-03-17T23:46:25.661 & 2013-03-17T23:47:22.218 &  2013-03-17T23:48:13.174 &  2013-03-17T23:48:58.659 \\        
2013-03-17 & E7      &  2013-03-18T10:07:18.278 & 2013-03-18T10:08:15.224 &  2013-03-18T10:09:06.720 &  2013-03-18T10:09:52.956 \\        
2013-04-01 & N2437   &  2013-04-01T23:41:01.172 & 2013-04-01T23:41:57.648 &  2013-04-01T23:42:48.813 &  2013-04-01T23:43:34.128 \\        
2013-04-01 & E5      &  2013-04-01T23:48:53.414 & 2013-04-01T23:49:50.642 &  2013-04-01T23:50:42.027 &  2013-04-01T23:51:28.462 \\        
2013-04-01 & N2818   &  2013-04-02T04:01:05.166 & 2013-04-02T04:02:02.762 &  2013-04-02T04:02:54.838 &  2013-04-02T04:03:41.083 \\        
2013-04-02 & N2437   &  2013-04-02T23:46:57.255 & 2013-04-02T23:47:54.762 &  2013-04-02T23:48:46.278 &  2013-04-02T23:49:31.803 \\        
2013-04-02 & E5      &  2013-04-03T00:01:52.088 & 2013-04-03T00:02:53.324 &  2013-04-03T00:03:43.989 &  2013-04-03T00:04:29.674 \\        
2013-04-02 & N2818   &  2013-04-03T04:10:05.557 & 2013-04-03T04:11:03.454 &  2013-04-03T04:11:56.079 &  2013-04-03T04:12:42.765 \\        
2013-04-11 & N2437   &  2013-04-11T23:28:10.963 & 2013-04-11T23:29:07.989 &  2013-04-11T23:29:59.225 &  2013-04-11T23:30:45.010 \\        
2013-04-11 & E5      &  2013-04-11T23:36:55.921 & 2013-04-11T23:37:52.268 &  2013-04-11T23:38:43.694 &  2013-04-11T23:39:29.329 \\        
2013-04-11 & L104    &  2013-04-12T06:01:31.694 & 2013-04-12T06:02:31.300 &  2013-04-12T06:03:23.616 &  2013-04-12T06:04:10.372 \\        
2013-04-12 & N2437   &  2013-04-12T23:13:55.704 & 2013-04-12T23:14:52.810 &  2013-04-12T23:15:44.256 &  2013-04-12T23:16:30.751 \\        
2013-04-12 & E5      &  2013-04-12T23:22:58.885 & 2013-04-12T23:23:55.211 &  2013-04-12T23:24:45.617 &  2013-04-12T23:25:32.472 \\        
2013-04-12 & N2818   &  2013-04-13T02:19:36.408 & 2013-04-13T02:20:34.395 &  2013-04-13T02:21:26.570 &  2013-04-13T02:22:12.556 \\        
2013-04-12 & E5      &  2013-04-13T03:59:33.650 & 2013-04-13T04:00:30.827 &  2013-04-13T04:01:22.243 &  2013-04-13T04:02:08.007 \\        
2013-04-12 & N5139   &  2013-04-13T05:50:16.948 & 2013-04-13T05:51:14.274 &  2013-04-13T05:52:05.731 &  2013-04-13T05:52:51.955 \\        
2013-04-12 & MarkA   &  2013-04-13T09:22:57.299 & 2013-04-13T09:23:55.405 &  2013-04-13T09:24:47.811 &  2013-04-13T09:25:34.046 \\        
2013-04-13 & N2818   &  2013-04-13T23:20:54.713 & 2013-04-13T23:21:51.769 &  2013-04-13T23:22:43.515 &  2013-04-13T23:23:31.401 \\        
2013-04-13 & L101    &  2013-04-13T23:27:46.890 & 2013-04-13T23:28:45.227 &  2013-04-13T23:29:37.303 &  2013-04-13T23:30:23.658 \\        
2013-04-13 & E5      &  2013-04-13T23:34:27.046 & 2013-04-13T23:35:24.083 &  2013-04-13T23:36:14.909 &  2013-04-13T23:36:59.524 \\        
2013-04-13 & N2818   &  2013-04-14T05:21:20.757 & 2013-04-14T05:22:18.134 &  2013-04-14T05:23:09.359 &  2013-04-14T05:23:55.485 \\        
2013-04-13 & MarkA   &  2013-04-14T09:24:13.086 & 2013-04-14T09:25:11.023 &  2013-04-14T09:26:02.879 &  2013-04-14T09:26:49.654 \\        
2013-05-06 & N2818   &  2013-05-06T23:30:34.324 & 2013-05-06T23:31:35.670 &  2013-05-06T23:32:28.976 &  2013-05-06T23:33:17.502 \\        
2013-05-06 & PG1323  &  2013-05-06T23:37:43.993 & 2013-05-06T23:38:43.400 &  2013-05-06T23:39:36.016 &  2013-05-06T23:40:22.301 \\        
2013-05-06 & E7      &  2013-05-07T04:44:11.329 & 2013-05-07T04:45:11.516 &  2013-05-07T04:46:05.412 &  2013-05-07T04:46:54.028 \\        
2013-05-06 & E7      &  2013-05-07T09:44:42.317 & 2013-05-07T09:45:42.644 &  2013-05-07T09:46:37.430 &  2013-05-07T09:47:26.115 \\        
2013-05-09 & N2818   &  2013-05-09T23:09:18.695 & 2013-05-09T23:10:18.092 &  2013-05-09T23:11:12.137 &  2013-05-09T23:12:00.153 \\        
2013-05-09 & L98     &  2013-05-09T23:16:32.503 & 2013-05-09T23:17:32.410 &  2013-05-09T23:18:25.066 &  2013-05-09T23:19:12.512 \\        
2013-05-09 & E7      &  2013-05-10T06:28:32.743 & 2013-05-10T06:29:33.239 &  2013-05-10T06:30:27.096 &  2013-05-10T06:31:17.801 \\        
2013-05-09 & E7      &  2013-05-10T09:27:54.353 & 2013-05-10T09:28:54.239 &  2013-05-10T09:29:48.205 &  2013-05-10T09:30:36.921 \\        
2013-05-11 & N2818   &  2013-05-11T22:59:15.069 & 2013-05-11T23:00:14.996 &  2013-05-11T23:01:09.222 &  2013-05-11T23:01:57.747 \\        
2013-05-11 & L98     &  2013-05-11T23:06:25.488 & 2013-05-11T23:07:23.995 &  2013-05-11T23:08:18.021 &  2013-05-11T23:09:09.067 \\        
2013-05-11 & E7      &  2013-05-12T09:59:28.061 & 2013-05-12T04:52:21.060 &  2013-05-12T04:53:23.328 &  2013-05-12T04:54:15.054 \\        
2013-05-11 & E7      &         --               & 2013-05-12T10:00:27.938 &  2013-05-12T10:01:21.553 &  2013-05-12T10:02:08.908 \\        
2013-05-26 & N2818   &  2013-05-26T22:50:30.106 & 2013-05-26T22:51:28.873 &  2013-05-26T22:52:21.088 &  2013-05-26T22:53:07.393 \\        
2013-05-26 & L98     &  2013-05-26T22:58:23.418 & 2013-05-26T22:59:19.995 &  2013-05-26T23:00:11.430 &  2013-05-26T23:00:57.816 \\        
2013-05-27 & N2818   &  2013-05-27T22:53:14.066 & 2013-05-27T22:54:11.412 &  2013-05-27T22:55:03.838 &  2013-05-27T22:55:53.354 \\     
2013-05-27 & L98     &  2013-05-27T23:01:38.672 & 2013-05-27T23:02:35.479 &  2013-05-27T23:03:27.094 &  2013-05-27T23:04:12.659 \\ 
\label{tab:log}
\end{longtable}
\end{small}
\begin{minipage}{18.cm}
Notes -- The filenames are defined by the execution date and time
(down to the 1~ms level). Here, only images from the FORS2 detector 1 are
shown for reasons of clarity. The corresponding image names from detector
2 are obtained by adding 1~ms to the execution time. Files can be
downloaded from the ESO archive: 
  http://archive.eso.org/eso/eso\_archive\_main.html.
The number of images per chip are: 271 (Filter $B$ ), 274 (Filter $V$), 272 (Filter $R$), and 273 (Filter $I$).
\end{minipage}
\label{lastpage}
\end{onecolumn}
\twocolumn


\end{document}